\documentclass[iop]{emulateapj}

\usepackage{verbatim}
\usepackage{url}

\def\E{{\cal E}}
\def\F{{\cal F}}

\newcommand{\simgt}{\,\hbox{\lower0.6ex\hbox{$\sim$}\llap{\raise0.6ex\hbox{$>$}}}\,}
\newcommand{\simlt}{\,\hbox{\lower0.6ex\hbox{$\sim$}\llap{\raise0.6ex\hbox{$<$}}}\,}

\begin{document}

\title{A Parameter Space Exploration of Galaxy Cluster Mergers I: Gas Mixing and the Generation of Cluster Entropy}

\author{J. A. ZuHone}
\affil{Harvard-Smithsonian Center for Astrophysics, Cambridge, MA 02138}
\begin{abstract}
We present a high-resolution set of adiabatic binary galaxy cluster merger simulations using FLASH. These are the highest-resolution simulations to date of such mergers using an AMR grid-based code with Eulerian hydrodynamics. In this first paper in a series we investigate the effects of merging on the entropy of the hot intracluster gas, specifically with regard to the ability of merging to heat and disrupt cluster ``cool-cores.'' We find, in line with recent works, that the effect of fluid instabilities that are well-resolved in grid-based codes is to significantly mix the gases of the two clusters and to significantly increase the entropy of the gas of the final merger remnant. This result is characteristic of mergers over a range of initial mass ratio and impact parameter. In line with this, we find that the kinetic energy associated with random motions is higher in our merger remnants which have high entropy floors, indicating the motions have efficiently mixed the gas and heated the cluster core with gas of initially high entropy. We examine the implications of this result for the maintenance of high entropy floors in the centers of galaxy clusters and the derivation of the properties of dark matter from the thermal properties of the X-ray emitting gas.
\end{abstract}


\section{Introduction}\label{sec:intro}

The growth of structure in the universe is thought to proceed in a bottom-up fashion, with smaller structures merging into larger structures \citep[e.g.,][]{whi93}. The largest structures that have formed to date via this process are clusters of galaxies. Clusters of galaxies are important astrophysical objects, both in their own right and for their use as cosmological probes. Galaxy clusters contain a good representation of the different forms and kinds of matter within the universe. The bulk of the mass in clusters of galaxies is comprised of dark matter \citep{zwy37,bah77}, which is believed to be largely collisionless. The bulk of the baryonic material in clusters of galaxies is comprised of a hot diffuse plasma called the intracluster medium (ICM) \citep[e.g.,][]{sar88}. The remaining (and smallest) component of mass is that of the galaxies themselves. Galaxy clusters provide an opportunity to witness the interplay between these different kinds and forms of matter in close quarters. Additionally, clusters have become very useful as probes of cosmology. The number density of clusters of a given mass at a given redshift is a sensitive function of the parameters of the standard big bang cosmological model. Accurate measurements of the density and temperature structure of galaxy clusters enable determination of cluster masses (under the assumptions of spherical symmetry and hydrostatic equilibrium) which in turn helps to constrain the values of cosmological parameters \citep{kita96,voit05}.

Extensive study of both of these aspects of astrophysics with galaxy clusters requires state-of-the-art observations and simulations. The latest generation of X-ray telescopes ({\it Chandra}, {\it XMM-Newton}) has provided high-resolution observations of the ICM that have made it possible to probe sensitively the density, temperature, and metallicity structure of the gas. These observations have made possible the discovery of a myriad of features in the ICM such as shocks, cold fronts, and bubbles, to name a few. In addition, such precise measurements have made hydrostatic estimates of cluster masses a useful tool for constraining cosmological parameters on a par with other methods \citep{vik09b}. In tandem with these advances in observation, astrophysical simulation codes have taken advantage of large-scale computing resources and innovative algorithms to model clusters of galaxies with unprecedented resolution and with the ability to include detailed models of the relevant physical processes occurring in clusters. 

One of the most fundamental of these physical processes that shapes the life of clusters is merging. The merging process is responsible for forming the clusters and greatly influences the state of the matter within a cluster throughout its lifetime. Mergers drive shocks into the intracluster medium, heating and compressing the gas and driving turbulent motions. As a result of merging the kinetic properties of the dark matter are also significantly altered. Mergers also drive galaxy activity, producing new bursts of star formation as the gas is compressed, or in some cases possibly inhibiting it \citep{fuj99,bek03}. Many observed clusters of galaxies show signs of current or recent merging. Two prominent examples of recent note are 1E~0657-56 (the ``Bullet Cluster'') \citep{mar02,clo06}, and Abell 520 \citep{mar05,mah07}. Explaining the myriad of observable consequences of mergers using simulations is an active area of research.

From the perspective of simulations, there are two ways to study the merging process. The first is by studying mergers that occur in the context of a cosmological simulation, ``out in the wild.'' The advantage of this approach is that it captures as accurately as possible the merging process as it occurs in the real universe, with many mergers of many different mass ratios and impact parameters going on at the same time, in the context of the overall cosmological expansion. The second computational approach to studying mergers is by focusing on particular merger scenarios in isolation via simulations of idealized cluster mergers. These simulations involve the setup of two or more idealized, spherically symmetric clusters in a localized setting and with an initial orbital trajectory. Though this method is useful for studying individual actual mergers as seen in observations \citep[e.g.][]{tak06,spr07,aka08,mas08,ran08,zuh09a,zuh09b}, an additional application has been to explore a space over parameters such as the mass ratio and initial impact parameter of the progenitor clusters or by varying the physics of the cluster ICM or dark matter. 

Several previous investigations of merger parameter spaces \citep[e.g.][]{roe97,ric01,gom02,rit02,mcc07,poo06,poo07,poo08,tak10} have emphasized the effects of merging on cluster morphology, the thermodynamics of the gas, and observables. We build on this work with a new set of simulations. The set of simulations presented here is the highest-resolution parameter study of idealized binary cluster mergers performed with an AMR-based PPM code to date, with a resolution 3-4 times higher than the most recent investigation \citep{ric01} based on such a formalism. 

In this first of two papers, we use these simulations to examine the effect of merging on the entropy of the cluster gas. Observational surveys of clusters have shown that although in nearly all clusters the entropy (defined here as $S \equiv k_BTn_e^{-2/3}$) follows a power-law profile with radius, the central core entropies of clusters vary \citep{llo00,pon03,voit05c,fal07,mor07,cav09,pra09}. These observed entropy profiles are well-fit by an entropy profile of the form
\begin{equation}
S(r) = S_0 + S_1{\left(r \over {0.1r_{200}}\right)}^\alpha
\label{eqn:entropy}
\end{equation}
Where $S_0$ is the value for the core entropy and $\alpha \sim 1.0-1.3$. In particular, \citet{cav09} and \citet{pra09} showed that clusters exhibit a bimodal distribution of central core entropies. \citet{cav09} found a low-entropy peak at $S_0 \sim 15~{\rm keV~cm}^2$, a high-entropy peak at $S_0 \sim 150~{\rm keV~cm}^2$, and very few clusters with core entropies $S_0 \sim 30-50~{\rm keV~cm}^2$. Similarly, \citet{pra09} found peaks at $S_0 \sim 3$ and $S_0 \sim 75~{\rm keV~cm}^2$. These core entropies are higher than what would be expected from a simple ``cooling flow'' scenario \citep[e.g.][]{fab77,fab94}, where the core gas is allowed to cool and compress unabated until it turns into stars or molecular clouds. High-resolution {\it XMM-Newton}\/ spectroscopy \citep{pet01,pet03} and {\it Chandra}\/ spectral imaging \citep{dav01} have shown that there is indeed little gas below $T\simeq 1$ keV in the cores of clusters with some of the highest cooling rates.

Since cooling via X-ray radiation is directly observed, but the expected central entropies from unabated cooling is not, clusters with high entropy floors require a heating mechanism.  Proposed candidates for heating of cluster cores include magnetic field reconnection \citep{sok90}, thermal conduction due to electron collisions (e.g., Narayan \& Medvedev 2001; Fabian, Voigt, \& Morris 2002; Zakamska \& Narayan 2003) and turbulent conduction (e.g., Cho et al.\ 2003; Voigt \& Fabian 2004), and heating by cosmic rays (e.g., Inoue \& Sasaki 2001, Colafrancesco \&
Marchegiani 2008); a recent review can be found in \citet{petersonfabian06}.
The currently favored mechanism is heating by the central AGN (e.g.,
B\"ohringer et al.\ 1993; Binney \& Tabor 1995; McNamara et al.\ 2001, 2005;
Fabian et al.\ 2006; Forman et al.\ 2007; for a recent review see McNamara
\& Nulsen 2007). The AGN explosions blow the bubbles in the ICM
and inject energy into the ICM in the form of relativistic particles as well
as mechanical energy.  However, the precise mechanism by which this energy
heats the central ICM is still unclear. A fine balance between AGN
explosions and cooling is required to avoid the complete blow-up of the cool
cores, which gave rise to ``feedback'' models, where the cooling flow itself
feeds the AGN. 

Thermal conduction is another particularly attractive idea, because it
taps the vast reservoir of thermal energy in the gas just outside the cool
core, while automatically ensuring that the core will not be overheated,
since the heat influx decreases with diminishing temperature gradient. The
classic plasma conductivity via Coulomb collisions was shown to be
insufficient even at its full Spitzer value (e.g., Zakamska \& Narayan
2003). It has a strong temperature dependence and decreases right where it
is most needed, and tangled magnetic fields should further suppress it
(as was indeed observed outside the cool cores, Markevitch et al.\ 2003a). Additionally, magnetohydrodynamic instabilities such as the heat-flux-driven buoyancy instability (HBI) may suppress heat conduction to cluster cores from the cluster outskirts altogether by aligning magnetic field lines perpendicular to the radial temperature gradients in clusters \citep{qua08, par08, bog09, par09}, though recent works have shown that modest amounts of turbulence in cluster cores may work to keep the fields randomized and sustain some conduction of heat to the cluster core \citep{rus10, par10}. Though the effects of AGN and conduction may account for the lack of systems with extremely low central cooling times and entropies, it is also unclear whether or not either effect can explain the highest entropy floors seen in some clusters, which can reach central entropy values of $S_0 \sim 100$ or higher.

Previous investigations of cluster mergers \citep[e.g.,][]{rit02, mcc07, poo08} have indicated that so-called cluster ``cool cores'' (characterized by temperature inversions, high gas densities, and low central entropies and cooling times) are largely resilient to the heating that occurs as a result of merging. The central entropies and cooling times increase for a short time after the core passage, but the heating is not enough to offset cooling, which quickly reestablishes a cool core within a few Gyr \citep{poo08}. \citet{mcc07}, in a suite of adiabatic idealized merger simulations, determined that the ICM is heated in two major stages: the first is when the shocks are generated at the initial core passage and the second is at a later stage when material is shock-heated as it recollapses onto the merger remnant. Though the overall entropy of the merger remnant was increased with respect to the progenitor systems consistent with the self-similar prediction, the power-law behavior of the entropy profile was maintained all the way down into the central core (see Figure 5 of that paper).  It is notable that these SPH merger simulations also resulted in little mixing of the cluster gases, since such mixing could provide a source of heating to cluster cores if the gases are of very different entropies. 

It is known that the various methods for solving the equations of hydrodynamics result in different degrees of mixing. While previous investigations of mixing due to merging \citep[e.g.,][]{rit02, poo08} have employed the Lagrangian ``smoothed-particle hydrodynamics'' (SPH) formulation for solving the Euler equations, our simulations have been performed using an Eulerian PPM formulation. The two different approaches to hydrodynamics differ in their ability to appropriately model turbulence and the onset of fluid instabilities \citep[see, e.g.][]{dol05,age07,wad08}. The amount of mixing of the gas is also dependent on the precise nature of the physics of the ICM, as the presence of magnetic fields, turbulence, and dissipative processes will affect the mixing of the gas. The extent to which the ICM mixes as a result of mergers and its effect on the thermodynamic properties of the cluster gas is therefore dependent the physical processes operating in the ICM as well as the chosen algorithm for solving the hydrodynamic equations. Therefore, it is important as a baseline to characterize the mixing process and the effects of such mixing on the thermodynamics of the cluster gas in the simplest model for ICM, that of an inviscid, unmagnetized gas, in the context of an Eulerian, PPM-based hydrodynamics code. Doing so is the goal of this paper. 

This paper is organized as follows. In Section \ref{sec:sims} we describe the simulation method, setup, and rationale for the initial conditions. In Section \ref{sec:results} we present the results of the simulations, focusing on the degree of mixing of the intracluster gas components and the alteration of the thermodynamic state of the cluster gas as a result. In Section \ref{sec:disc} we discuss the implications of these results for different approaches to hydrodynamics for galaxy cluster simulations, the survival of cool cores in cluster mergers, and the use of X-ray observables to determine properties of the dark matter. In Section \ref{sec:conc} we summarize this work and suggest avenues for future investigations. Throughout this paper we assume a spatially flat $\Lambda$CDM cosmology with $h = 0.7$, $\Omega_{\rm m} = 0.3$, and $\Omega_b = 0.02h^{-2}$. 

\section{Simulations}\label{sec:sims}

\subsection{Method}\label{sec:method}

We performed our simulations using FLASH, a parallel hydrodynamics/$N$-body astrophysical simulation code developed at the Center for Astrophysical Thermonuclear Flashes at the University of Chicago \citep{fry00}. FLASH uses adaptive mesh refinement (AMR), a technique that places higher resolution elements of the grid only where they are needed. In our case we are interested in capturing sharp ICM features like shocks and ``cold fronts'' accurately, as well as resolving the inner cores of the cluster dark matter halos. As such it is particularly important to be able to resolve the grid adequately in these regions. AMR allows us to do so without needing to have the whole grid at the same resolution. FLASH solves the Euler equations of hydrodynamics using the Piecewise-Parabolic Method (PPM) of \citet{col84}, which is ideally suited for capturing shocks. FLASH also includes an $N$-body module which uses the particle-mesh method to solve for the forces on gravitating particles. The gravitational potential is computed using a multigrid solver included with FLASH \citep{ric08}.

\begin{deluxetable*}{ccccccccc}
\tabletypesize{\scriptsize}
\tablecaption{Initial Cluster Parameters\label{tab:ICs}}
\tablewidth{0pt}
\tablehead{
\colhead{Cluster} & \colhead{$M_{200}$} & \colhead{$r_{200}$} & \colhead{$c_{200}$} & \colhead{$f_g$} & \colhead{$T_X$} & \colhead{$S_0$} & \colhead{$S_1$} & \colhead{} \\ 
\colhead{} & \colhead{($M_{\odot}$)} & \colhead{(kpc)} & \colhead{} & \colhead{} & \colhead{(keV)} & \colhead{(keV cm$^{2}$)} & \colhead{(keV cm$^{2}$)} & \colhead{$N_p$}
}
\startdata
C1 & $6 \times 10^{14}$ & 1552.25 & 4.5 & 0.1056 & 4.97 & 9.62 & 192.40 & 5,000,000 \\
C2 & $2 \times 10^{14}$ & 1076.27 & 4.7 & 0.0879 & 2.42 & 5.08 & 101.60 & 1,684,119 \\
C3 & $6 \times 10^{13}$ & 720.49  & 5.1 & 0.0686 & 1.10 & 2.73 &  54.60 & 513,137 \\
\enddata
\end{deluxetable*}

\begin{table}[thdp]
\caption{Initial Merger Parameters\label{tab:SimGrid}}
\begin{center}
\begin{tabular}{ccccc}
\hline
\hline
Simulation & $R$ & $b$ (kpc) & $b/r_{200,1}$ & $\lambda$ \\
\hline
S1 & 1:1  & 0.0    & 0.0 & 0.0 \\
S2 & 1:1  & 464.43 & 0.3 & 0.013 \\
S3 & 1:1  & 932.28 & 0.6 & 0.026 \\
S4 & 1:3  & 0.0    & 0.0 & 0.0 \\
S5 & 1:3  & 464.43 & 0.3 & 0.009 \\
S6 & 1:3  & 932.28 & 0.6 & 0.018 \\
S7 & 1:10 & 0.0    & 0.0 & 0.0 \\
S8 & 1:10 & 464.43 & 0.3 & 0.003 \\
S9 & 1:10 & 932.28 & 0.6 & 0.006 \\
\hline
\end{tabular}
\end{center}
\end{table}

\subsection{Initial Conditions}\label{sec:ICs}
For these idealized cluster merger simulations, we choose initial conditions based on information gleaned from cosmological simulations and cluster observations. Since we are interested in the effects of merging on relaxed systems, our initial clusters are configured to be systems consistent with observed relaxed clusters and cluster scaling relations. For the temperature normalization, we choose clusters that lie along the $M_{500}-T_X$ relation of \citet{vik09a}, which is
\begin{equation}
M_{500} = M_0E(z)^{-1}(T_X/5~{\rm keV})^\alpha
\end{equation}
with $M_0 = 3.02 \times 10^{14} h^{-1} M_\odot$ and $\alpha = 1.53$. The dependence on the redshift $z$ is encapsulated in the function $E(z) \equiv H(z)/H_0$. For the total gas mass within $r_{500}$, we use the relation \citep[also from][]{vik09a}
\begin{equation}
f_g (h/0.72)^{1.5} = 0.125 + 0.037~{\rm log_{10}} M_{15}
\end{equation}
where $M_{15}$ is the cluster total mass, $M_{500}$, in units of $10^{15} h^{-1} M_\odot$. 

For our grid of simulations we have chosen a set of masses representative of galaxy clusters and groups. The chosen masses are given in terms of $M_{200}$, corresponding to the mass within which the average density is $200\rho_{\rm crit}(z)$. The mass $M_{200}$ is related to the radius $r_{200}$ via
\begin{equation}
M_{200} = {4\pi \over 3}{[200\rho_{\rm crit}(z)]}r_{200}^3
\end{equation}
$M_{200}$ is calculated from $M_{500}$ from the assumed value of $c_{200}$ and the assumed form of the density profile.

For the total mass distribution (gas and dark matter) we choose an NFW \citep{NFW97} profile:
\begin{equation}
\rho_{\rm tot}(r) = {\rho_s \over {r/r_s{(1+r/r_s)}^2}}.
\end{equation}
Where the scale density $\rho_s$ and scale radius $r_s$ are determined by the following constraints:
\begin{eqnarray}
r_s &=& \frac{r_{200}}{c_{200}} \\
\rho_s &=& \frac{200}{3}c_{200}^3\rho_{\rm crit}(z){\left[{\rm log}(1+r/r_s) - {r/r_s \over {1 + r/r_s}}\right]}^{-1}
\end{eqnarray}
The NFW functional form is carried out to $r_{200}$. For radii $r > r_{200}$ the mass density follows an exponential profile:
\begin{equation}
\rho_{\rm tot}(r) = {\rho_{\rm s} \over {c_{200}(1+c_{200})^2}}{\left({r \over r_{200}}\right)^\kappa}{\rm exp}{\left(-{r-r_{200} \over r_{\rm decay}}\right)}
\end{equation}
where $\kappa$ is set such that the density and its first derivative are continuous at $r = r_{200}$ and $r_{\rm decay} = 0.1r_{200}$.

The total mass distribution uniquely determines the gravitational potential, and via the constraint of hydrostatic equilibrium of the gas within this potential, the pressure profile is uniquely determined. To determine the physical quantities of density and temperature, more information is needed. Since we are here interested in the effect of mergers on the entropy profile of clusters, we make use of Equation \ref{eqn:entropy} to set the gas properties in our initial systems. We start off with small core entropies ($S_0$, see Table \ref{tab:ICs}) in order to make our models consistent with relaxed, ``cool-core'' galaxy clusters, and set $\alpha = 1.1$. With this information in hand, we write the condition for hydrostatic equilibrium as follows:
\begin{eqnarray}
{dP \over dr} &=& -\rho_g{d\Phi \over dr} \\
{k_B \over {\mu}m_p}{d({\rho_g}T) \over dr} &=& -\rho_g{d\Phi \over dr} \\
{k_B \over {\mu}m_p}{d \over dr}\left[{T{\left(T \over S\right)}^{3/2}}\right] &=& -\left({T \over S}\right)^{3/2}{d\Phi \over dr}
\end{eqnarray}
Following other authors \citep{poo06}, we solve this differential equation by imposing $T(r_{200}) = \frac{1}{2}T_{200}$, where
\begin{equation}
k_BT_{200} \equiv {GM_{200}{\mu}m_p \over 2r_{200}}
\end{equation}
is the so-called ``virial temperature'' of the cluster. With the temperature profile in hand we may use it and the entropy to determine the gas density profile, and so also determine the dark matter density profile via $\rho_{\rm DM} = \rho_{\rm tot} - \rho_g$.  

After the radial profiles are determined it remains to set up the distribution of positions and velocities for the dark matter particles. Here we follow the procedure outlined in \citet{kaz06}. For the particle positions a random deviate $u$ is uniformly sampled in the range [0,1] and the function $u = M_{\rm DM}(r)/M_{\rm DM}(r_{\rm max})$ is inverted to give the radius of the particle from the center of the halo. For the particle velocities, the procedure is less trivial. Many previous investigations have made use of the ``local Maxwellian approximation.'' In this procedure at a given radius the particle velocity is drawn from a Maxwellian distribution with dispersion $\sigma^2(r)$, where the latter quantity has been derived from solving the Jeans equation \citep{bin87}. It has been shown that this approach is not sufficient to accurately represent the velocity distribution functions of dark matter halos with a central cusp such as the NFW profile \citep{kaz04}. To accurately realize particle velocities, we choose to directly calculate the distribution function via the Eddington formula \citep{edd16}:
\begin{equation}
\F(\E) = \frac{1}{\sqrt{8}\pi^2}\left[\int^\E_0{d^2\rho \over d\Psi^2}{d\Psi \over \sqrt{\E - \Psi}} + \frac{1}{\sqrt{\E}}\left({d\rho \over d\Psi}\right)_{\Psi=0} \right]
\end{equation}
where $\Psi = -\Phi$ is the relative potential and $\E = \Psi - \frac{1}{2}v^2$ is the relative energy of the particle. We tabulate the function $\F$ in intervals of $\E$ interpolate to solve for the distribution function at a given energy. Particle speeds are chosen from this distribution function using the acceptance-rejection method. Once particle radii and speeds are determined, positions and velocities are determined by choosing random unit vectors in $\Re^3$. The simulation parameters for each simulated cluster are given in Table \ref{tab:ICs}.

\begin{figure*}
\begin{center}
\plotone{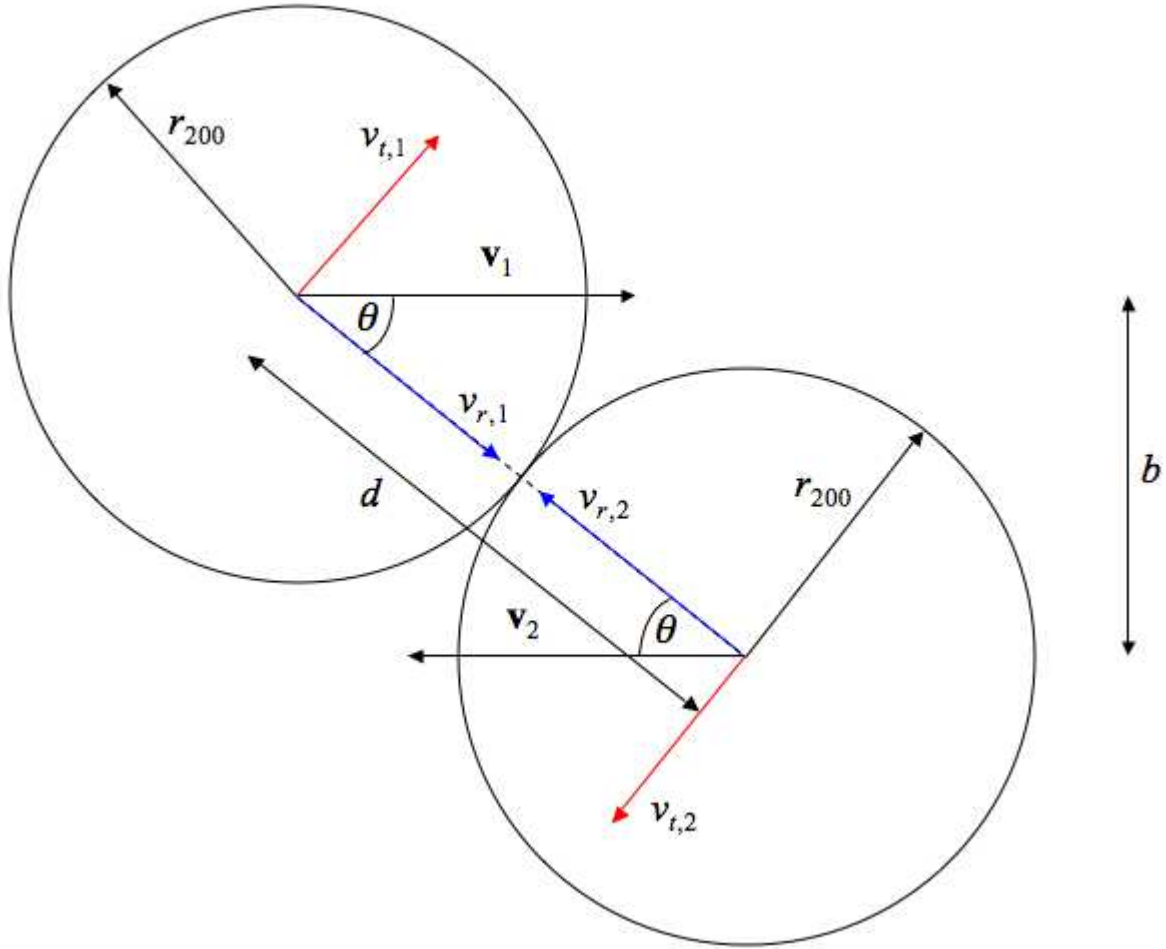}
\caption{Schematic representation of initial merger geometry.\label{fig:schematic}}
\end{center}
\end{figure*}

For all of the simulations, we set up the two clusters within a cubical computational domain of width $L = 10h^{-1}$ Mpc on a side. The distance between the cluster centers is given by the sum of their respective $r_{200}$. \citet{vit02} demonstrated from cosmological simulations that the average infall velocity for merging clusters is $v_{\rm in}(r_{\rm vir}) = 1.1V_c$, where $V_c = \sqrt{GM_{\rm vir}/r_{\rm vir}}$ is the circular velocity at the virial radius $r_{\rm vir}$ for the primary cluster. For all of our simulations, this is chosen as the initial relative velocity ($v_{\rm in} \approx 1200$ km/s). The clusters are initialized in the $x-y$ coordinate plane at $z$ = 0.

For this investigation we explore a parameter space in mass ratio of the clusters $R$ and in initial impact parameter $b$. The set of simulations in this parameter space is detailed in Table \ref{tab:SimGrid}. Taking the virial mass of our primary system as $M_{200} = 6 \times 10^{14} M_\odot$, we perform mergers involving virial mass ratios of $R$ = 1:1, 1:3, and 1:10. We also vary the impact parameter $b$ for our runs. \citet{vit02} also demonstrated that for merger encounters the average tangential velocity is $v_\perp \simeq 0.4V_C$. Instead of splitting our initial velocity into radial and tangential components, we choose to initialize our systems with impact parameters that result in relative tangential velocities consistent with the results of \citet{vit02}. Since the velocity $v_\perp$ represents an average velocity, we select a range of tangential velocities consistent with this relation. Figure \ref{fig:schematic} shows the initial merger geometry and how the tangential and radial components of the velocity are related to the initial choice of impact parameter. Finally, for each merger, we detail the angular momentum parameter $\lambda$ = $J|E|^{1/2}/GM^{5/2}$ \citep{pee69} as a measure of the angular momentum of each simulation in Table \ref{tab:SimGrid}.

The highest AMR resolution of this set of simulations is ${\Delta}x$ = 5$h^{-1}$~kpc, which is $\sim$3-4 times higher than the resolution in \citet{ric01}. The number of dark matter particles in each simulation is given in Table \ref{tab:ICs}. 

\begin{figure*}
\begin{center}
\plotone{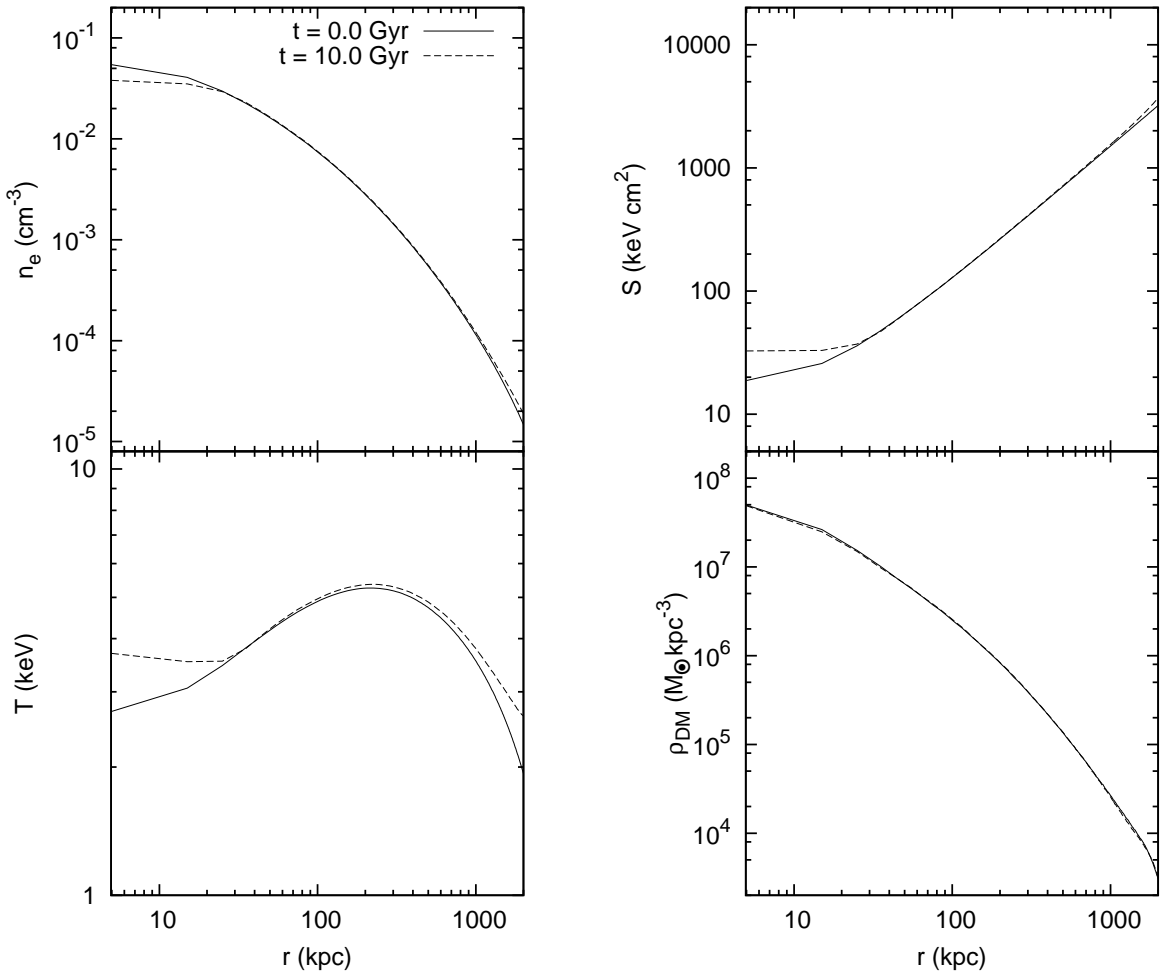}
\caption{Radial profiles of the electron number density, gas temperature, gas entropy, and dark matter density at the epochs $t$ = 0.0~Gyr and $t$ = 10.0~Gyr of a simulation of a single galaxy cluster in isolation.\label{fig:stability}}
\end{center}
\end{figure*}

Since we will be concerned mainly with the effects of binary merging on cluster cores, it is important to ensure that the dramatic changes that will be seen to occur in the simulations are the result of merging and not evolution of the isolated primary cluster initial conditions. To this end we have run a single-cluster test, evolving the primary cluster in isolation for the same 10~Gyr as our merger simulations. Figure \ref{fig:stability} shows the radial profiles of gas density, gas temperature, dark matter density, and gas entropy at the two epochs $t$ = 0.0~Gyr and $t$ = 10.0~Gyr. There is some moderate evolution in the profiles, particularly in the innermost few zones of the cluster core and in the cluster outskirts. The former is a consequence of force smoothing, (a known numerical effect due to the inability to resolve the gravitational force on scales smaller than the grid resolution). It is important to note that though there is some increase in the central entropy due to this effect, this increase is insignificant when compared to the increase that results from the mergers, as will be seen below. The evolution in the cluster outskirts is a consequence of the tapering off of the cluster profiles beyond $r_{200}$, but has a very small effect on the profiles only in these regions and does not affect the cluster core. 

\section{Results}\label{sec:results}

\subsection{Qualitative Description: Entropy Maps}

Over the course of a galaxy cluster merger, entropy is generated via processes that dissipate kinetic energy into heat energy, such as shocks and turbulence, and processes that bring into contact gases of different entropies and mix them. Here we show maps of the ICM entropy throughout the merger to get a broader picture of the major entropy-generating events during cluster mergers. Each set of maps are taken from slices through the $z = 0$ coordinate plane, the plane of the cluster centers' mutual orbit. 

\subsubsection{Equal-Mass Mergers}

\begin{figure*}
\begin{center}
\plotone{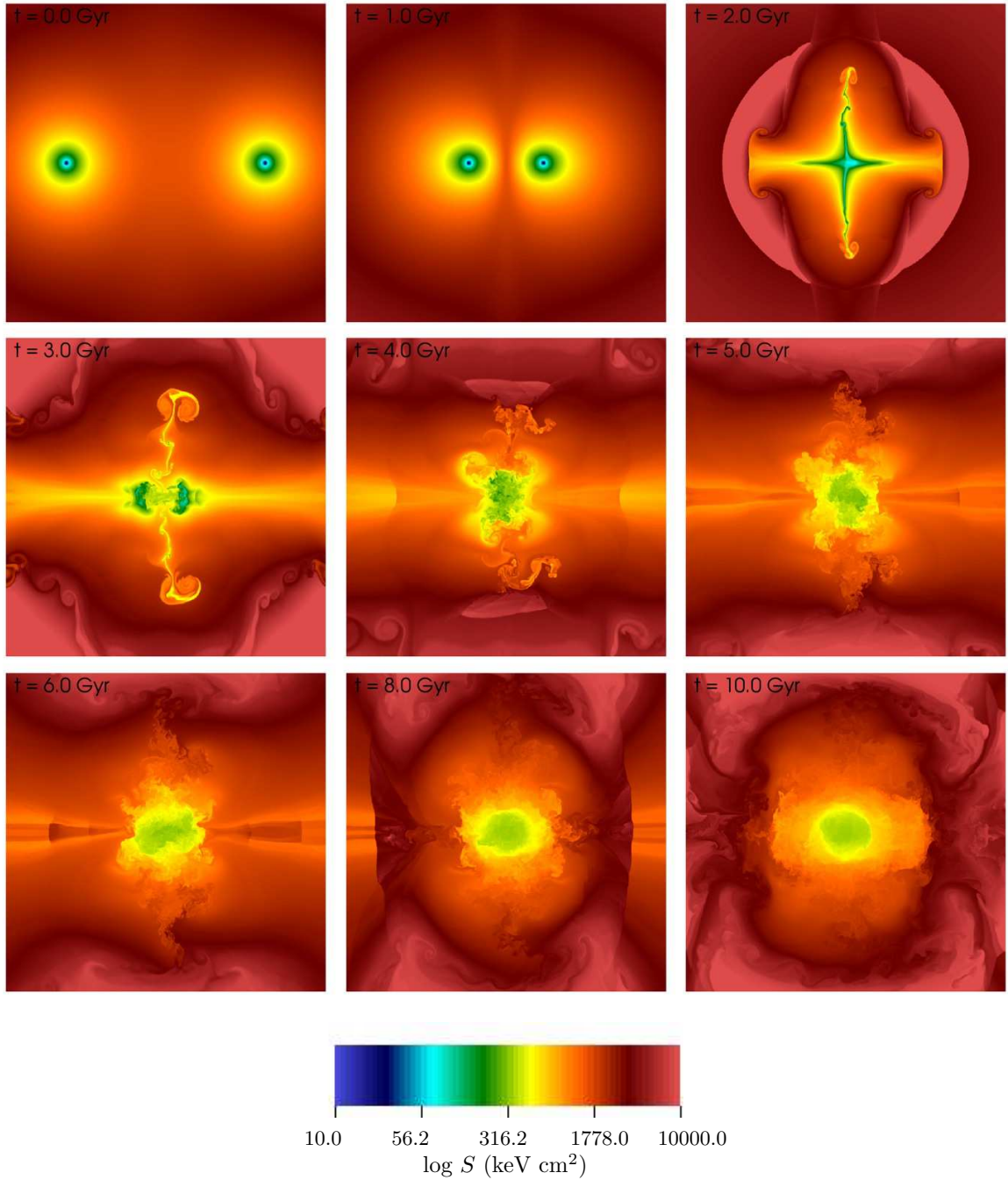}
\caption{Slices of entropy through the $z = 0$ coordinate plane for simulation S1 ($R$ = 1:1, $b$ = 0~kpc). The epochs shown are $t$ = 0.0, 1.0, 2.0, 3.0, 4.0, 5.0, 6.0, 8.0, and 10.0~Gyr. Each panel is 5~Mpc on a side.\label{fig:entr_S1}}
\end{center}
\end{figure*}

For our equal-mass mergers, there is no distinction between the primary and the secondary system, so either cluster could be chosen as the primary without any loss of specificity (in all discussions that follow of the equal-mass mergers, it is assumed that the cluster whose center of mass has $x < 0$ is the primary). In the head-on case (simulation S1, Figure \ref{fig:entr_S1}), the cluster cores approach each other, driving equal and opposite shocks into the gas. As the dark matter cores pass each other, the shocked gas in between the clusters forms a flattened low-entropy ``pancake''-like structure ($t \approx$ 1.6~Gyr). As the shocks depart, this gas expands and cools adiabatically for a time, but eventually falls back toward the center of the system. The gas at high radii perpendicular to the merger axis develops Rayleigh-Taylor instabilities and is quickly mixed in with the surrounding gas. Meanwhile, the central gas is dragged back and forth by the oscillating dark matter cores ($t \approx$ 3-4~Gyr). This process drives more shocks into the gas, heating it further. Eventually the gas forms a core with a roughly constant entropy of $S \sim $ 300~keV~cm$^2$ with a radius of $r \sim$ 300~kpc.

\begin{figure*}
\begin{center}
\plotone{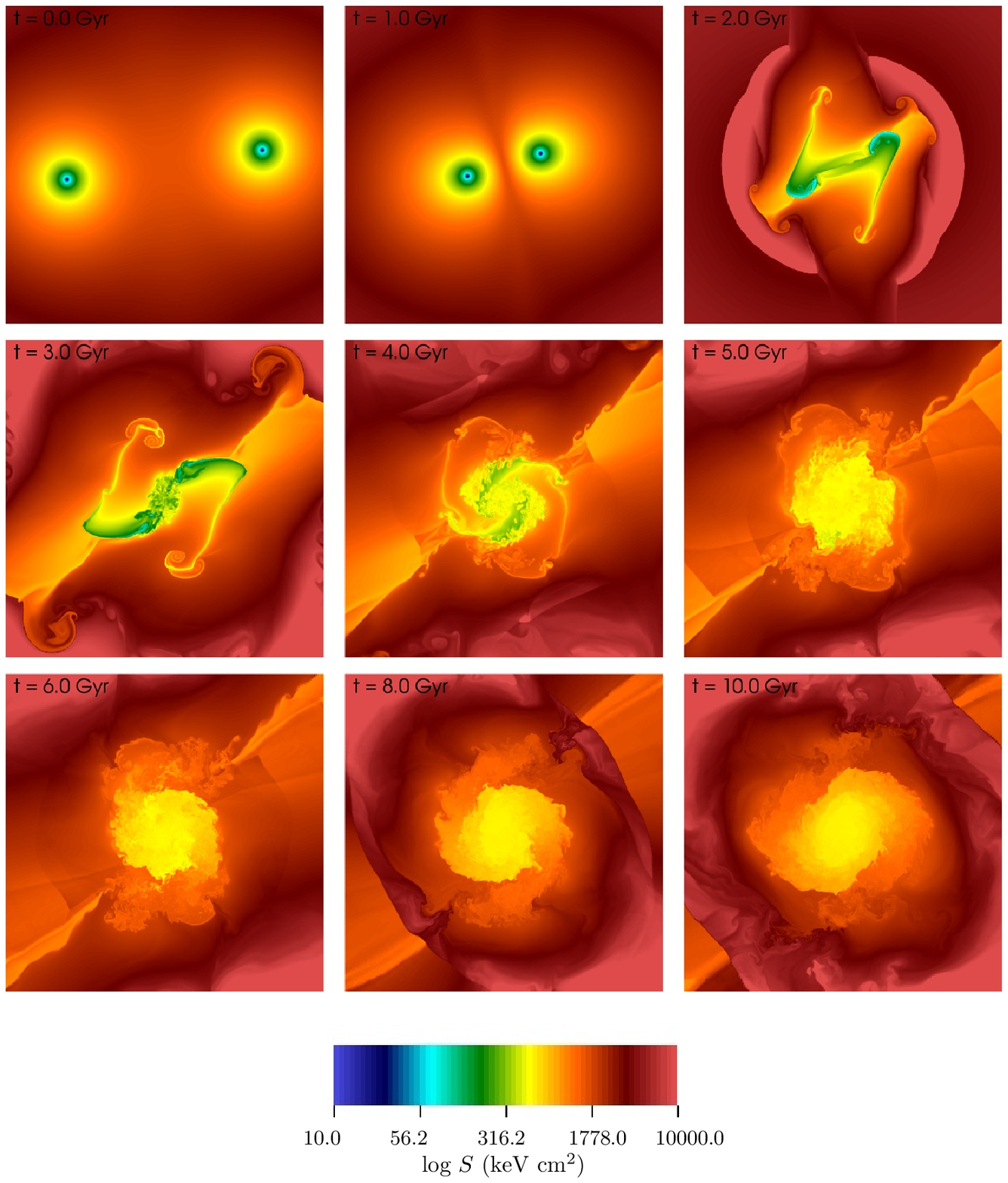}
\caption{Slices of entropy through the $z = 0$ coordinate plane for simulation S2 ($R$ = 1:1, $b$ = 464~kpc). The epochs shown are the same as in \ref{fig:entr_S1}. Each panel is 5~Mpc on a side.\label{fig:entr_S2}}
\end{center}
\end{figure*}

\begin{figure*}
\begin{center}
\plotone{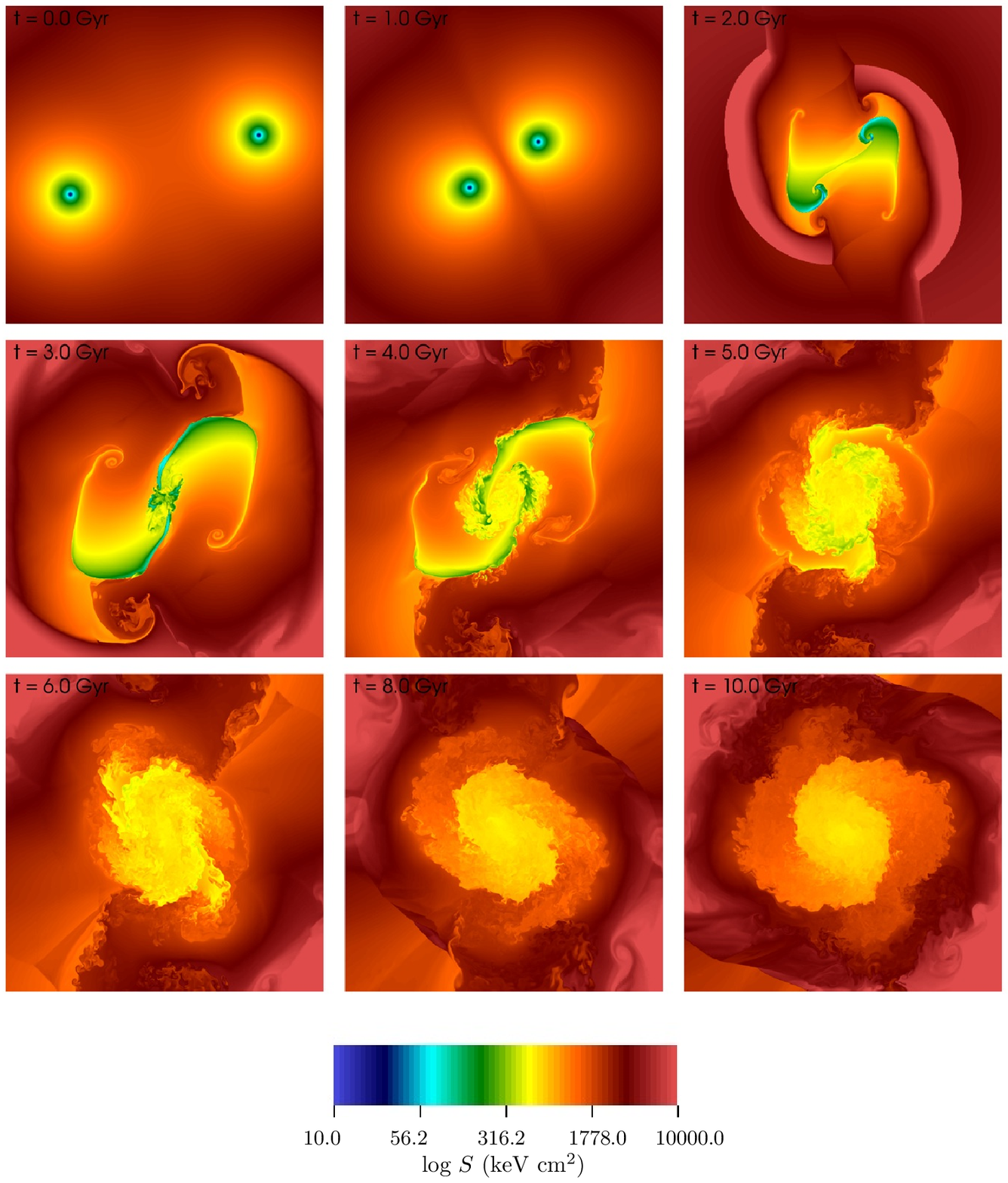}
\caption{Slices of entropy through the $z = 0$ coordinate plane for simulation S3 ($R$ = 1:1, $b$ = 932~kpc). The epochs shown are the same as in \ref{fig:entr_S1}. Each panel is 5~Mpc on a side.\label{fig:entr_S3}}
\end{center}
\end{figure*}

\begin{figure*}
\begin{center}
\plotone{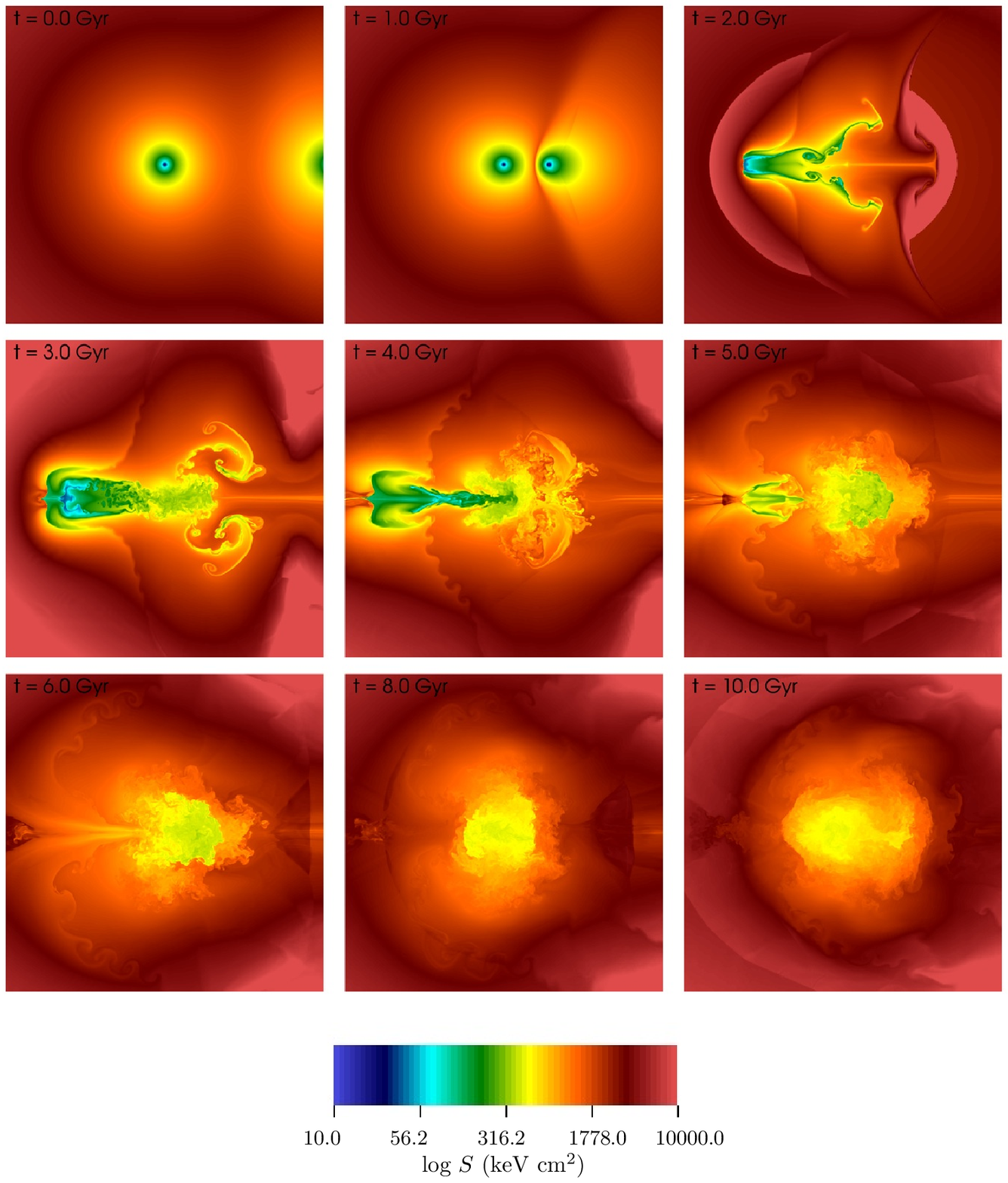}
\caption{Slices of entropy through the $z = 0$ coordinate plane for simulation S4 ($R$ = 1:3, $b$ = 0~kpc). The epochs shown are the same as in \ref{fig:entr_S1}. Each panel is 5~Mpc on a side.\label{fig:entr_S4}}
\end{center}
\end{figure*}

\begin{figure*}
\begin{center}
\plotone{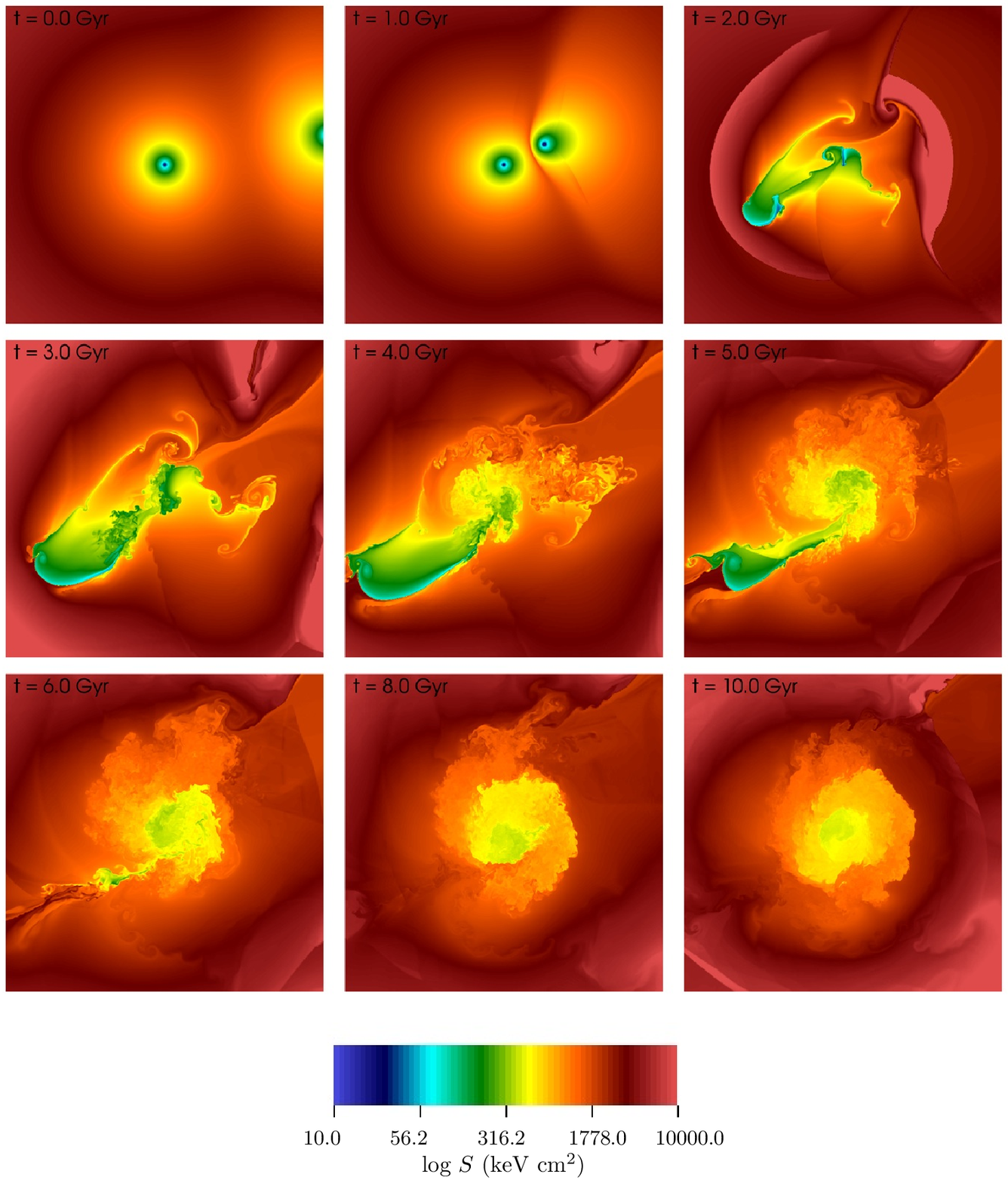}
\caption{Slices of entropy through the $z = 0$ coordinate plane for simulation S5 ($R$ = 1:3, $b$ = 464~kpc). The epochs shown are the same as in \ref{fig:entr_S1}. Each panel is 5~Mpc on a side.\label{fig:entr_S5}}
\end{center}
\end{figure*}

\begin{figure*}
\begin{center}
\plotone{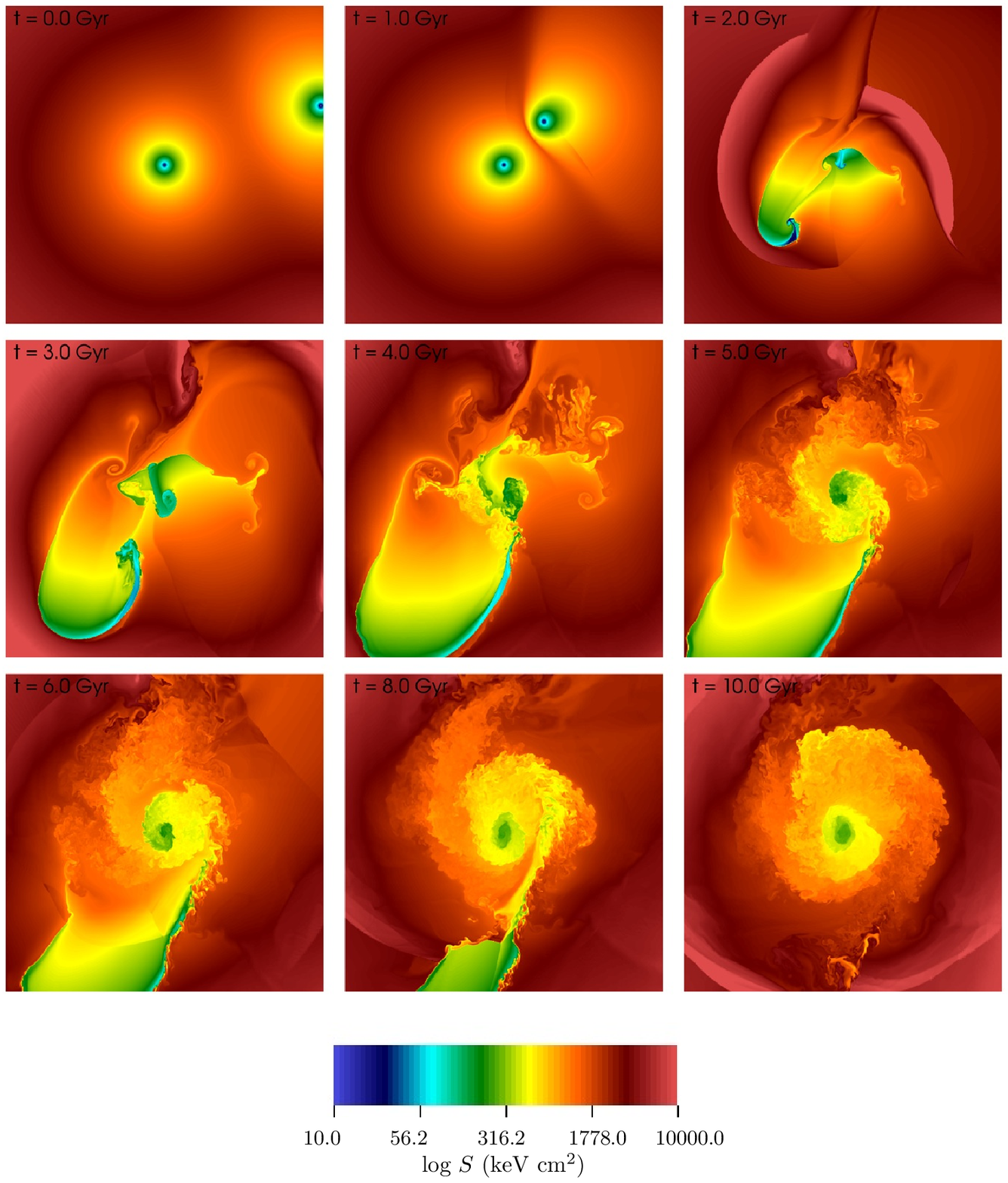}
\caption{Slices of entropy through the $z = 0$ coordinate plane for simulation S6 ($R$ = 1:3, $b$ = 932~kpc). The epochs shown are the same as in \ref{fig:entr_S1}. Each panel is 5~Mpc on a side.\label{fig:entr_S6}}
\end{center}
\end{figure*}

\begin{figure*}
\begin{center}
\plotone{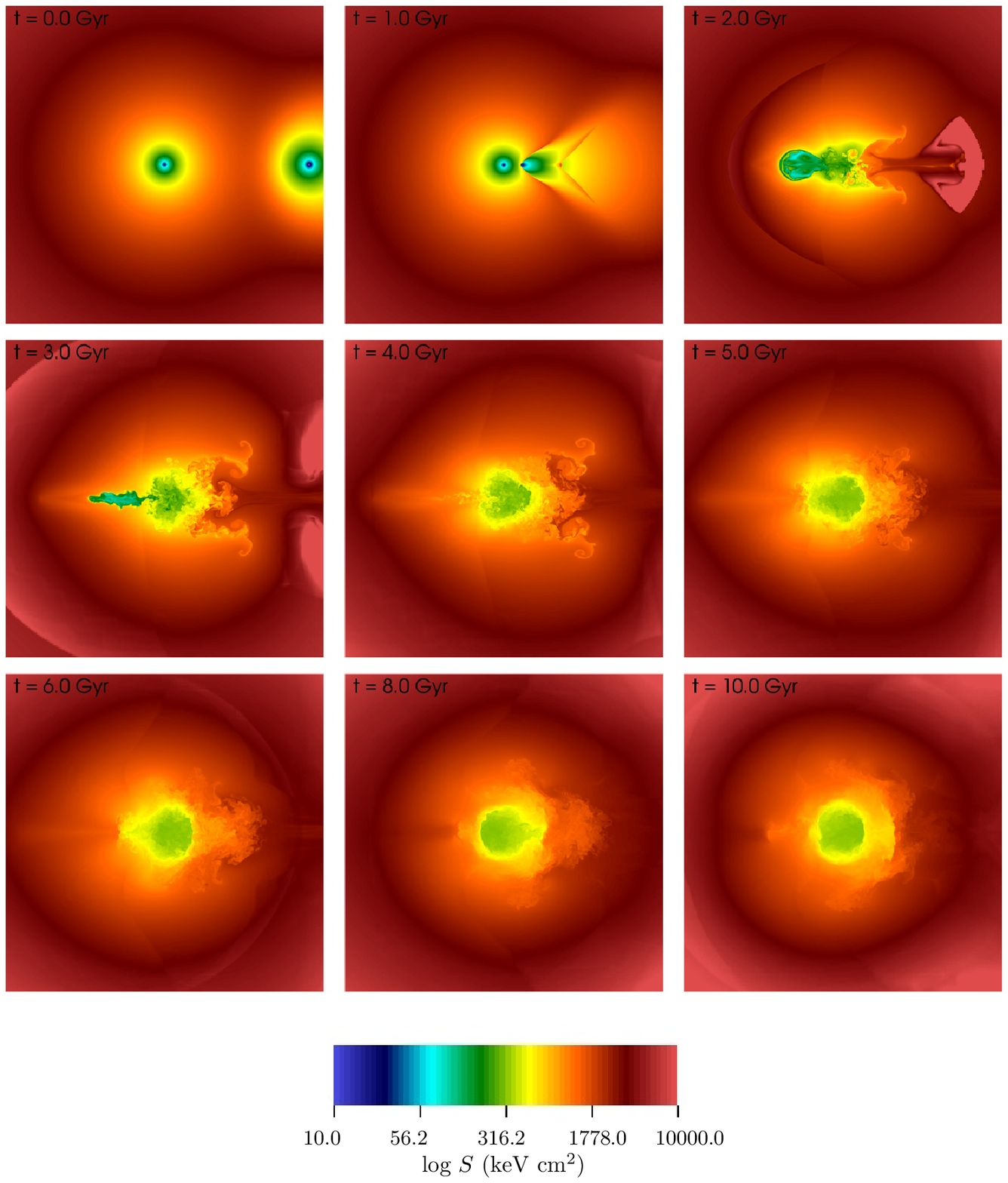}
\caption{Slices of entropy through the $z = 0$ coordinate plane for simulation S7 ($R$ = 1:10, $b$ = 0~kpc). The epochs shown are the same as in \ref{fig:entr_S1}. Each panel is 5~Mpc on a side.\label{fig:entr_S7}}
\end{center}
\end{figure*}

\begin{figure*}
\begin{center}
\plotone{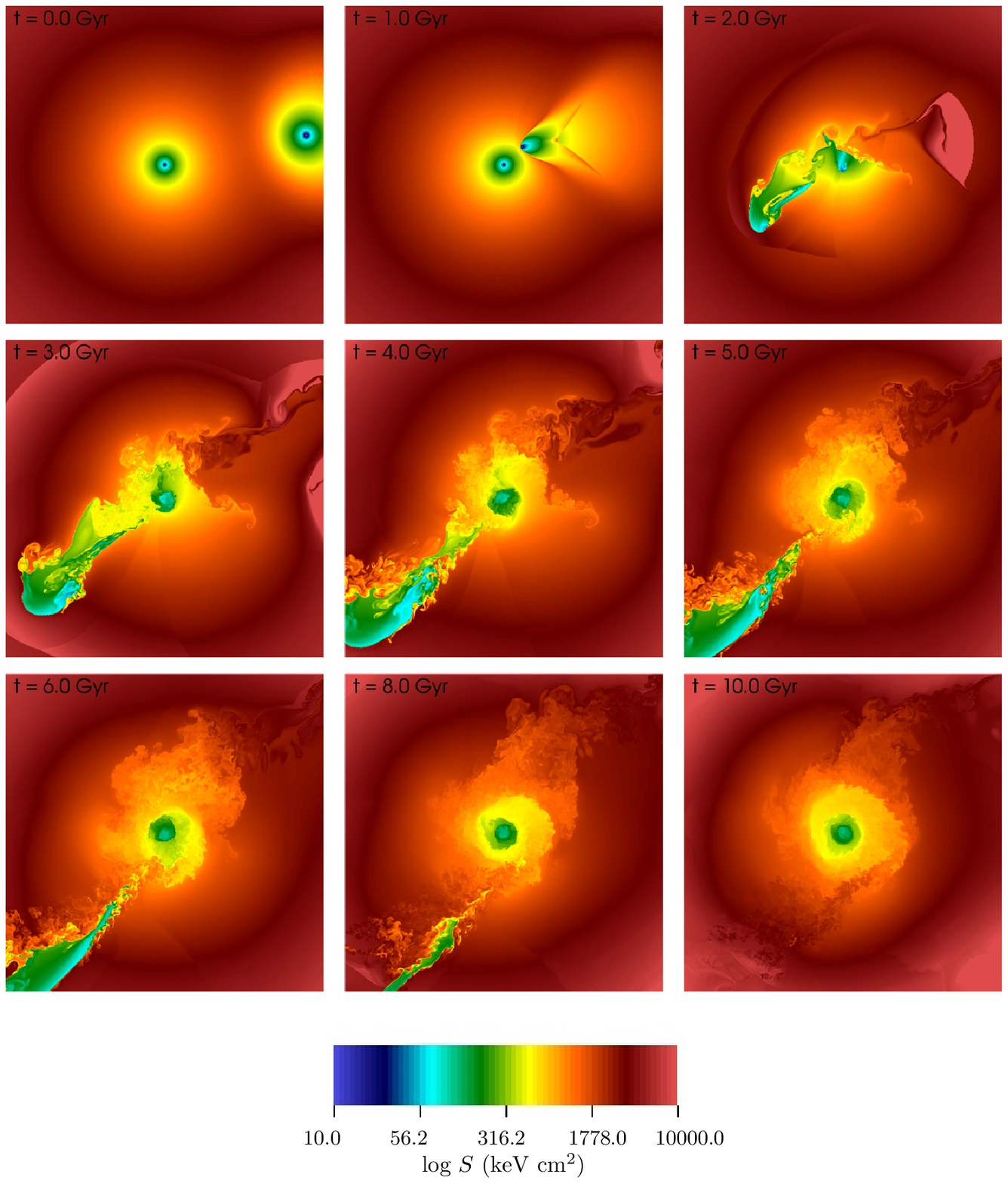}
\caption{Slices of entropy through the $z = 0$ coordinate plane for simulation S8 ($R$ = 1:10, $b$ = 464~kpc). The epochs shown are the same as in \ref{fig:entr_S1}. Each panel is 5~Mpc on a side.\label{fig:entr_S8}}
\end{center}
\end{figure*}

\begin{figure*}
\begin{center}
\plotone{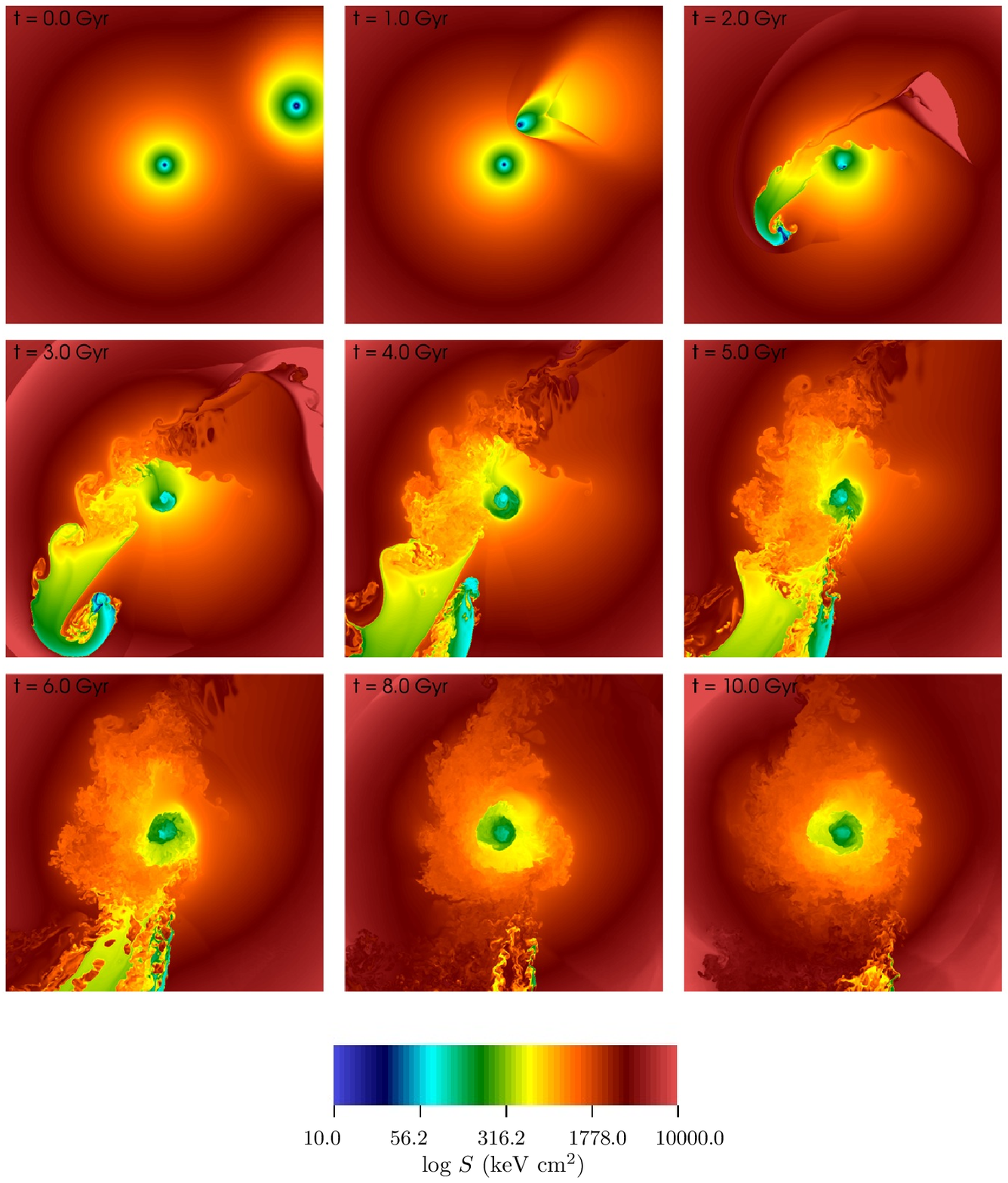}
\caption{Slices of entropy through the $z = 0$ coordinate plane for simulation S9 ($R$ = 1:10, $b$ = 932~kpc). The epochs shown are the same as in \ref{fig:entr_S1}. Each panel is 5~Mpc on a side.\label{fig:entr_S9}}
\end{center}
\end{figure*}

\begin{figure*}
\begin{center}
\plotone{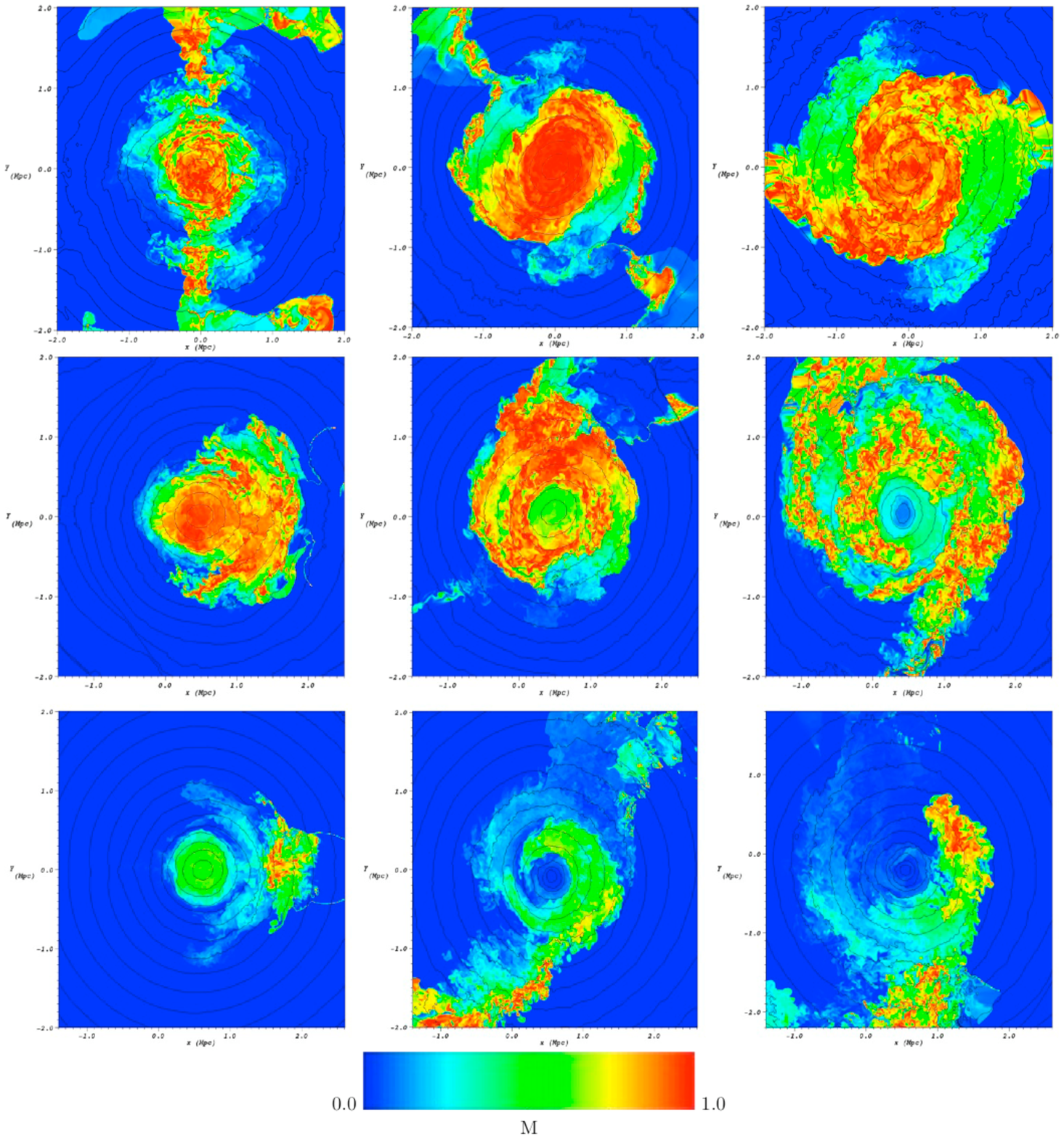}
\caption{The degree of gas mixing (as defined in equation \ref{eq:mixing}) in each simulation, at a time $t$ = 10 Gyr. At the top, from left to right, are S1, S2, and S3, in the middle, from left to right, are S4, S5, and S6, and at the bottom, from left to right, are S7, S8, and S9. Each panel is 4 Mpc on a side. \label{fig:mixing_figure}}
\end{center}
\end{figure*}

The off-center equal-mass cases (simulations S2 and S3, Figures \ref{fig:entr_S2} and \ref{fig:entr_S3}) are very similar to each other. In these cases the gas cores do not collide at the first core passage but sideswipe, creating a bridge of stripped, low-entropy gas that stretches between the two dark matter cores ($t \approx$ 2~Gyr). The gas remaining in the cores is initially held back by ram pressure as it moves outward, but as it moves out to lower density regions the effect of this pressure is weakened and the gas ``slingshots'' through and out the other side of the potential well. This gas expands adiabatically, curving around the dark matter cores and cooling, forming two large plumes of gas. The heads of these plumes fall back along with the dark matter cores to the center of the system by $t \approx$ 3~Gyr and are heated. The plumes themselves are stable for a few Gyr but eventually (by $t \approx$ 4~Gyr) they have become mixed in with the surrounding gas due to the Kelvin-Helmholtz instability. The central gas in these cases forms a roughly constant-entropy core of $S \sim$~500~keV~cm$^2$ with a radius of $r \sim$ 400~kpc.

\subsubsection{Unequal-Mass Mergers}

The unequal mass mergers have a qualitatively different evolution, as a result of the asymmetry in mass, which results in the formation of similar but somewhat different structures. In the head-on, 1:3 mass-ratio case (simulation S4, Figure \ref{fig:entr_S4}), the denser, lower-entropy core of the secondary cluster punches through the less dense core of the primary ($t \approx$ 1.5~Gyr), disrupting it completely. As the secondary's gas moves forward, a conical stream of low-entropy gas that has been stripped from the core trails it, falling into the gravitational potential well of the primary cluster ($t \approx$ 2.0~Gyr). As this gas falls in, the coherent structures are ripped apart by fluid instabilities and mixed in with the surrounding, higher-entropy ICM. At the same time, the secondary's core, which had been initially trailing the dark matter core due to the ram pressure of the primary's gas, now encounters regions of lower density. This causes the gas to fall back into the dark matter core, which then overshoots the potential minimum and begins to climb out of it on the other side, expanding and cooling adiabatically ($t \approx$ 2.0-3.0~Gyr). At this point, the secondary's dark matter core has already begun to fall back towards the primary's. By $t$ = 4.0~Gyr the gas remaining from the secondary's core has formed a low-entropy, collimated stream that flows into the coalescing dark matter halos. By the end of the simulation at $t$ = 10.0~Gyr this gas has thoroughly mixed in with the higher-entropy gas of the primary and has formed a roughly constant entropy core with $S_0 \sim$~450 keV~cm$^2$. 

In the off-center, unequal-mass cases (simulations S5 and S6, Figures \ref{fig:entr_S5} and \ref{fig:entr_S6}), a very different situation emerges. The secondary cluster does not disrupt the core of the primary cluster initially, but the gravitational disturbance caused by its passage sets up sloshing of the central core gas in the dark matter-dominated potential ($t \approx$ 2.0~Gyr). This is due to the fact that as both the dark matter and the gas are pulled toward the secondary cluster during the first core passage, the gas is held back by ram pressure and becomes displaced from the potential minimum. As the gas sloshes back into the potential minimum and then out again, it expands and cools, but as it comes into contact with higher entropy gas at larger radii it mixes with this gas and is heated. Meanwhile, the gas from the secondary cluster has formed a plume and stripped-gas bridge connecting back to the primary in much the same way as in the equal-mass cases. The bridge accretes cold gas back onto the primary cluster, followed by the plume which falls in which is mixed in with the sloshing central gas ($t \approx$ 4-6~Gyr). Throughout this time period the plume is relatively stable but is eventually broken up and mixed in with the surrounding gas. 

In our most extreme mass ratio cases, the effect of the subcluster on the main cluster is smallest, but the results are qualitatively similar to the 1:3 mass ratio cases, with the associated core disruption and accreting stream in the head-on case (simulation S7, Figure \ref{fig:entr_S7}) and subcluster plumes and main core sloshing in the off-center cases (simulations S8 and S9, Figures \ref{fig:entr_S8} and \ref{fig:entr_S9}). However, in these cases the onset of the effects of instabilities on these features is much sooner after the merger than in the previous cases. The Kelvin-Helmholtz instability begins to break up the plumes and sloshing core cold fronts almost immediately following the core passage. This is because the gravitational support from the subcluster is considerably weaker, making it much easier to strip the gas and providing less gravitational support against the onset of instability.

\subsection{ICM Mixing}\label{sec:mixing}

\begin{figure*}
\begin{center}
\includegraphics[width=0.3\textwidth]{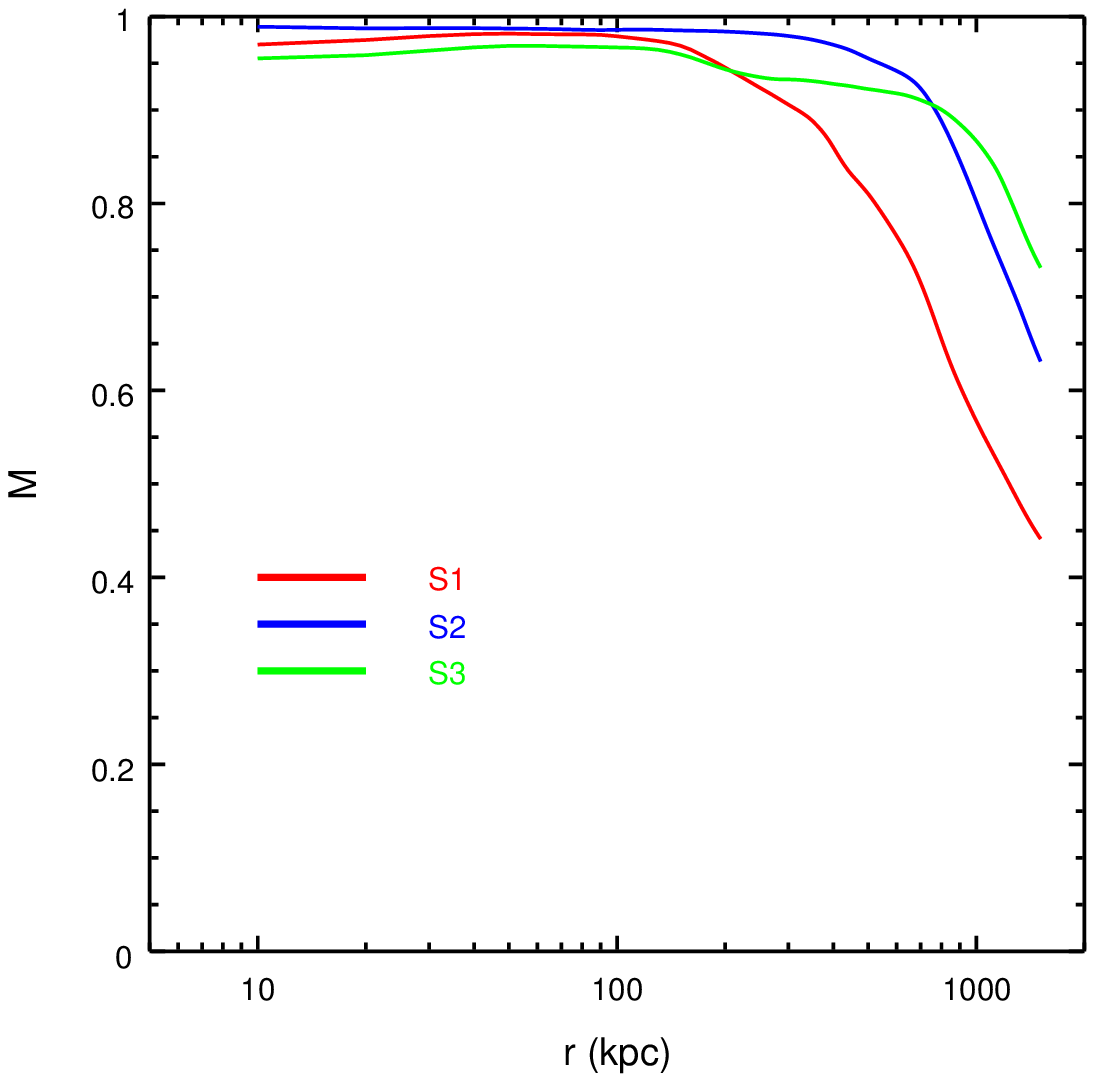}
\enspace
\includegraphics[width=0.3\textwidth]{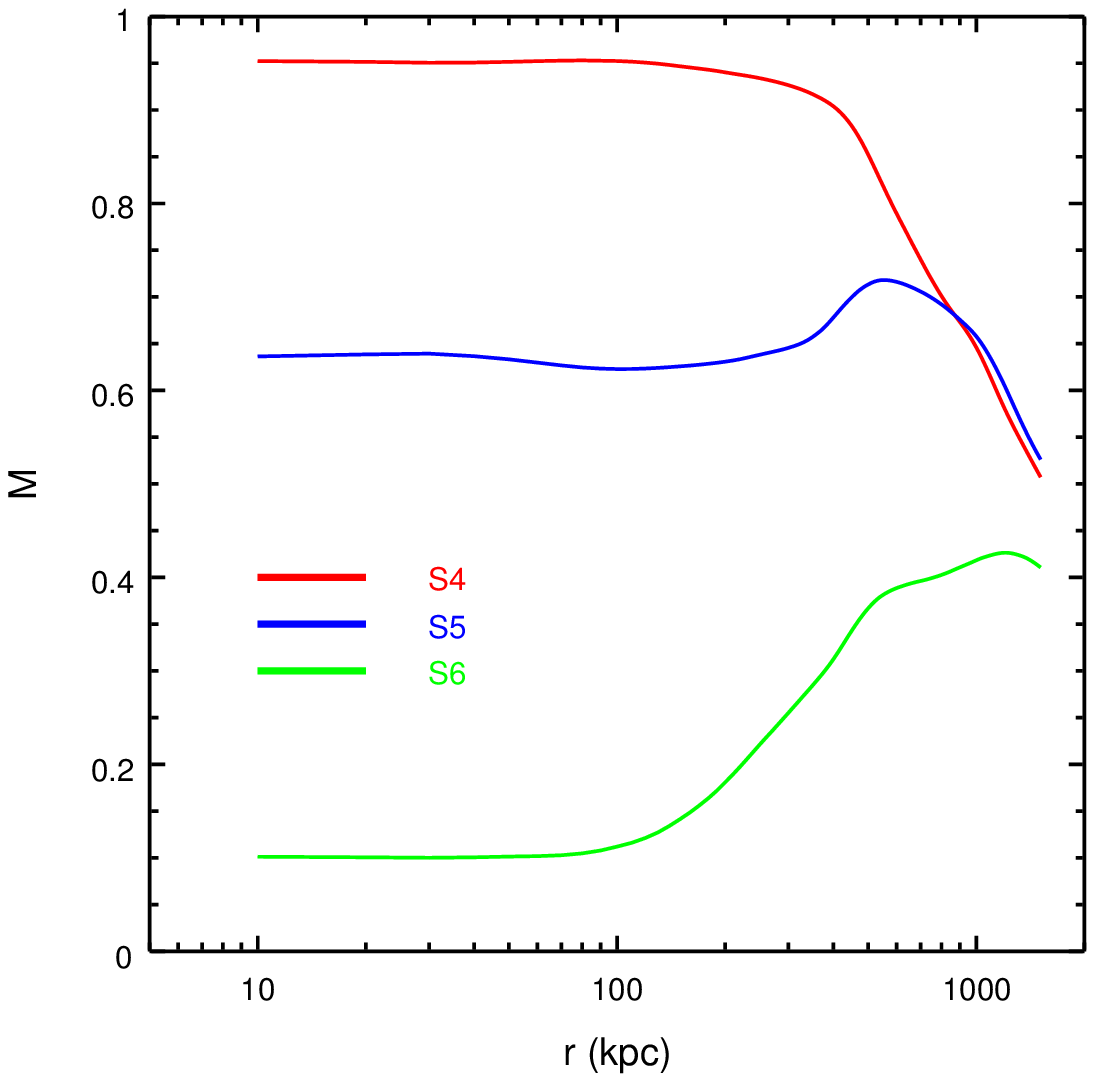}
\enspace
\includegraphics[width=0.3\textwidth]{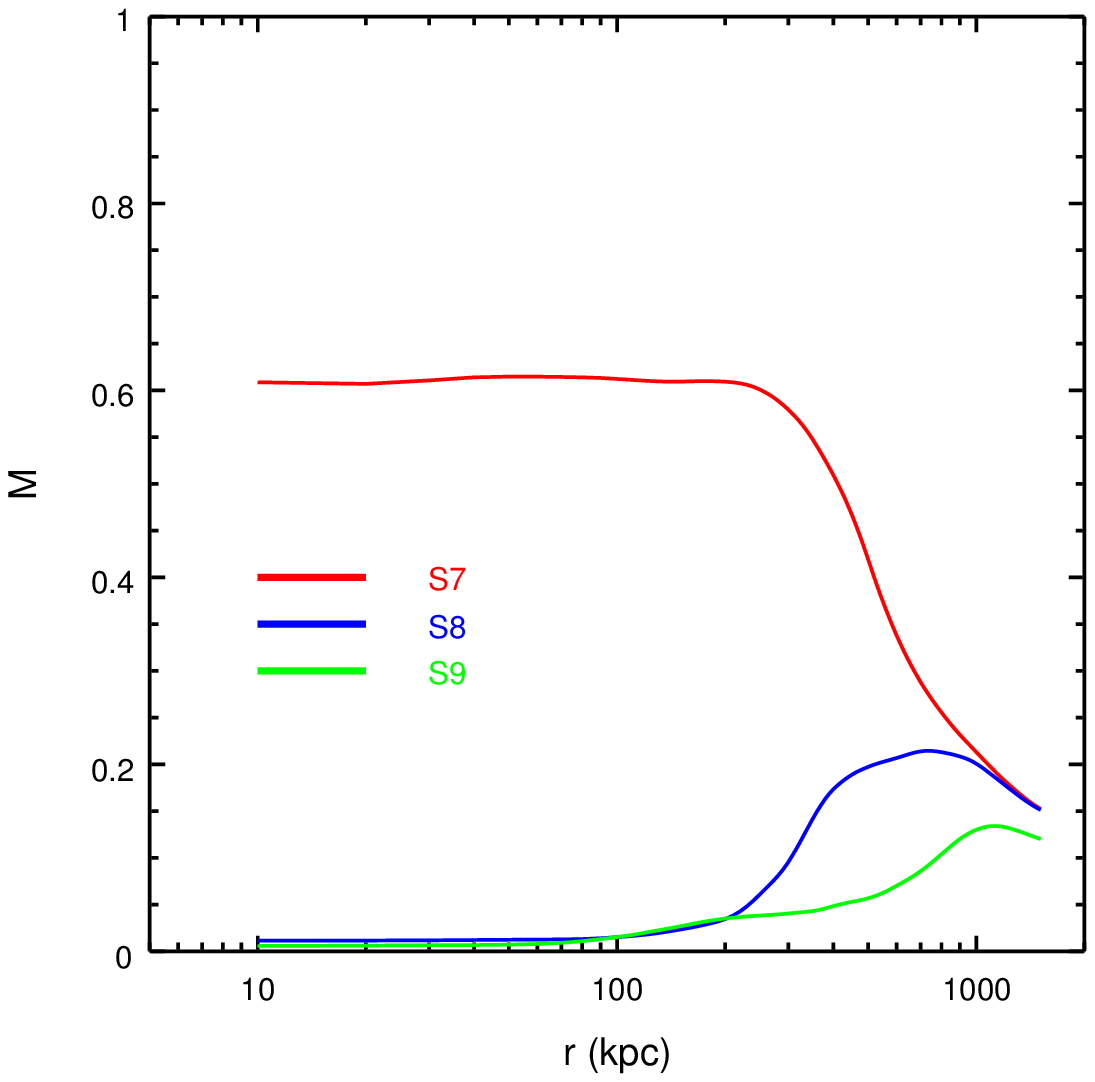}
\caption{Gas mixing profiles at $t$ = 10 Gyr. Left: Gas mixing profiles for the 1:1 mass-ratio simulations. Center: Gas mixing profiles for the 1:3 mass-ratio simulations. Right: Gas mixing profiles for the 1:10 mass-ratio simulations.\label{fig:mix_profile}}
\end{center}
\end{figure*}

\begin{figure*}
\begin{center}
\plottwo{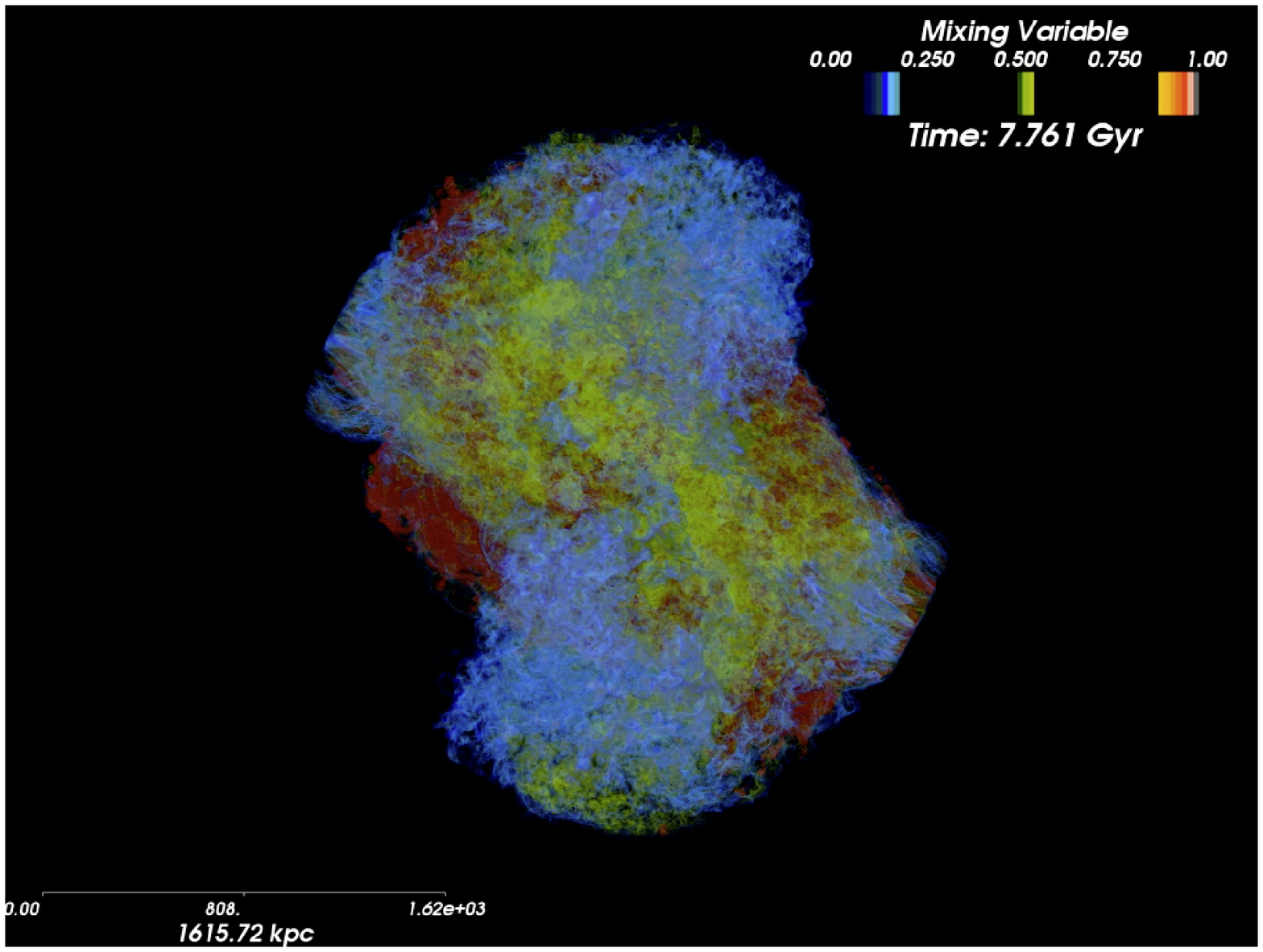}{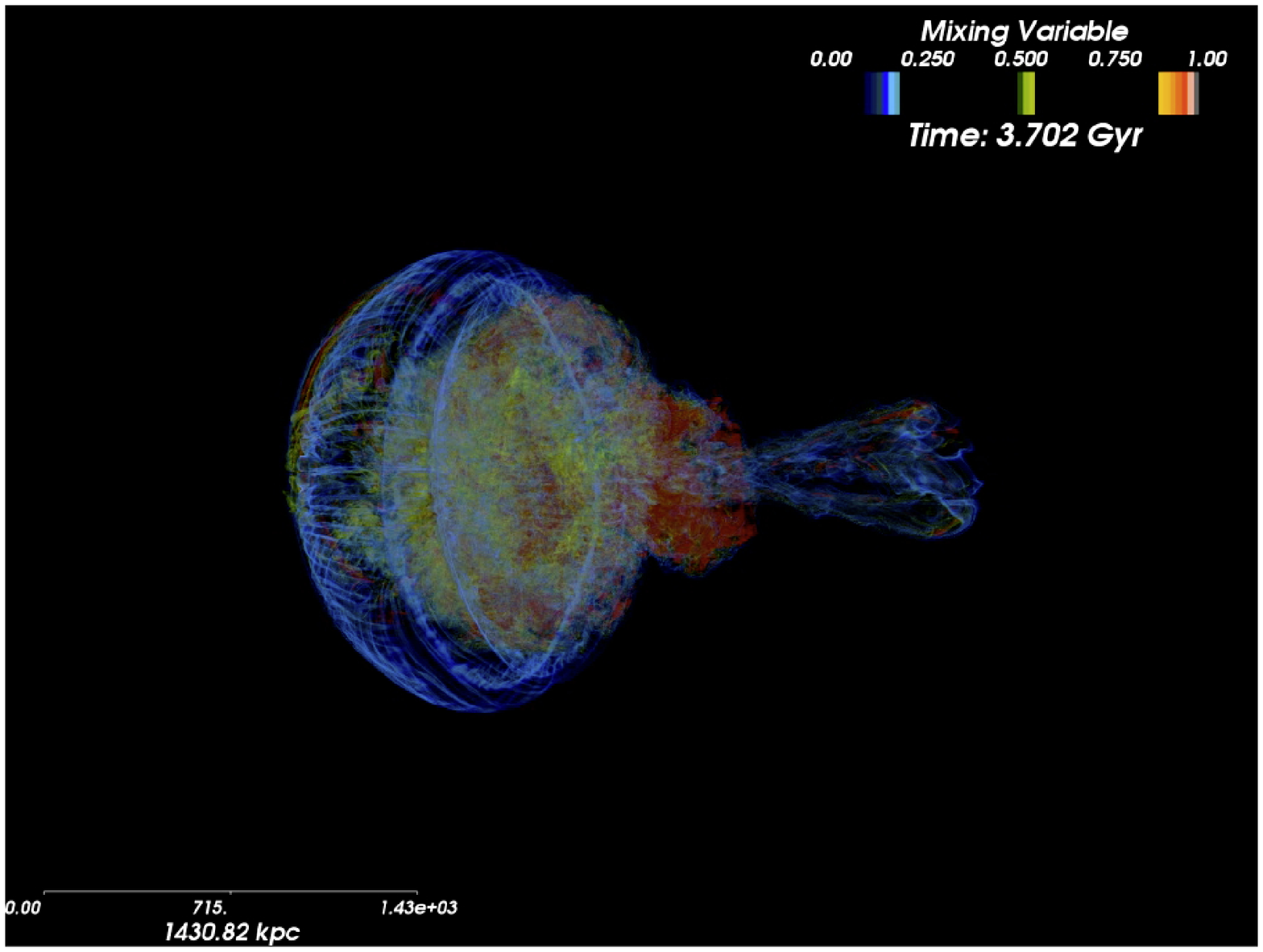}
\caption{The degree of gas mixing in 3-dimensional volume renderings, taken at particular snapshots. Full movies of gas mixing can be found at http://www.cfa.harvard.edu/~jzuhone/Research/Entropy\_and\_Mixing.html. Left: Simulation S3. Right: Simulation S4.\label{fig:mixing_movies}}
\end{center}
\end{figure*}

In order to investigate the degree of mixing of the gas of our initial systems, we have made use of ``mass scalars''. These quantities are defined as $Y_i = \rho_i/\rho$, where $\rho_i$ is the partial density, and are advected along with the fluid:

\begin{equation}
{\partial({\rho}{Y_i}) \over {\partial}t} + \nabla \cdot ({\rho}{Y_i}{\bf v}) = 0
\end{equation}

For the primary and secondary systems we define a mass scalar for each that is initially set equal to unity for radii within $r_{200}$. We then examine the degree of mixing of these two components at later times. To parameterize this mixing we follow \citet{rit02} and define the degree of mixing as

\begin{equation}
M = 1 - \left|\frac{\rho_1-\rho_2}{\rho_1+\rho_2}\right|
\end{equation}

Given the above definition for the mass scalars, this relation simply reduces to 

\begin{equation}
\label{eq:mixing}
M = 1 - \left|\frac{Y_1-Y_2}{Y_1+Y_2}\right|
\end{equation}

For example, a value of $M$ = 1.0 indicates equal amounts of the two cluster gas components in a cell ($Y_1/Y_2$ = 1), a value of $M$ = 0.5 indicates that the cluster gas is mixed in a ratio of $Y_1/Y_2 = 3$ or $Y_1/Y_2 = 1/3$, and a value of $M$ = 0.0 indicates the gas is not mixed at all, and that only the gas from one cluster is present.

Figure \ref{fig:mixing_figure} shows the resulting mixing $M$ with gas density contours overlaid at $t$ = 10 Gyr, in a slice though the $z = 0$ coordinate plane. Additionally, Figure \ref{fig:mix_profile} shows the mixing $M$ as a function of radius. For the equal mass mergers, the gas of the two clusters is well-mixed in all cases of varying impact parameter out to a radius of nearly $\sim$0.5 Mpc. In the head-on case (simulation S1), the gas is also significantly mixed along the plane perpendicular to the merger trajectory where the large ``pancakes'' of gas formed. The degree of mixing is more pronounced in the off-center cases (simulations S2 and S3). In these cases the fact that the cores slide pass each other and are stripped before finally merging into a single remnant allows for more mixing between the two than in the case where the collision is head-on. For the 1:3 and 1:10 mass ratio mergers, the gas is also mixed but it becomes less efficient as the impact parameter increases. In the head-on case with a 1:3 mass ratio (simulation S4), the disruption of the primary's core by the secondary allows for significant mixing of the two cluster components. The result is very different for the off-center cases, however. In simulation S5 ($b$ = 464 kpc) in the center of the cluster the two components are only mixed in a 3-to-1 ratio and in simulation S6 ($b$ = 932 kpc) the central gas is hardly mixed with gas from the other cluster at all. Most of the mixing in these cases occurs at radii $r \simgt$ 500~kpc, as the stripping of the secondary's gas occurred as it made its first core passage, and as this gas fell toward the center of the primary it was broken up by instabilities and mixed with the surrounding gas. Similarly, in the head-on case for the 1:10 mass ratio (simulation S7) the central gas is mixed in a 3-to-1 ratio, but in the off-center cases (simulations S8 and S9) the central gas is not mixed with gas from the other cluster at all. In these extreme cases, the low entropy core is not disrupted completely but only displaced from the center somewhat and as the system as a whole relaxes this material sinks back to the center of the newly-reconstituted potential well.

Figure \ref{fig:mixing_movies} shows snapshots of volume-rendered movies of gas mixing in our galaxy cluster merger simulations that can be found at http://www.cfa.harvard.edu/~jzuhone/Research/Entropy\\*\_and\_Mixing.html.

\subsection{Radial Profiles}\label{sec:profiles}

It is instructive to examine the final radial profiles of various quantities compared to the initial radial profiles for the same quantities of the primary cluster. Whereas our initial gas profiles represent typical cool-core clusters with peaked gas density profiles, a positive radial temperature gradient at low radii, and power-law entropy profiles almost all the way to the cluster center, the final profiles appear very different. 

We have constructed these radial profiles by first taking the cluster potential minimum as the cluster center. By visual inspection we find that at the end of each simulation this position corresponds well with the location of the gas and dark matter density peaks. Secondly, we divide the cluster into spherical radial annuli of width $\Delta{r} = 10.5h^{-1}$~kpc, wide enough so that there are enough points/particles in each annulus to adequately sample each relevant quantity, but narrow enough to sample the radial range adequately. We have tested the robustness of our profiles by varying the width of our radial annuli by a factor of 2-3x, for which we find no significant changes in our profiles. The velocity mean and velocity dispersion are estimated from the sample mean and variance of the cells for the gas and the particles for the dark matter. 

\begin{figure*}
\begin{center}
\includegraphics[width=0.3\textwidth]{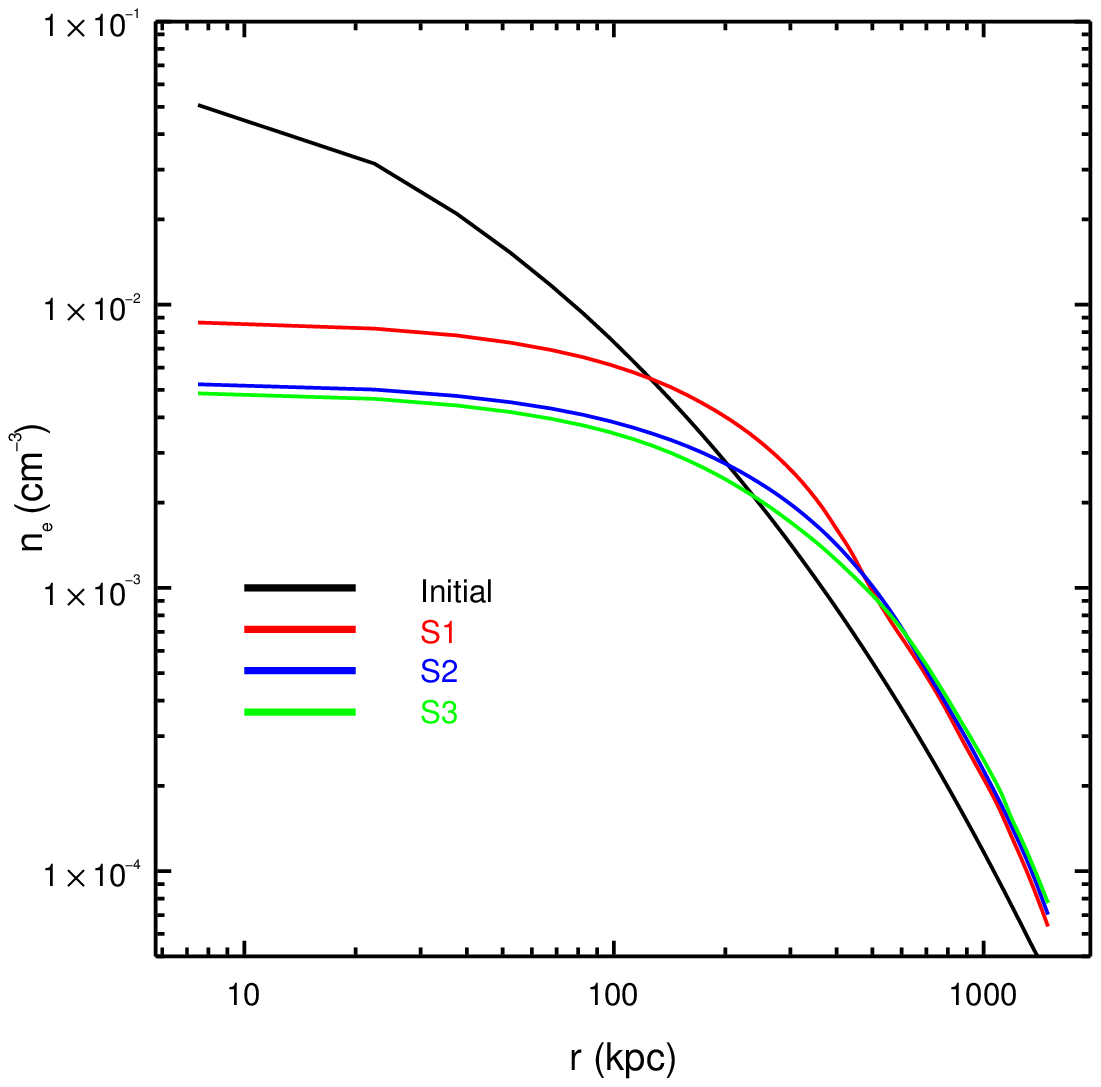}
\enspace
\includegraphics[width=0.3\textwidth]{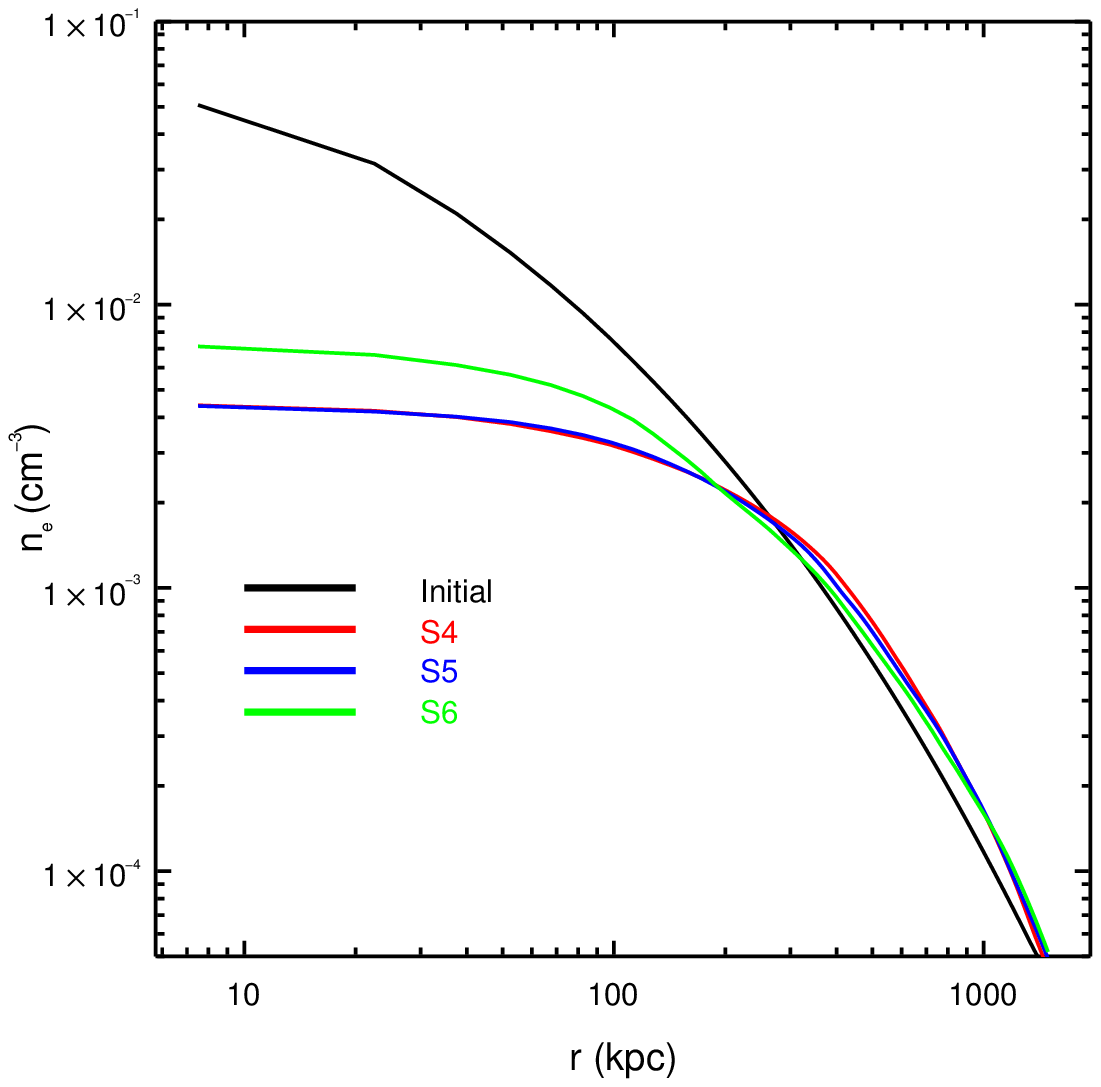}
\enspace
\includegraphics[width=0.3\textwidth]{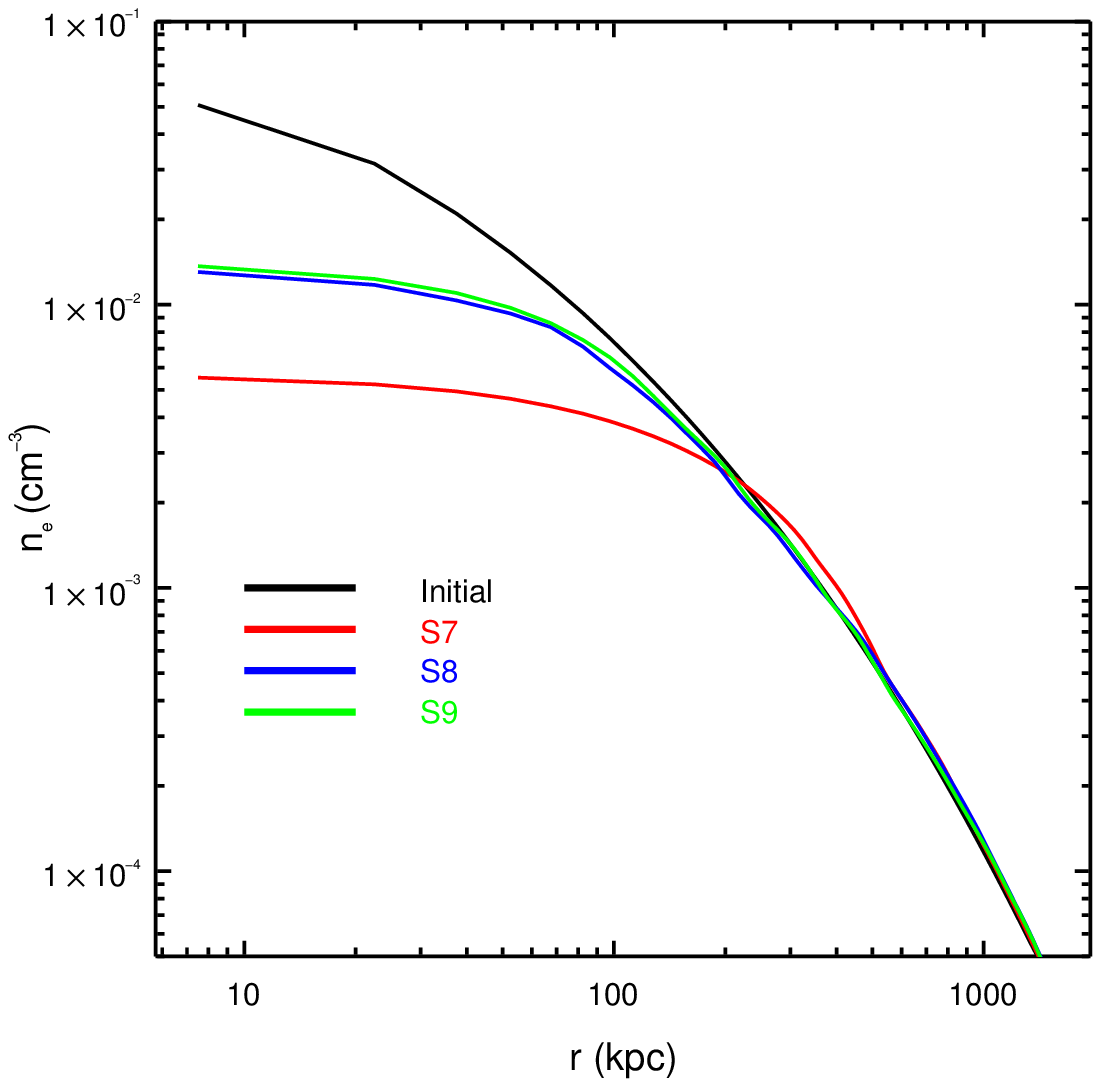}
\caption{Gas density profiles at $t$ = 10 Gyr, compared to the initial profile. Left: Gas density profiles for the 1:1 mass-ratio simulations. Center: Gas density profiles for the 1:3 mass-ratio simulations. Right: Gas density profiles for the 1:10 mass-ratio simulations.\label{fig:dens_profile}}
\end{center}
\end{figure*}

Figure \ref{fig:dens_profile} shows the final electron number density profile for each simulation, with the initial density profile of the primary shown for comparison. In all cases, the central density is reduced by a factor $\sim$3-10, and the gas mass is redistributed such that the resulting density profile is considerably flattened towards the center as compared to the initial density profile. Such gas density profiles are better fit by the traditional $\beta$-model \citep{cav76} than those of ``cool-core'' clusters, in which the gas density continues to rise towards the center.

\begin{figure*}
\begin{center}
\includegraphics[width=0.3\textwidth]{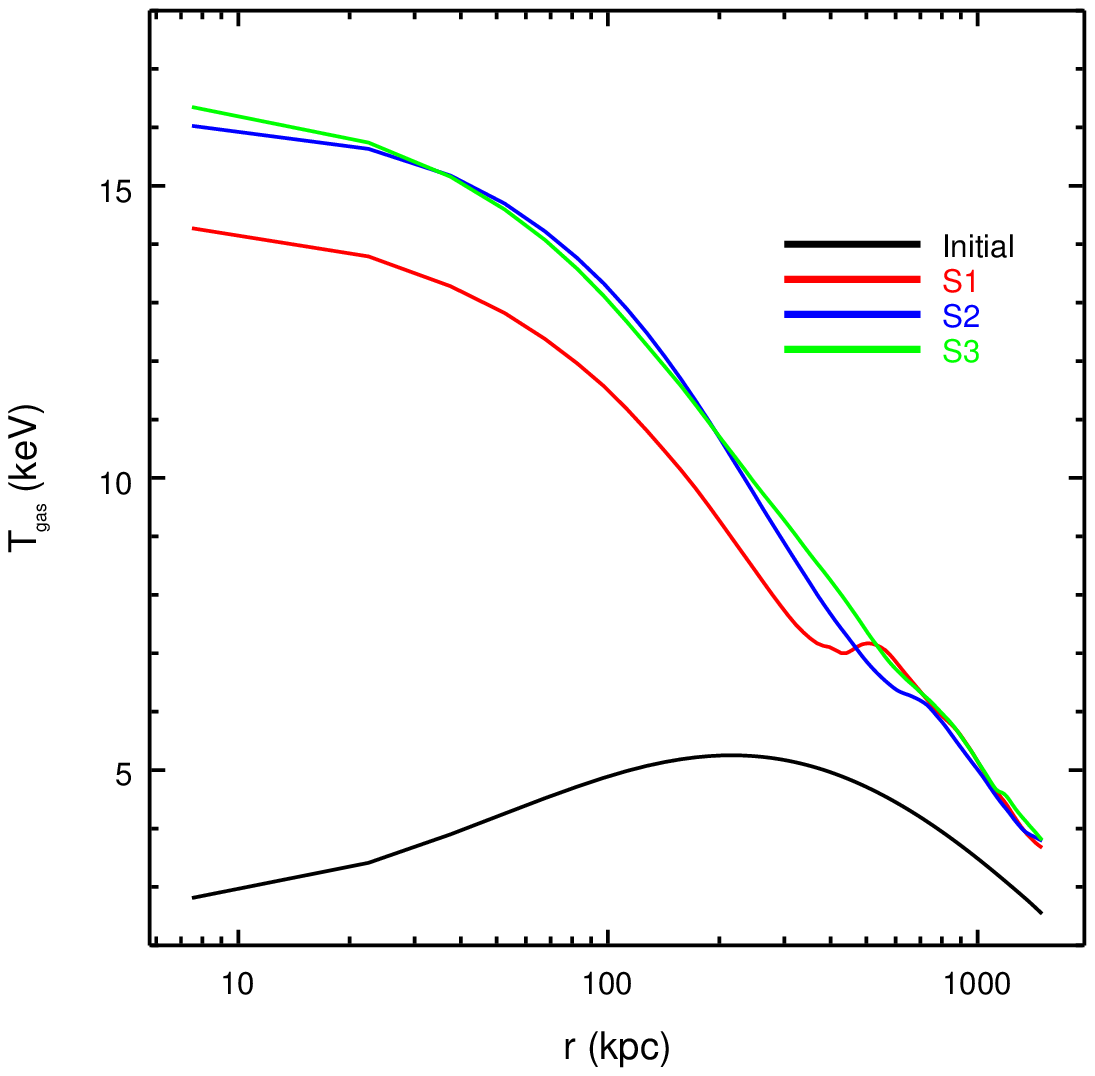}
\enspace
\includegraphics[width=0.3\textwidth]{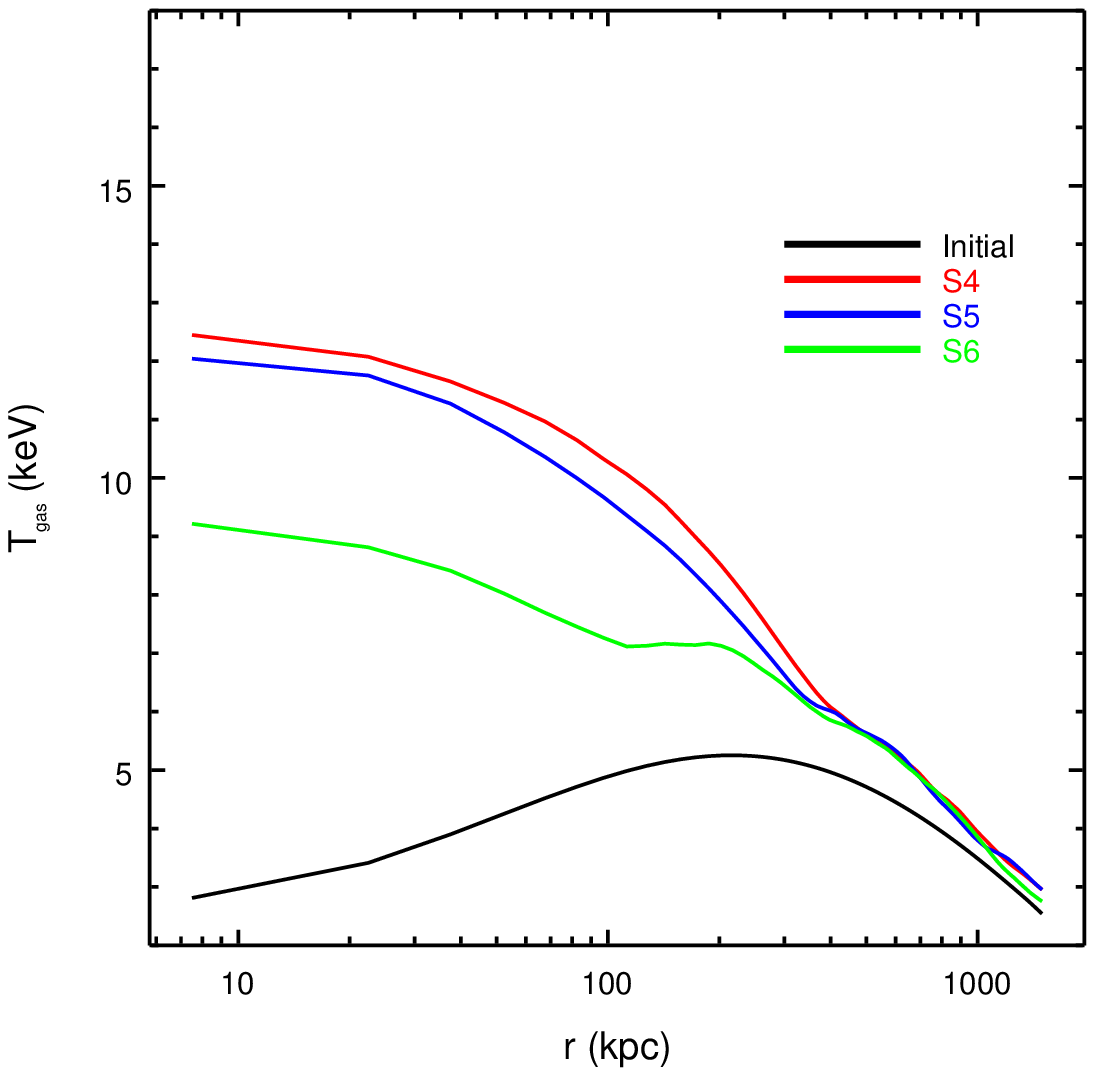}
\enspace
\includegraphics[width=0.3\textwidth]{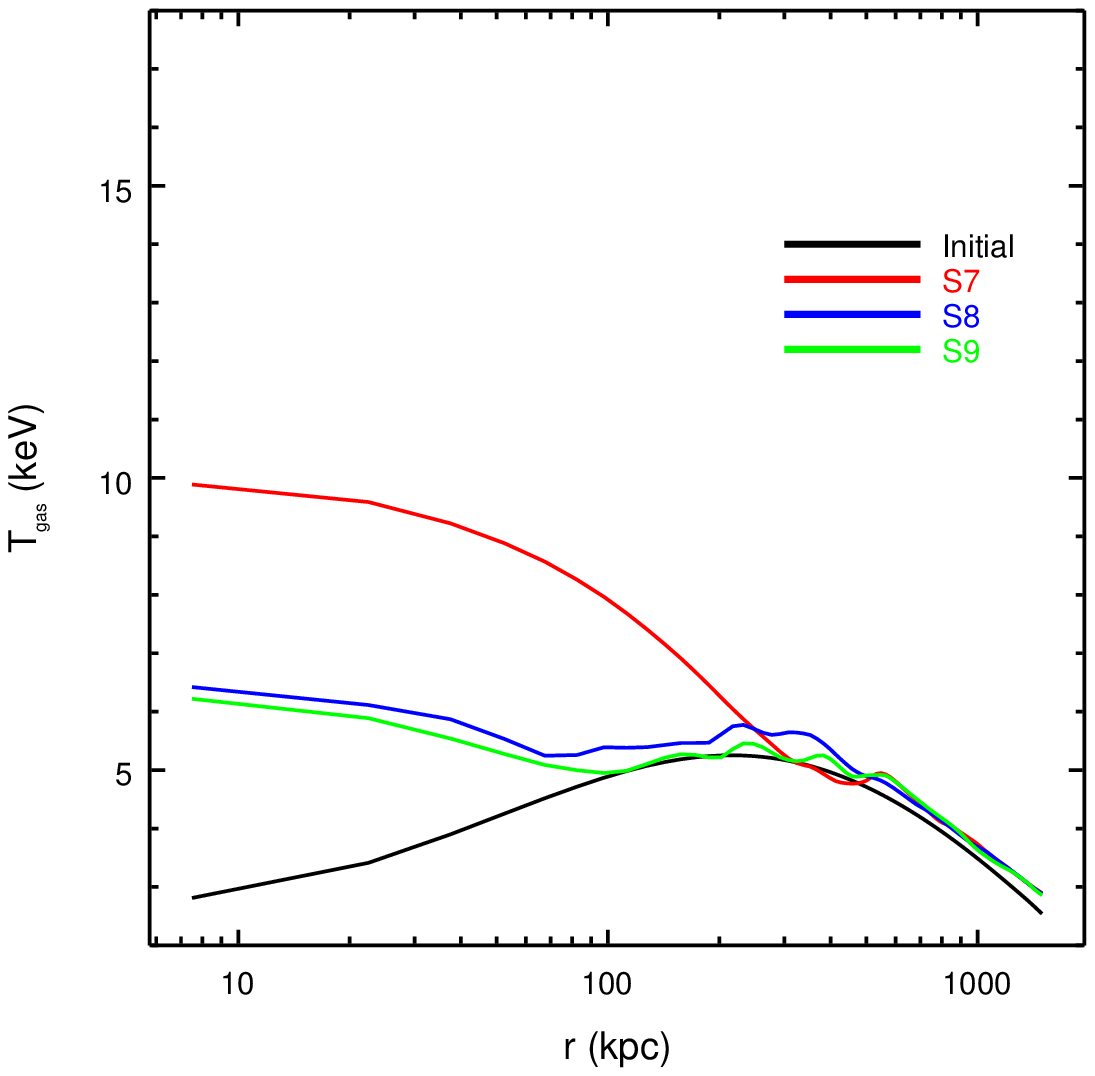}
\caption{Temperature profiles at $t$ = 10 Gyr, compared to the initial profile. Left: Temperature profiles for the 1:1 mass-ratio simulations. Center: Temperature profiles for the 1:3 mass-ratio simulations. Right: Temperature profiles for the 1:10 mass-ratio simulations.\label{fig:temp_profile}}
\end{center}
\end{figure*}

Figure \ref{fig:temp_profile} shows the final temperature profile for each simulation, with the initial temperature profile of the primary shown for comparison. The initial profile of the primary is typical of a ``cool-core'' cluster, featuring a temperature inversion with a negative radial gradient in the outskirts to a positive radial gradient at radii $r \simlt$ 200~kpc. Each merger converts the temperature profile from this form to one where the temperature rises almost continuously towards the center within approximately the same radial range. The differences are dramatic: in the equal-mass merger cases (simulations S1-S3) the central temperature is increased by nearly $\Delta{T} \approx$ 12~keV, with the 1:3 mass-ratio cases increasing it by $\Delta{T} \approx$ 7-9~keV and the 1:10 mass-ratio cases increasing it by $\Delta{T} \approx$ 3-7~keV.

\begin{figure*}
\begin{center}
\includegraphics[width=0.3\textwidth]{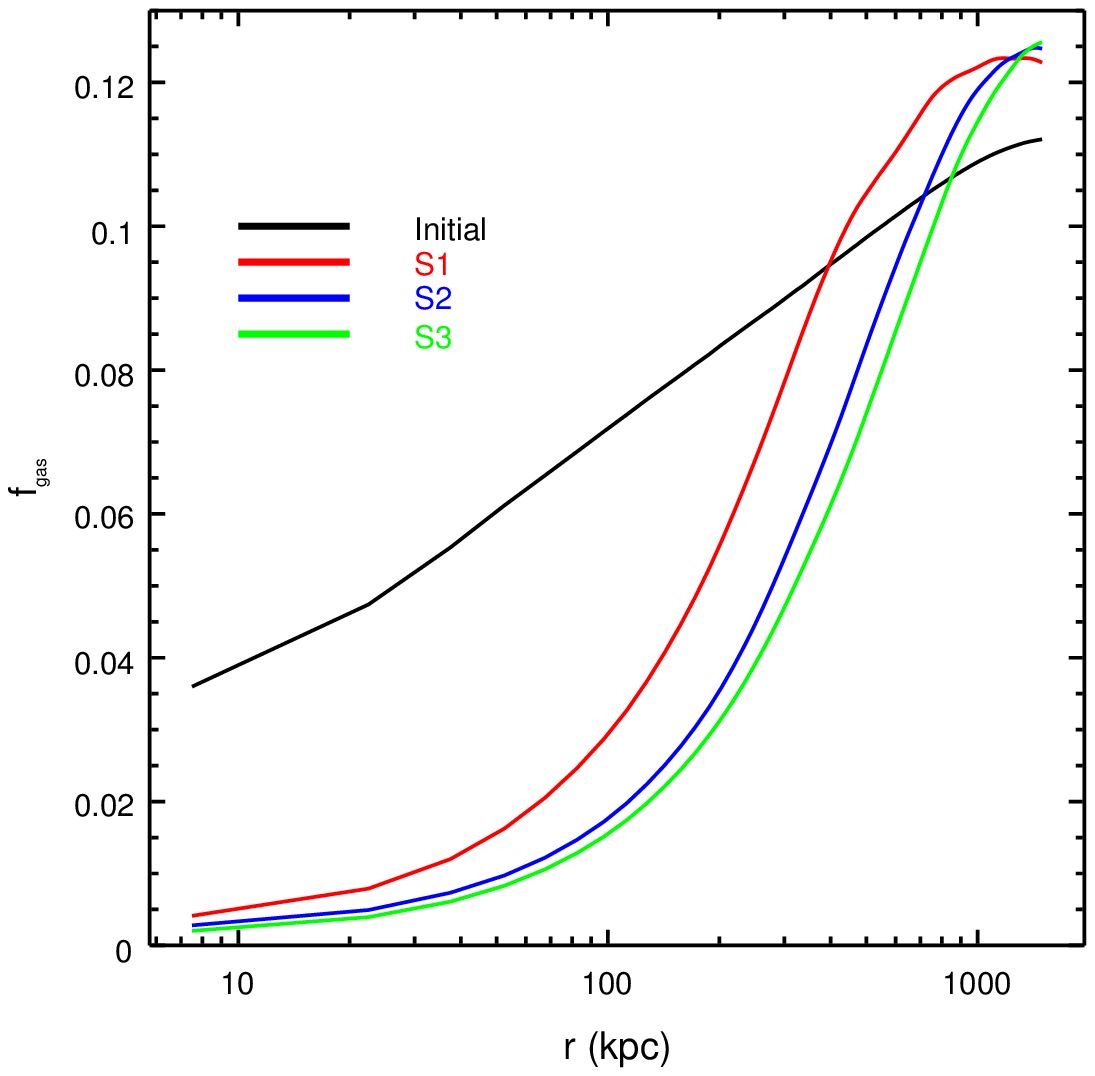}
\enspace
\includegraphics[width=0.3\textwidth]{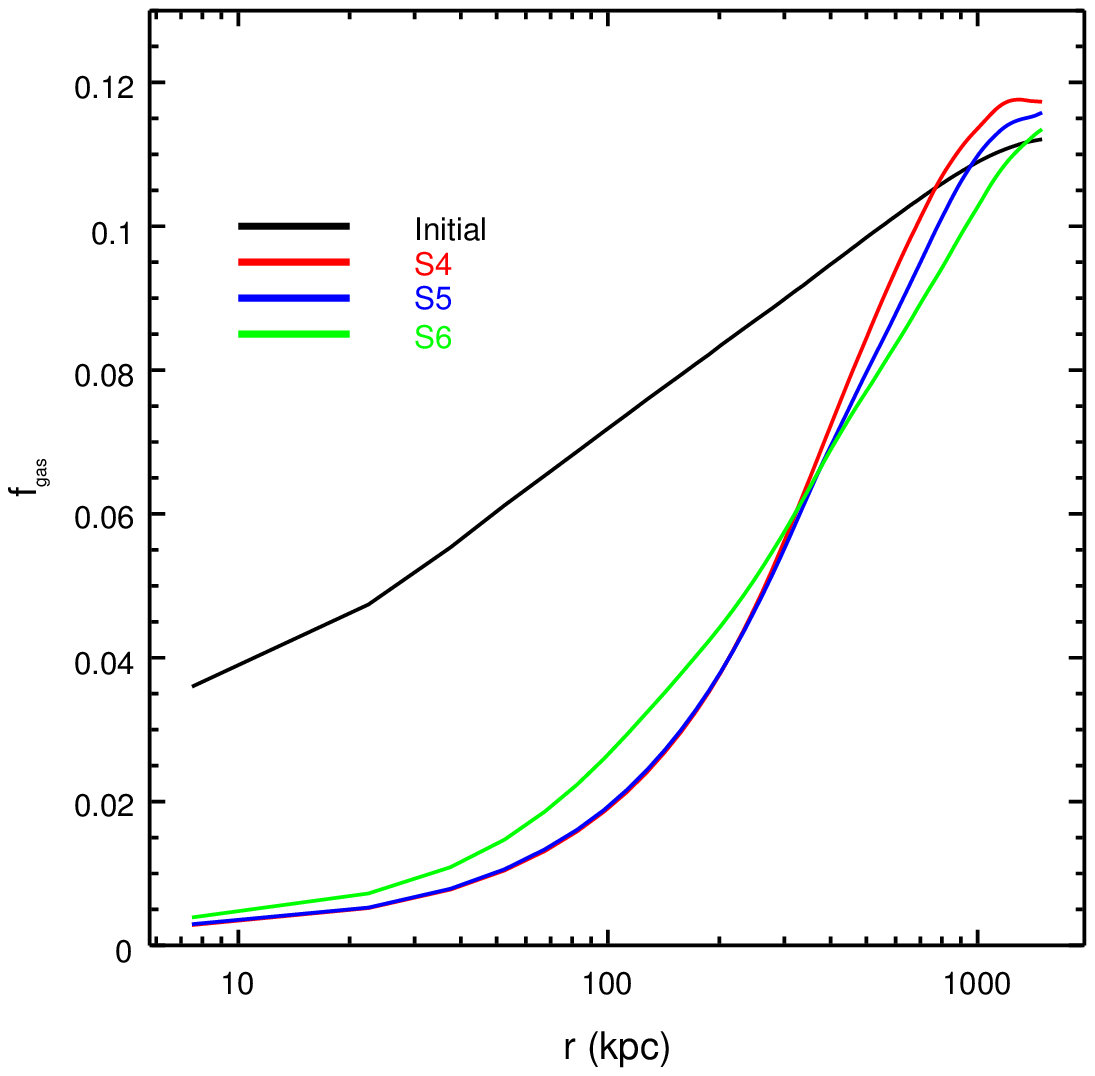}
\enspace
\includegraphics[width=0.3\textwidth]{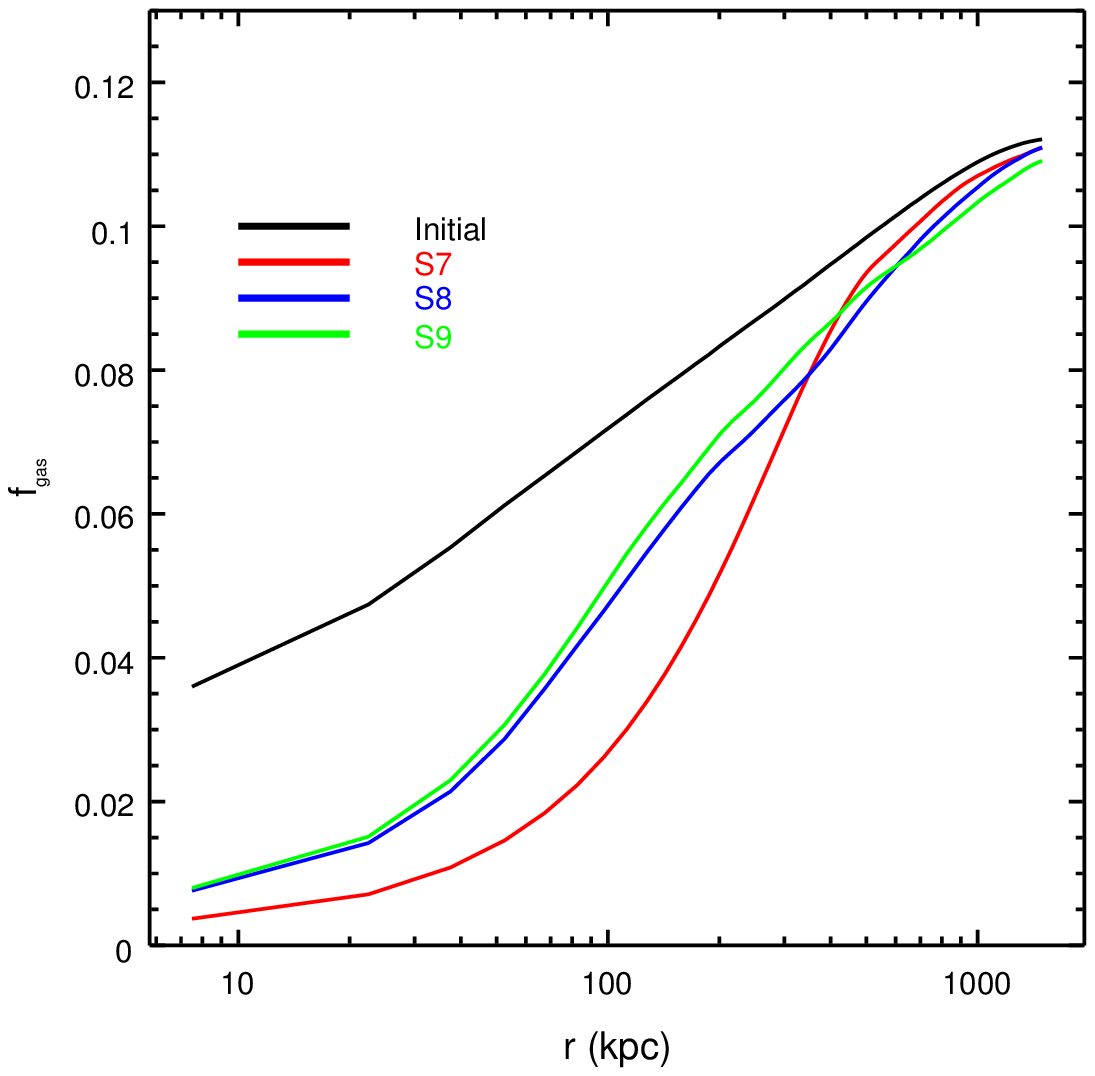}
\caption{Gas fraction profiles at $t$ = 10 Gyr, compared to the initial profile. Left: Gas fraction profiles for the 1:1 mass-ratio simulations. Center: Gas fraction profiles for the 1:3 mass-ratio simulations. Right: Gas fraction profiles for the 1:10 mass-ratio simulations.\label{fig:gfrac_profile}}
\end{center}
\end{figure*}

Figure \ref{fig:gfrac_profile} shows the final gas mass fraction ($f_{\rm gas}(<r) = M_{\rm gas}(<r)/M_{\rm tot}(<r)$) profiles for each simulation, with the initial gas fraction profile of the primary shown for comparison. Each merger considerably lowers the gas mass fraction in the central regions (by nearly a factor of 4 or more in each case), corresponding to the flattening of the gas profile, distributing it to larger radii. 

\begin{figure*}
\begin{center}
\includegraphics[width=0.3\textwidth]{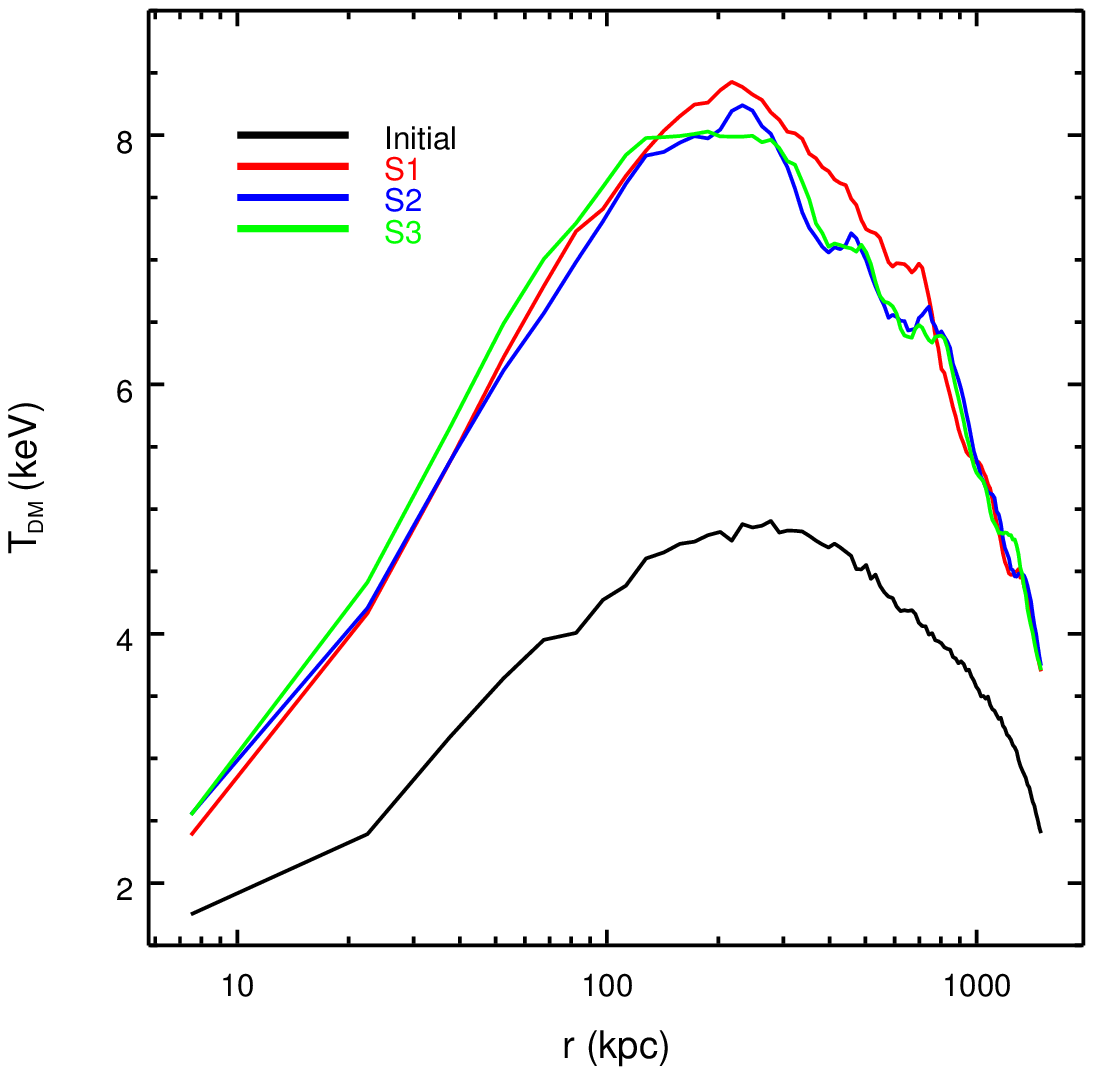}
\enspace
\includegraphics[width=0.3\textwidth]{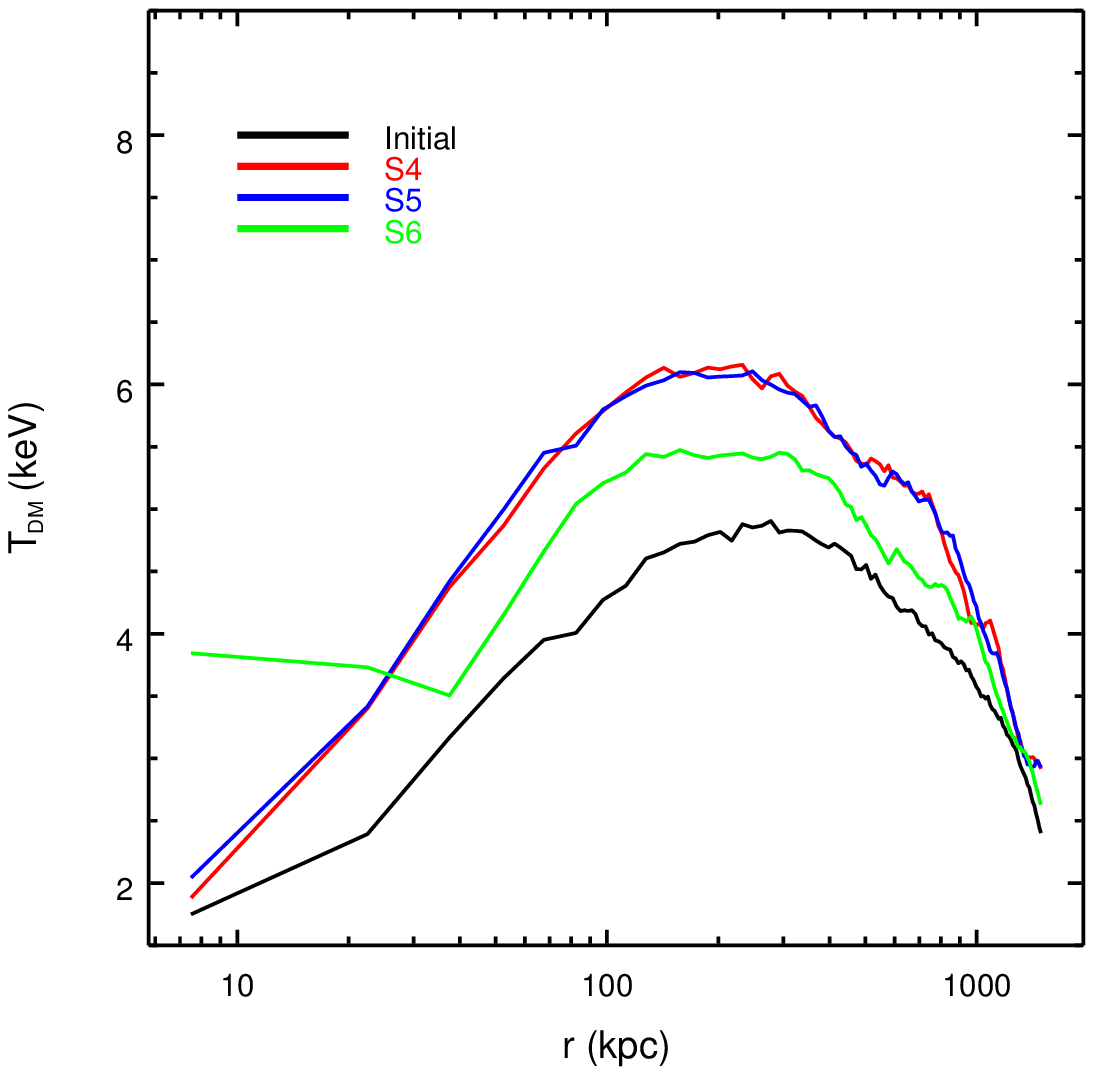}
\enspace
\includegraphics[width=0.3\textwidth]{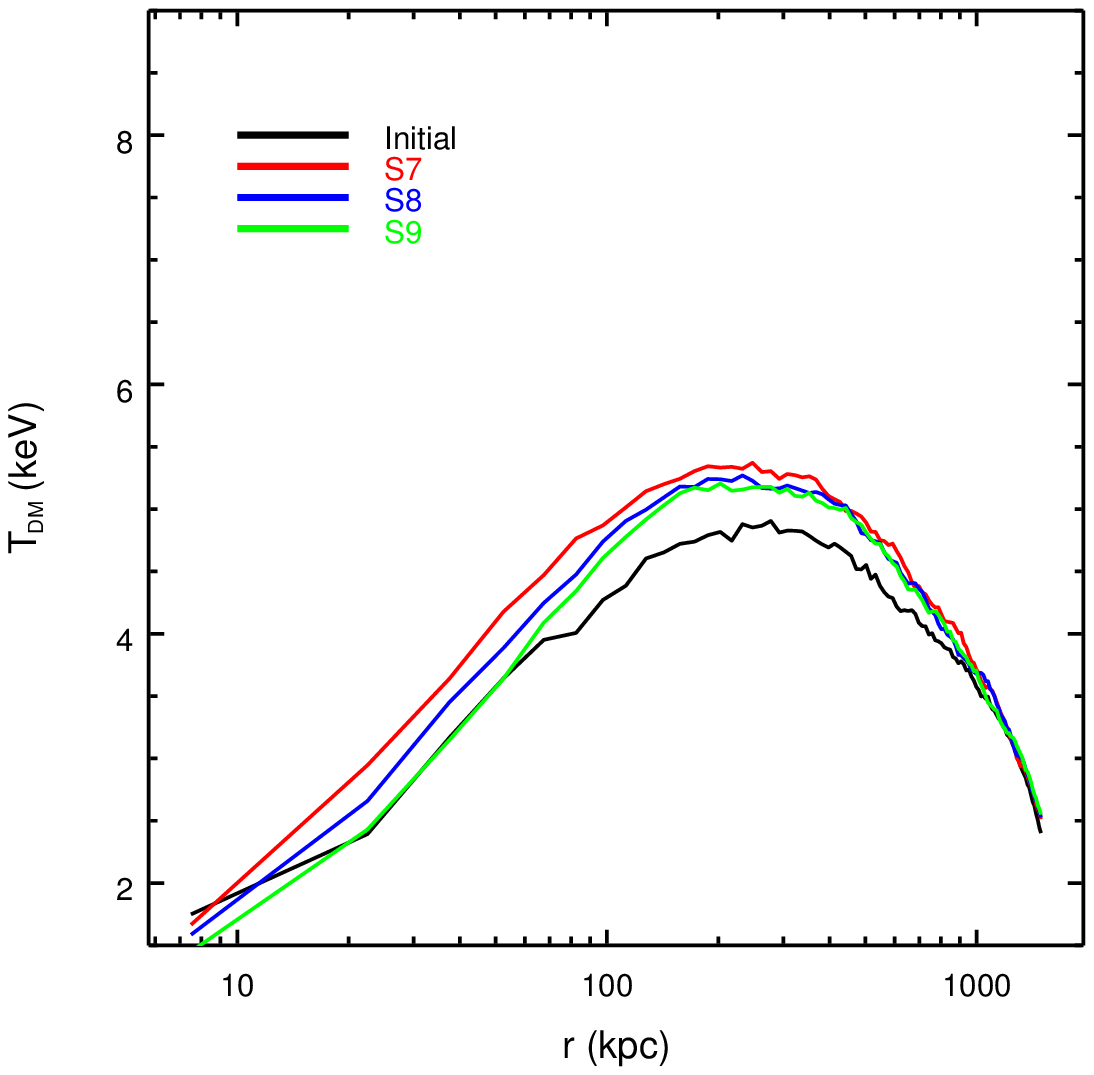}
\caption{DM temperature profiles at $t$ = 10 Gyr, compared to the initial profile. Left: Temperature profiles for the 1:1 mass-ratio simulations. Center: Temperature profiles for the 1:3 mass-ratio simulations. Right: Temperature profiles for the 1:10 mass-ratio simulations.\label{fig:tdm_profile}}
\end{center}
\end{figure*}

Recent works \citep[e.g.][]{han09,hos09,fal07} have suggested that the combination of X-ray observations with some simple assumptions about the form of the dark matter velocity dispersion profile might allow constraints to be put on quantities such as the magnitude of the dark matter velocity dispersion as well as the degree of its anisotropy. A simple way of achieving this is by defining a ``dark matter temperature'',
\begin{equation}
k_BT_{\rm DM} \equiv \frac{1}{3}{\mu}m_p\sigma_{\rm DM}^2,
\label{eqn:Tdm}
\end{equation}
where $\sigma_{\rm DM}$ is the 3-D velocity dispersion of the dark matter particles. This quantity was first introduced by \citet{ike04} to derive the dark matter velocity dispersion and compare it to the gas temperature. Figure \ref{fig:tdm_profile} shows the final dark matter temperature profiles compared to the initial profile. Unlike the gas temperature profiles, the shape of the profile is relatively unaffected by the merger though the overall normalization is increased. 

\begin{figure*}
\begin{center}
\includegraphics[width=0.3\textwidth]{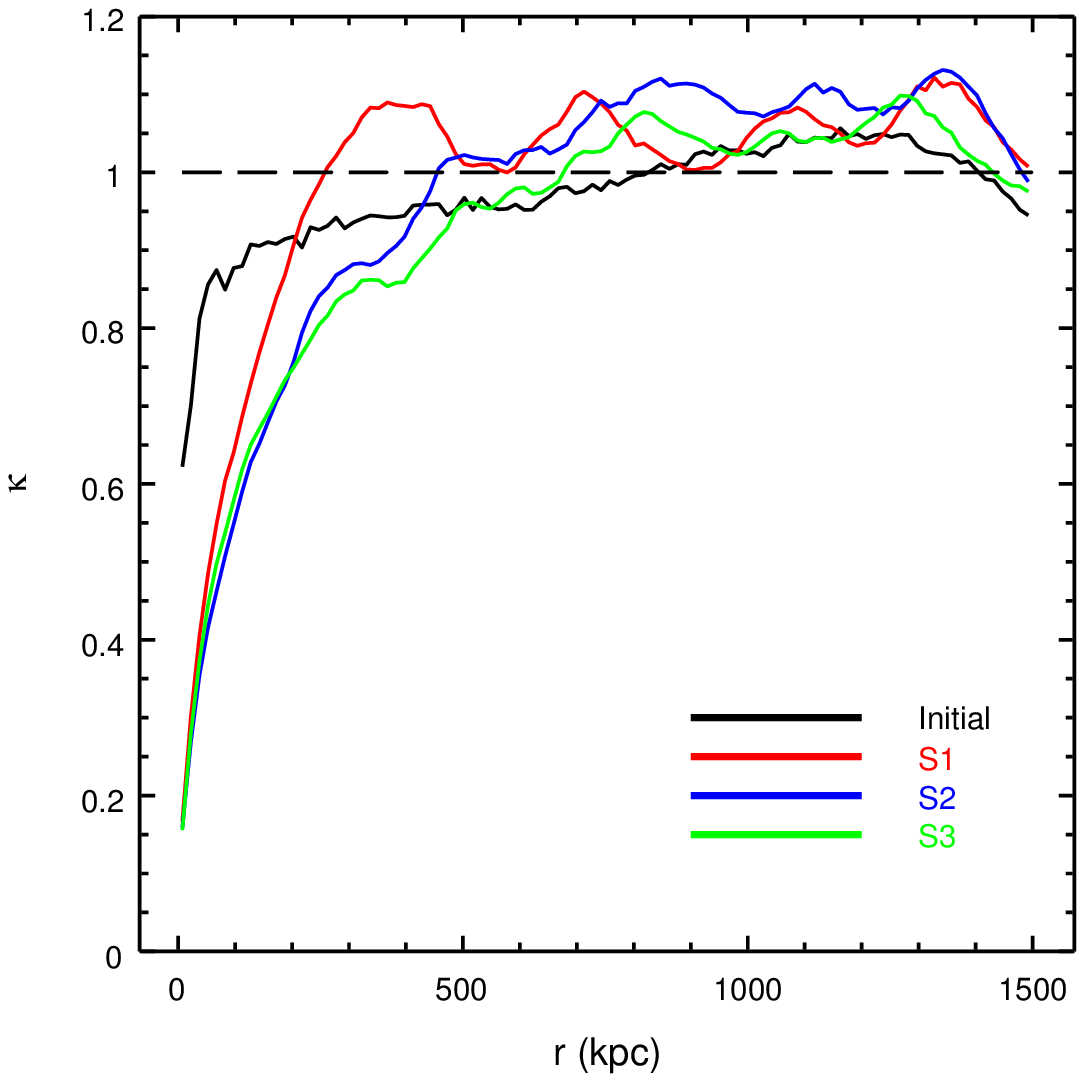}
\enspace
\enspace
\includegraphics[width=0.3\textwidth]{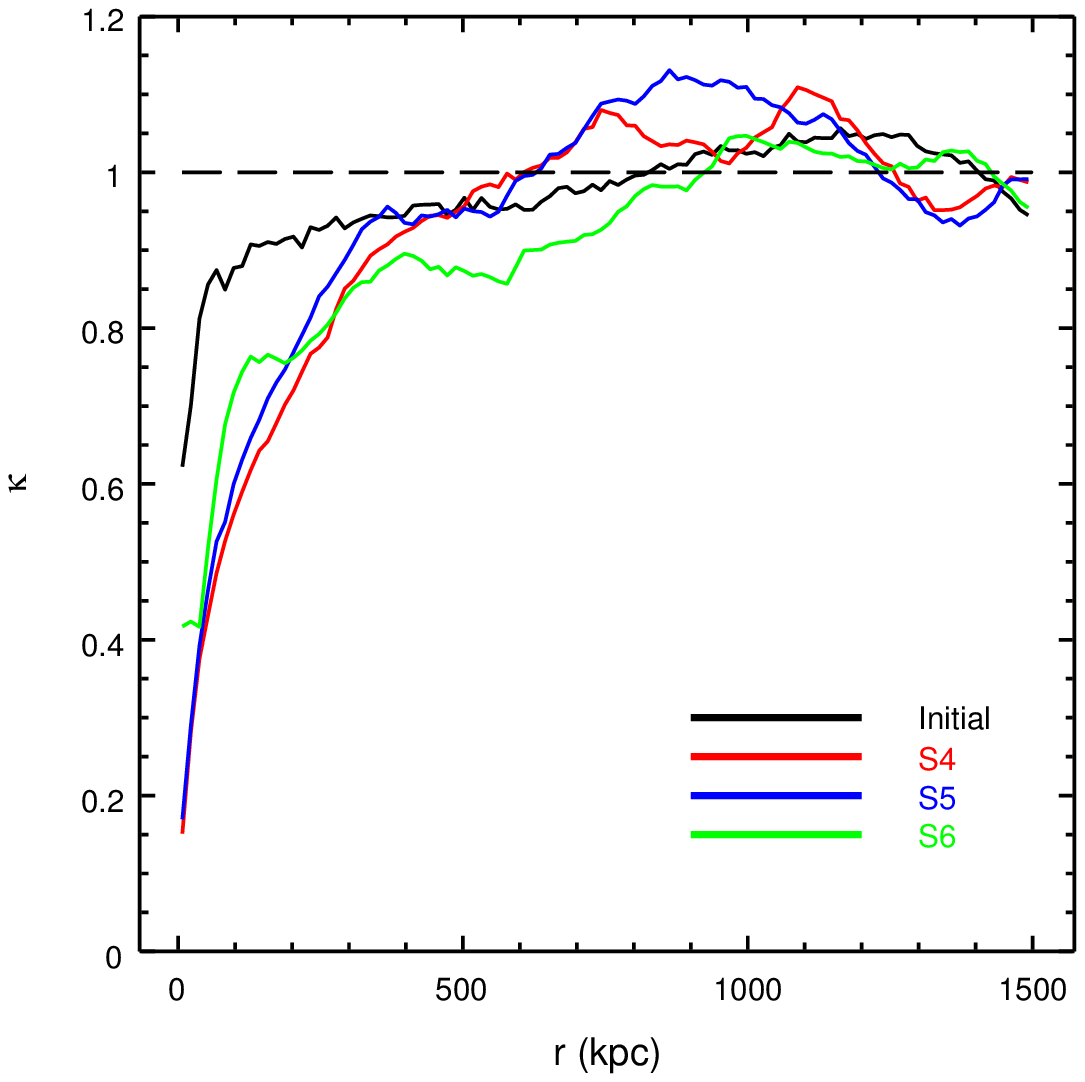}
\enspace
\enspace
\includegraphics[width=0.3\textwidth]{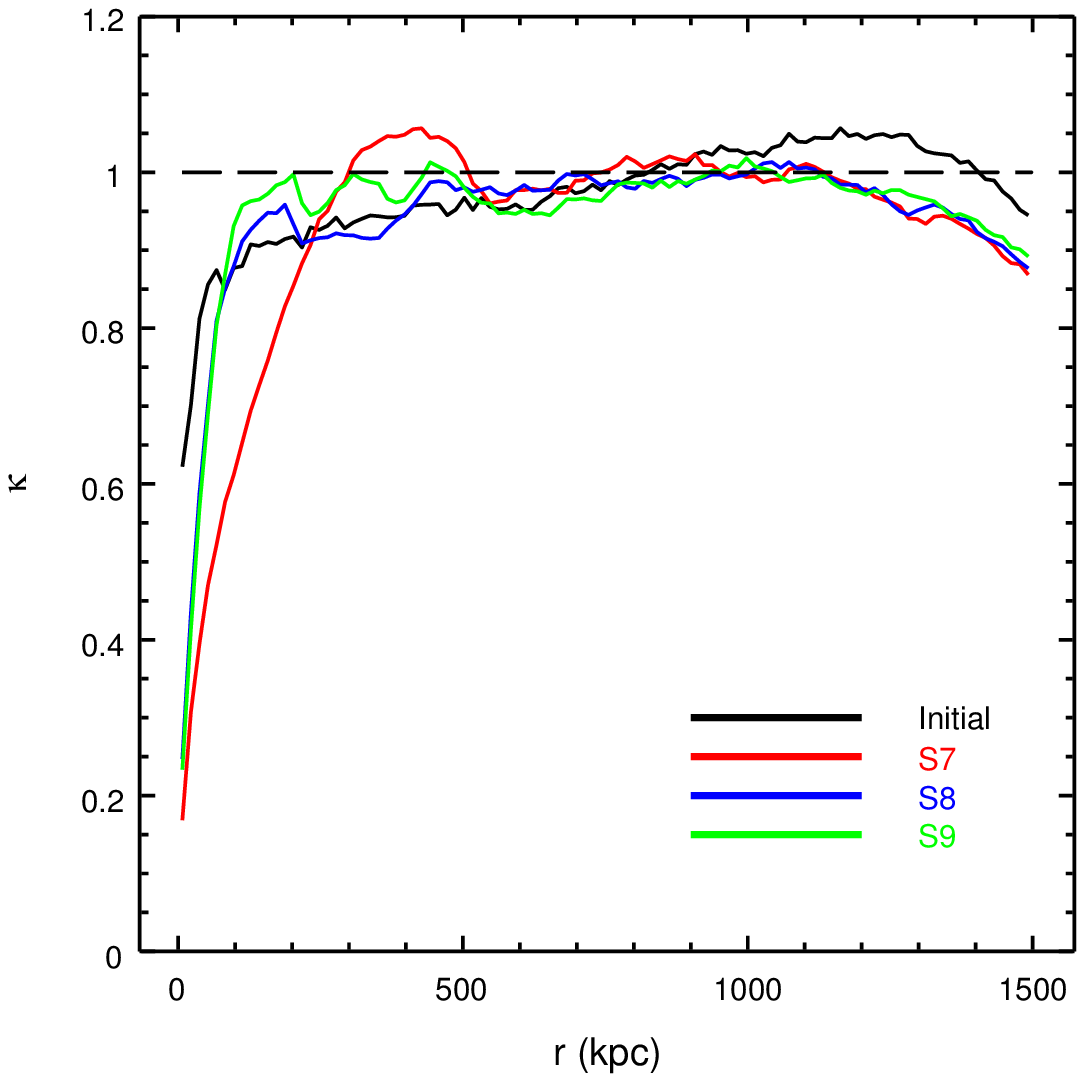}
\caption{Profiles of the ratio of the dark matter ``temperature'' to the gas temperature at $t$ = 10 Gyr, compared to the initial profile. Left: Temperature ratio profiles for the 1:1 mass-ratio simulations. Center: Temperature ratio profiles for the 1:3 mass-ratio simulations. Right: Temperature ratio profiles for the 1:10 mass-ratio simulations.\label{fig:kappa_profile}}
\end{center}
\end{figure*}

This temperature is not well-defined, due to the fact that there is no thermodynamic equilibrium for a collisionless gas (if it were valid, the Maxwellian assumption mentioned in Section \ref{sec:ICs} for generating initial conditions would be accurate), but it may still be phenomenologically useful. Assuming that the velocity of an individual particle is determined solely by the gravitational potential, the equivalence principle might lead one to expect a similar temperature profile between the dark matter and gas components of a galaxy cluster, e.g. $T_{\rm DM}(r) = {\kappa}T_g(r)$. In this way the observed gas temperature may be used to determine the kinetic properties of the dark matter. In the case of relaxed galaxy clusters, it may be sensible to assume $\kappa \approx 1$. Figure \ref{fig:kappa_profile} shows the temperature ratio radial profiles for each simulation.  For our initial relaxed primary cluster, the equality $T_{\rm gas} = T_{\rm DM}$ is valid to within 10\% for most of the radial range, excepting the cluster core within about 50-100~kpc. For the final merged clusters, over most of the radial range the assumption is still accurate to within 10\%, but within the inner 100-200~kpc the profiles begin to diverge. In the innermost regions of the final merger remnants, $T_{\rm DM}$ is only $\sim$20-30\% of $T_{\rm gas}$.

Our idealized progenitor clusters began with zero gas velocities relative to the rest frame of the cluster, so it is instructive to examine the velocities which are still present in our resultant systems at the end of our simulations. Figure \ref{fig:turb_profile} shows the total gas velocity dispersion profile, $\sigma_g = \sqrt(\sigma_r^2+\sigma_{\theta}^2+\sigma_{\phi}^2)$, for each simulation. Each merger converts a fraction of its energy into random motions of the gas, with the final velocity dispersion $\sigma_g$ ranging from 100-500~km s$^{-1}$. By far, the head-on mergers generate more turbulent motion than the off-center mergers, which may have significant velocity dispersions at larger radii $r \sim 300$~kpc but fail to penetrate the center of the primary, and the resulting dispersions are correspondingly less. Figure \ref{fig:rad_profile} shows the mean radial velocity profile, $\bar{v_r}$, for each simulation. By the end of each simulation, the mean radial velocities are quite small, with speeds $|\bar{v_r}| \simlt 50$~km s${^-1}$, with the largest velocities at the highest annuli of our profiles ($r \simgt 1$~Mpc). Figure \ref{fig:circ_profile} shows the circular velocity profile at each radius, $V_c = \sqrt(\bar{v_{\theta}}^2 + \bar{v_{\phi}}^2)$. Not surprisingly, head on-mergers generate the least amount of circular velocity ($V_c \simlt 100$~km s$^{-1}$), whereas off-center mergers induce more rotation, up to $V_c \sim 400-500$~km s$^{-1}$ in the strongest cases (equal mass mergers). The highest circular speeds at the end of each simulation are at radii $r \simgt 100$~kpc. Finally, Figure \ref{fig:mach_profile} shows the ratio of the velocity dispersion and the circular velocity to the sound speed of the gas squared; this provides a measure of the pressure support that may be provided by these motions. These values range from $\sim$10-20$\%$ over the radii $r \simlt$ 200~kpc, with the highest levels of support reaching up to $\sim$30-40$\%$ in the cluster outskirts. These levels of non-thermal pressure support from random and circular motions are consistent with previous results from cosmological simulations \citep[e.g.,][]{pif08,fan09,lau09}.

In our nine simulations we saw from the entropy maps that the final central entropy and the size of the entropy core varies as both the impact parameter and cluster mass ratio is varied. Figure \ref{fig:entr_profile} shows the final entropy profile for each simulation, with the initial entropy profile of the primary cluster shown for comparison. The combined effect of the heating and expansion of the central gas of the cluster is to raise the entropy of this central gas to a high adiabat. All of the final profiles exhibit a large entropy core where the entropy flattens out in the center, out to a radius of $r \sim 100-500$~kpc, and the value of the central entropy is significantly larger than the initial value, with $S_0 \sim$ 100-500 keV cm$^2$. The behavior of the profile outside of this central floor is close to power-law, essentially unchanged from the progenitor clusters, consistent with the observational and simulation works that show that the real difference between clusters in this regard is in the height of the entropy floor. 

Finally, the entropy of the gas in the ICM is closely related to the cooling time, the total energy of the gas divided by the energy loss rate. It is given by \citep{zhg03}:

\begin{equation}
t_c = 28.69~{\rm Gyr}\left({1.2 \over g}\right)\left({k_BT \over {\rm keV}}\right)^{\frac{1}{2}}\left({n_e \over {10^{-3}~{\rm cm}^{-3}}}\right)^{-1} 
\end{equation}

where $n_e$ is the electron number density, $k_BT$ is the gas temperature in keV, and $g$ is the frequency-averaged Gaunt factor. With some manipulation, the cooling time can be expressed as a function of entropy and electron density:

\begin{equation}
t_c = 28.69~{\rm Gyr}\left({1.2 \over g}\right)\left({S \over {100~{\rm keV~cm^2}}}\right)^{\frac{1}{2}}\left({n_e \over {10^{-3}~{\rm cm}^{-3}}}\right)^{-2/3} 
\label{eqn:tcool}
\end{equation}

In Figure \ref{fig:cool_profile} we show the cooling time profiles for the nine simulations at $t$ = 10~Gyr, with the age of the universe marked on the plot for comparison. All of the central cooling times have been significantly increased, by a factor of $\sim$6-20 over the initial value of $\sim$1~Gyr at the very center of the cluster. From Equation \ref{eqn:tcool}, we see that this is consistent with the large increases in entropy in the cluster cores.

\section{Discussion}\label{sec:disc}

\subsection{To Mix or Not to Mix?}\label{sec:mixing_disc}

In previous investigations of ICM mixing using controlled merger simulations, it has been reported that the ICM component in these simulations does not efficiently mix. \citet{rit02}, using SPH simulations, defined the degree of mixing as we have above and reported that the ICM only mixed well in the centers of merger remnants. \citet{poo08} investigated mixing from the perspective of metallicity gradients, and found no significant flattening of the metallicity profile (which would imply mixing of gas of different metallicities) following a merger in any of their simulations. \citet{ric01} did not investigate mixing, and in that case the reduced resolution of their simulations would have resulted in the suppression of smaller-wavelength unstable perturbations, lessening the amount of mixing. Additionally, we are concerned with the effects of merger heating on progenitor clusters with cool cores, whereas the initial systems in that study were more akin to our resultant systems. On the other hand, in an idealized setup of an infalling subcluster into a large cluster of galaxies, \citet{tak05} showed that as the subcluster falls into the core of the main cluster, its gas is thoroughly mixed due to Rayleigh-Taylor and Kelvin-Helmholtz instabilities. The resolution of the grid ($\Delta{x} = 2$~kpc) and the values of the chosen cluster masses in their simulations are similar to the values in our simulations.

\begin{figure*}
\begin{center}
\includegraphics[width=0.3\textwidth]{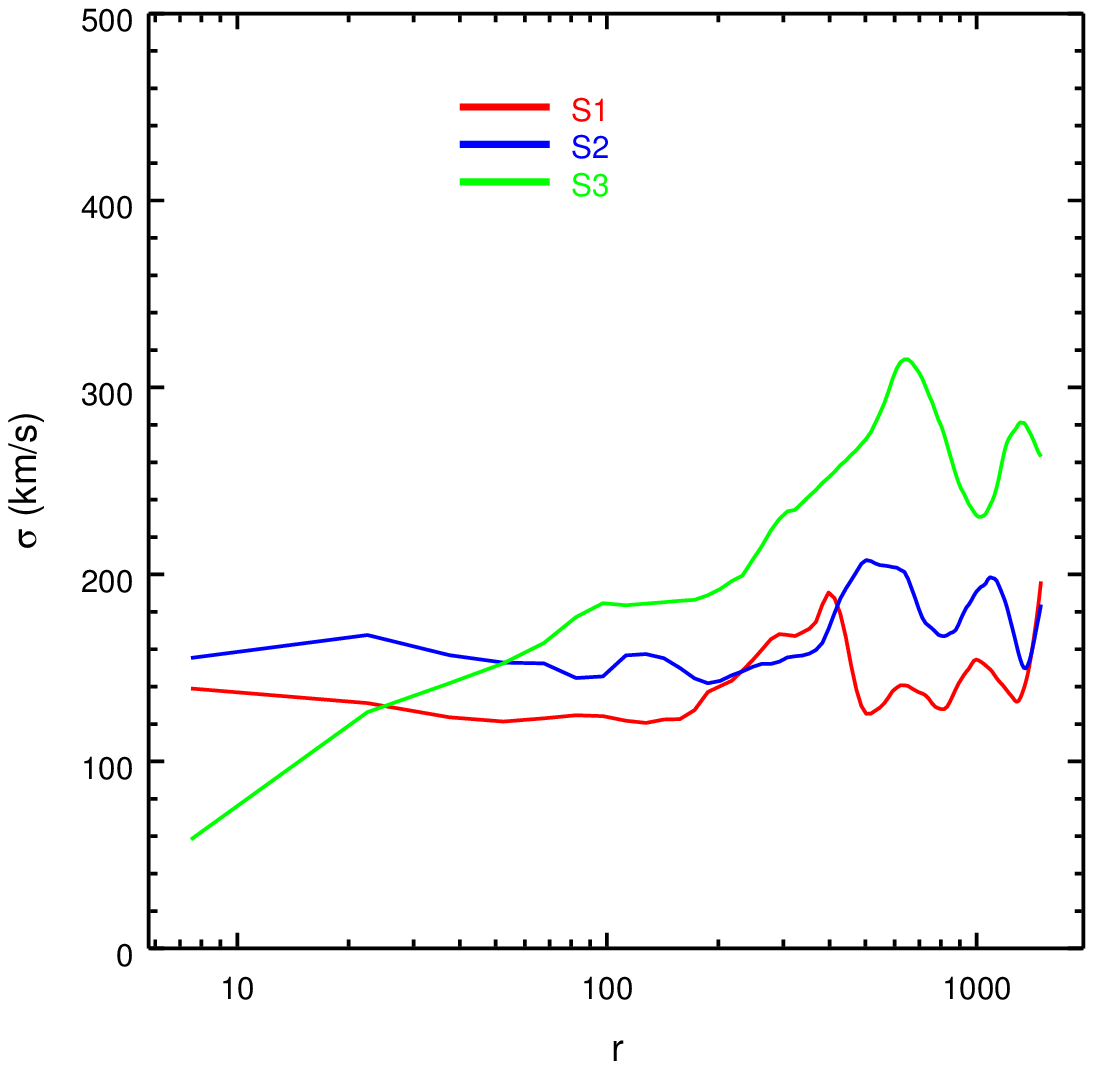}
\enspace
\includegraphics[width=0.3\textwidth]{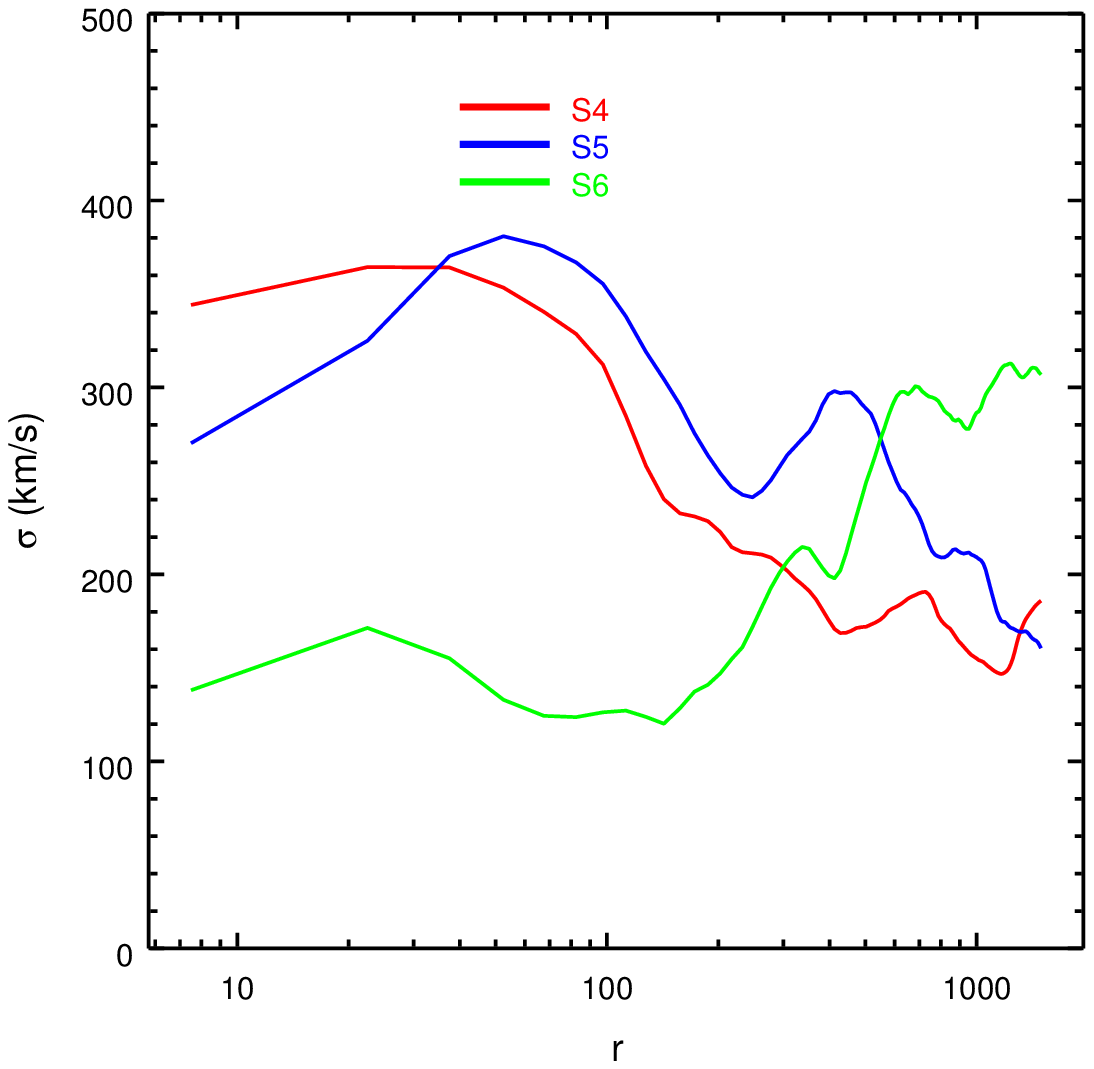}
\enspace
\includegraphics[width=0.3\textwidth]{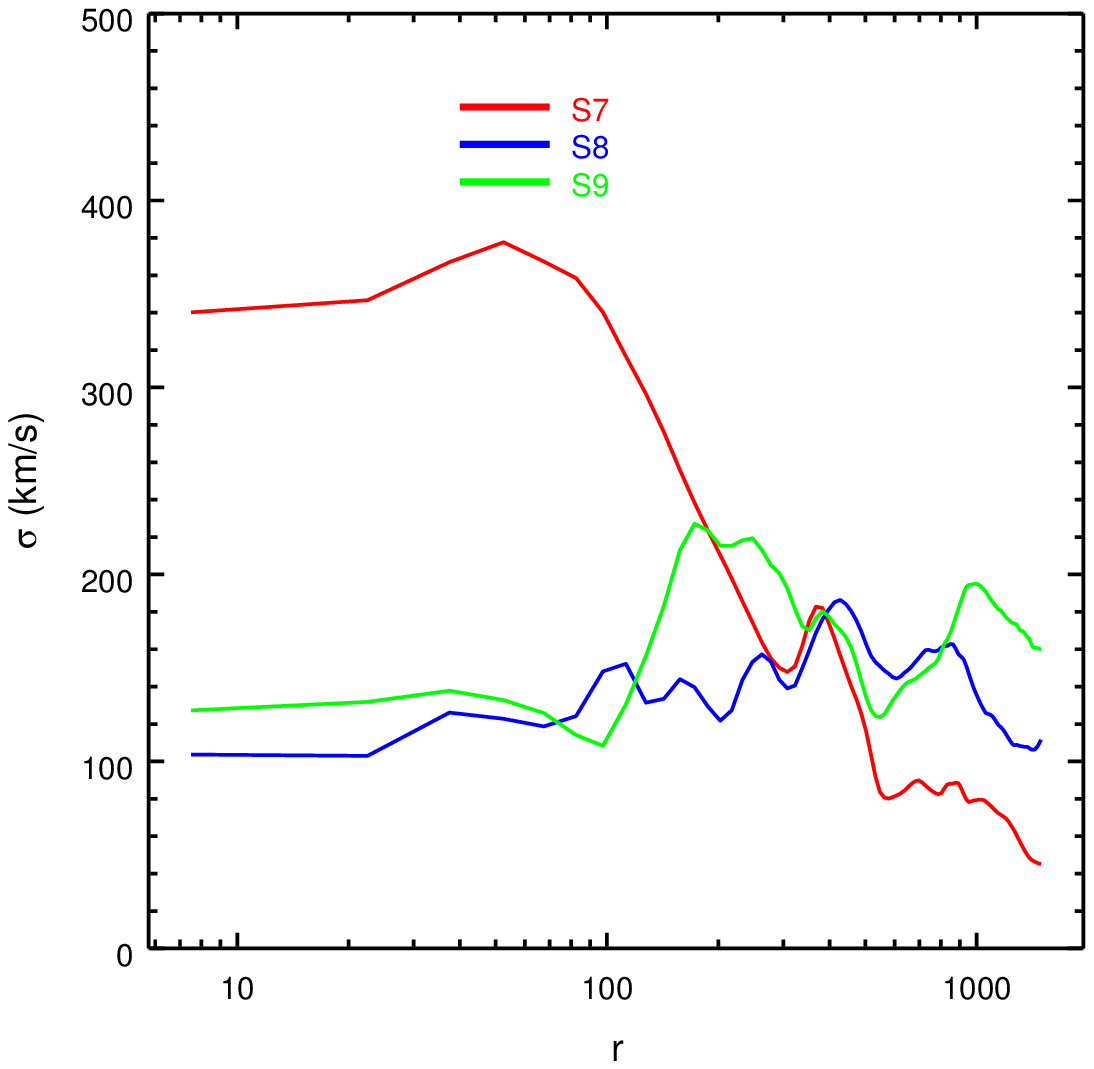}
\caption{Profiles of the gas velocity dispersion at $t$ = 10 Gyr. Left: Velocity dispersion profiles for the 1:1 mass-ratio simulations. Center: Velocity dispersion profiles for the 1:3 mass-ratio simulations. Right: Velocity dispersion profiles for the 1:10 mass-ratio simulations.\label{fig:turb_profile}}
\end{center}
\end{figure*}

\begin{figure*}
\begin{center}
\includegraphics[width=0.3\textwidth]{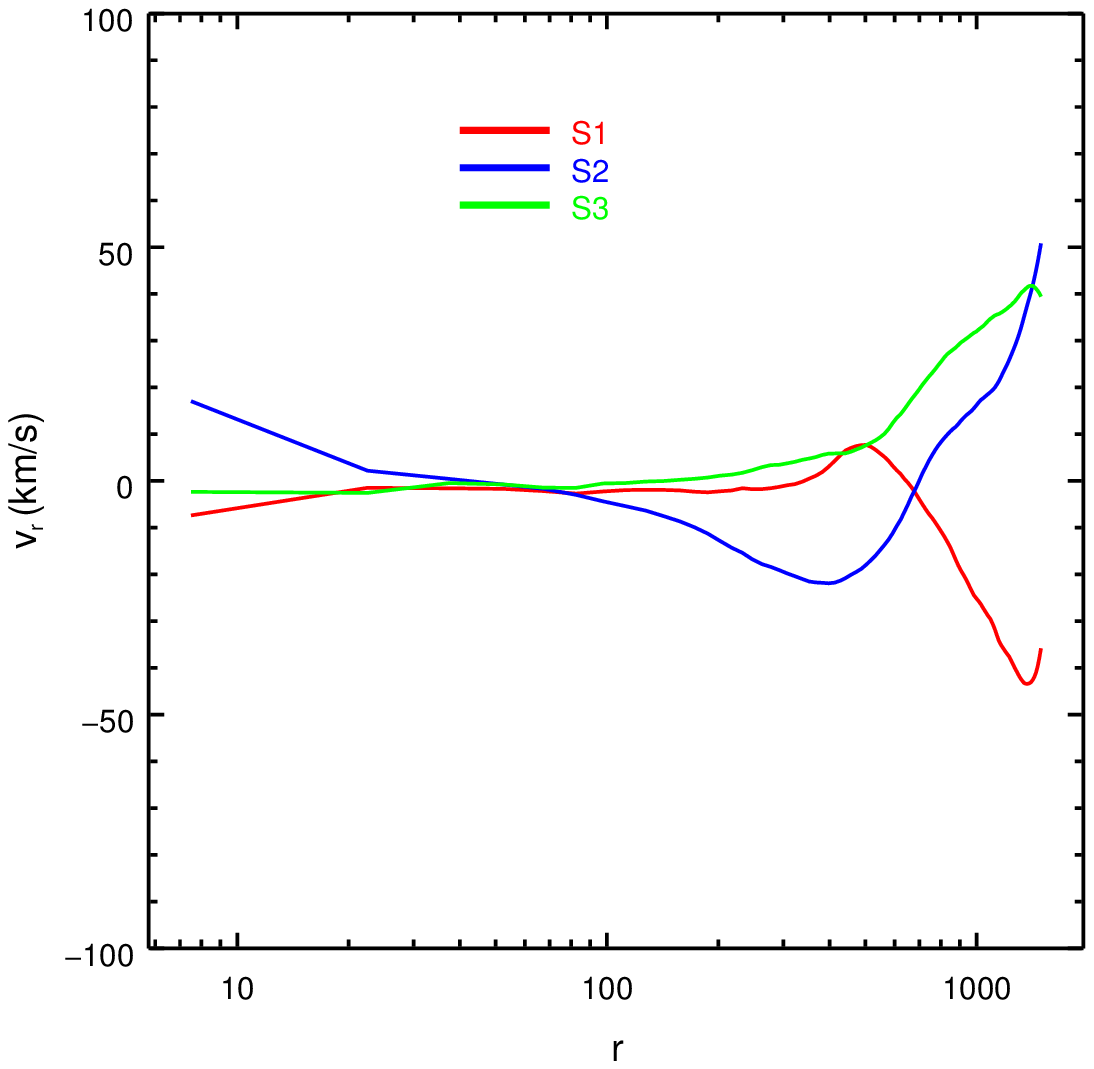}
\enspace
\includegraphics[width=0.3\textwidth]{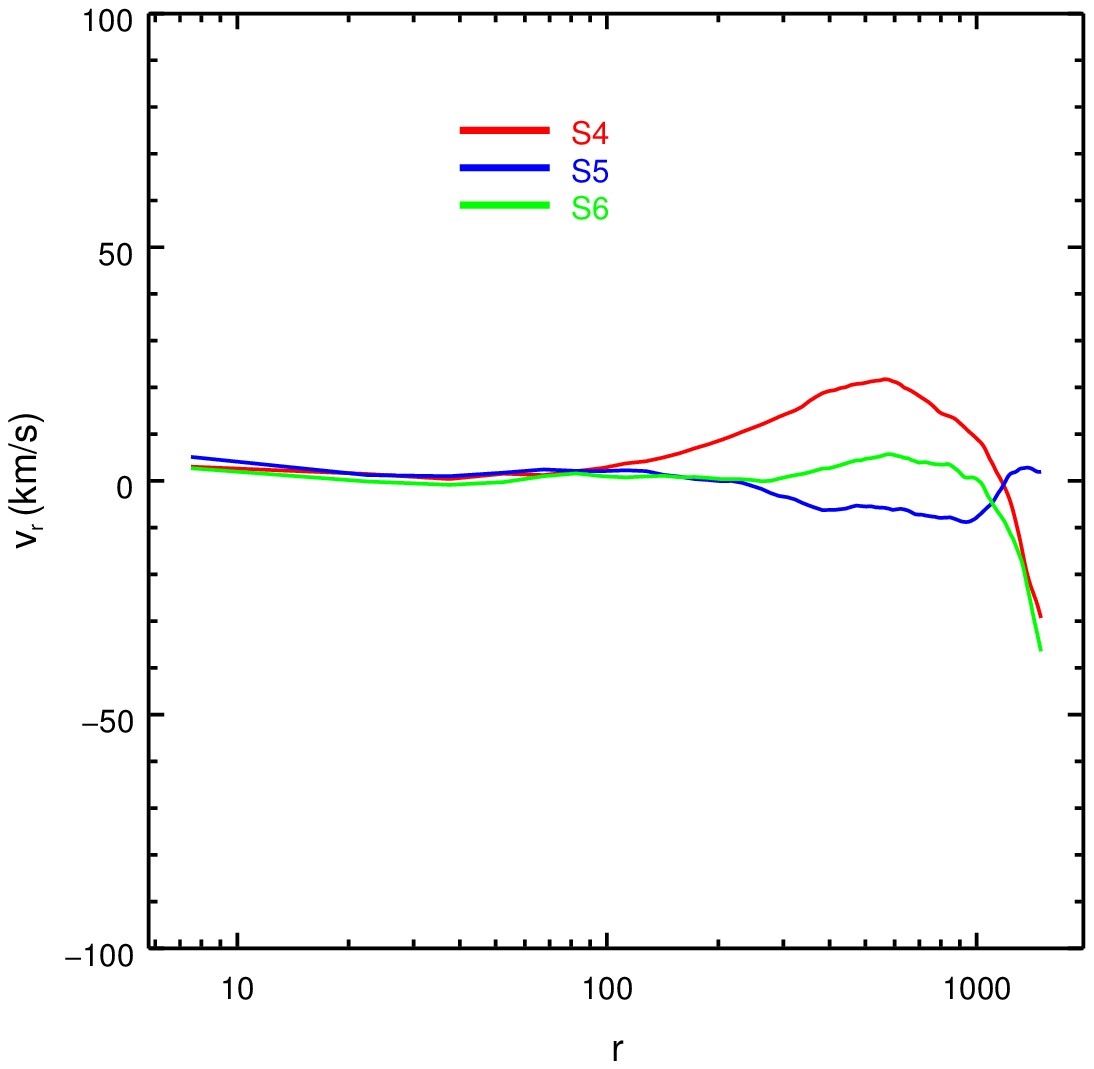}
\enspace
\includegraphics[width=0.3\textwidth]{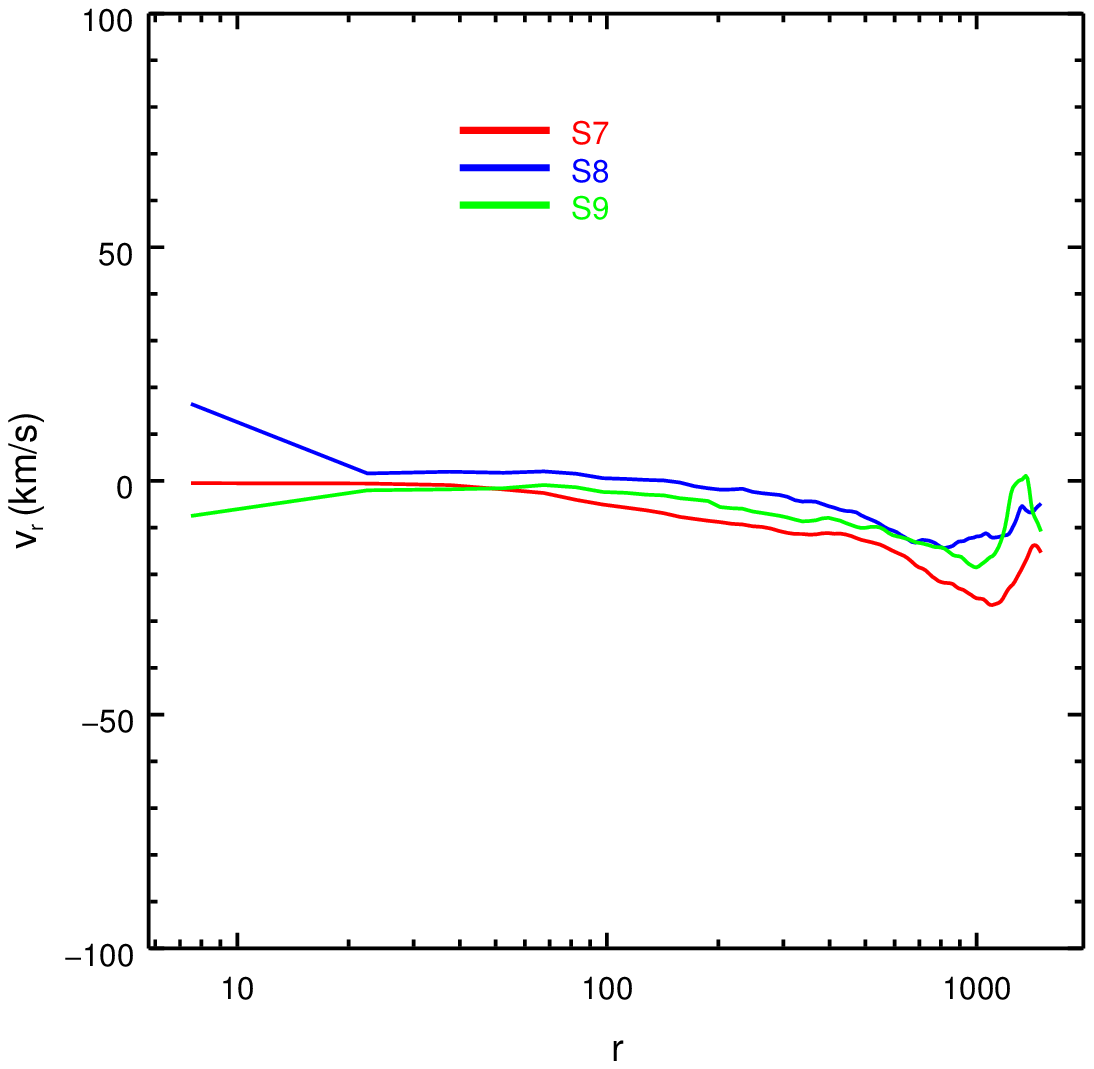}
\caption{Profiles of the gas radial velocity at $t$ = 10 Gyr. Left: Radial velocity profiles for the 1:1 mass-ratio simulations. Center: Radial velocity profiles for the 1:3 mass-ratio simulations. Right: Radial velocity profiles for the 1:10 mass-ratio simulations.\label{fig:rad_profile}}
\end{center}
\end{figure*}

\begin{figure*}
\begin{center}
\includegraphics[width=0.3\textwidth]{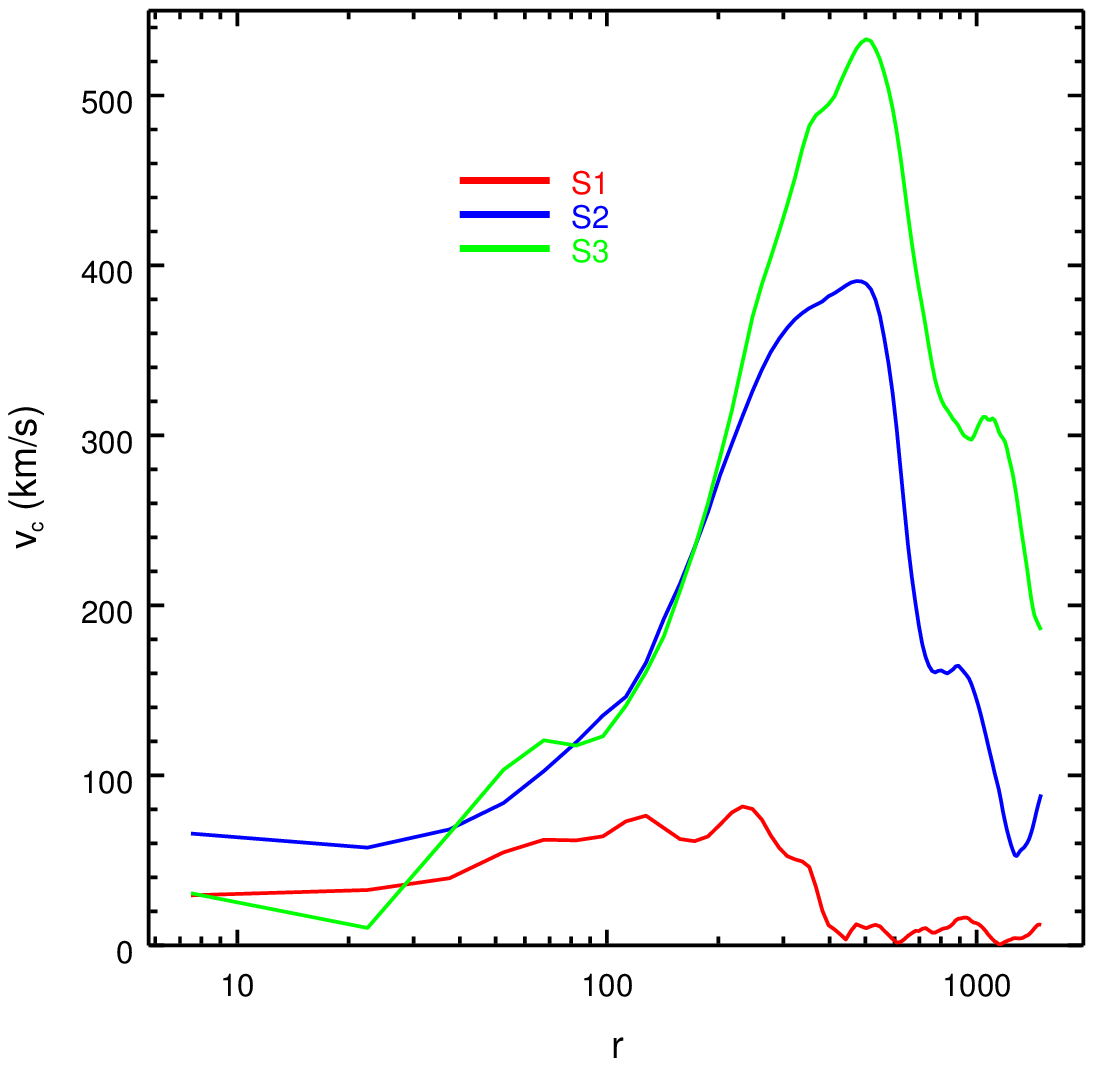}
\enspace
\includegraphics[width=0.3\textwidth]{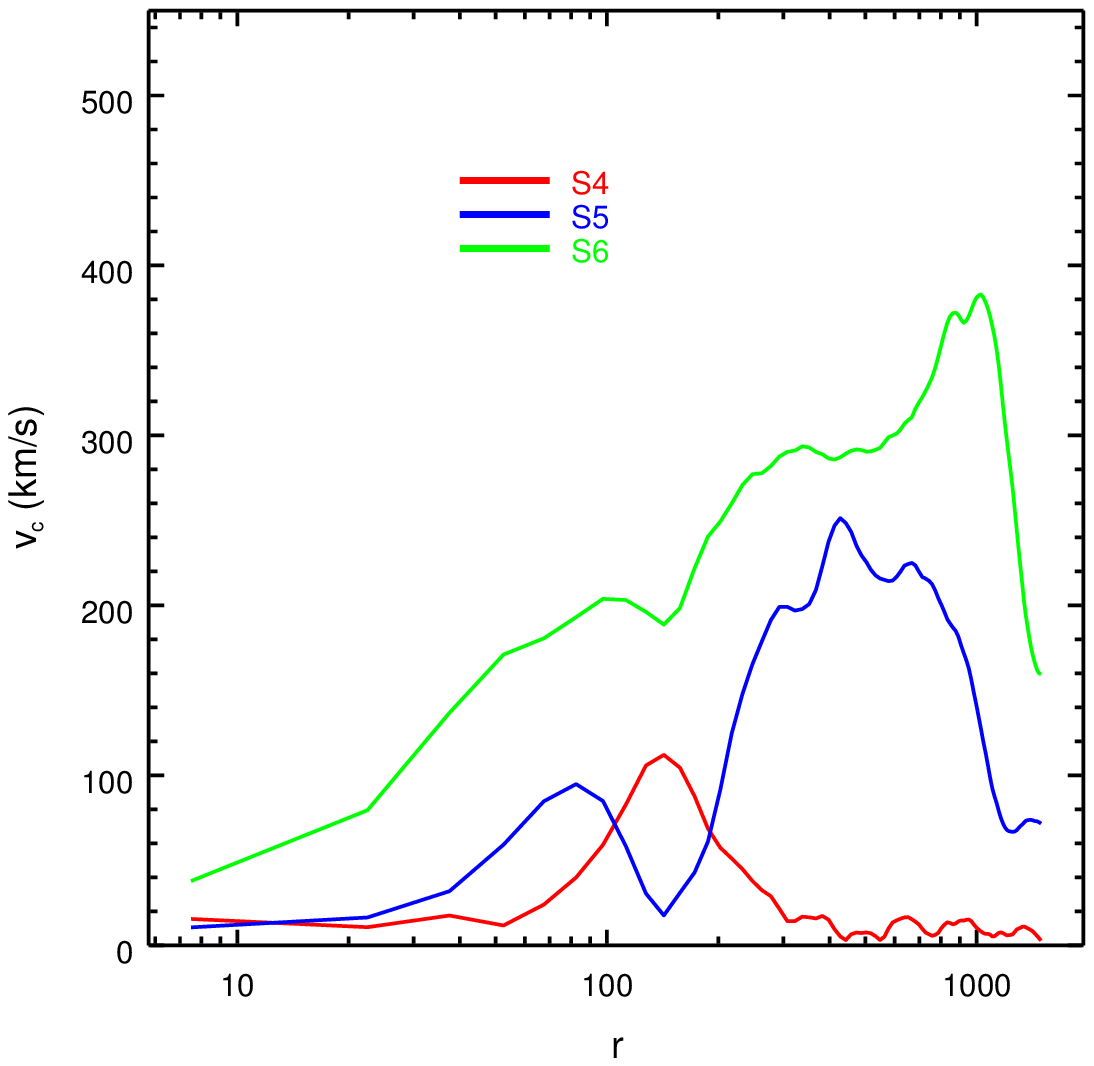}
\enspace
\includegraphics[width=0.3\textwidth]{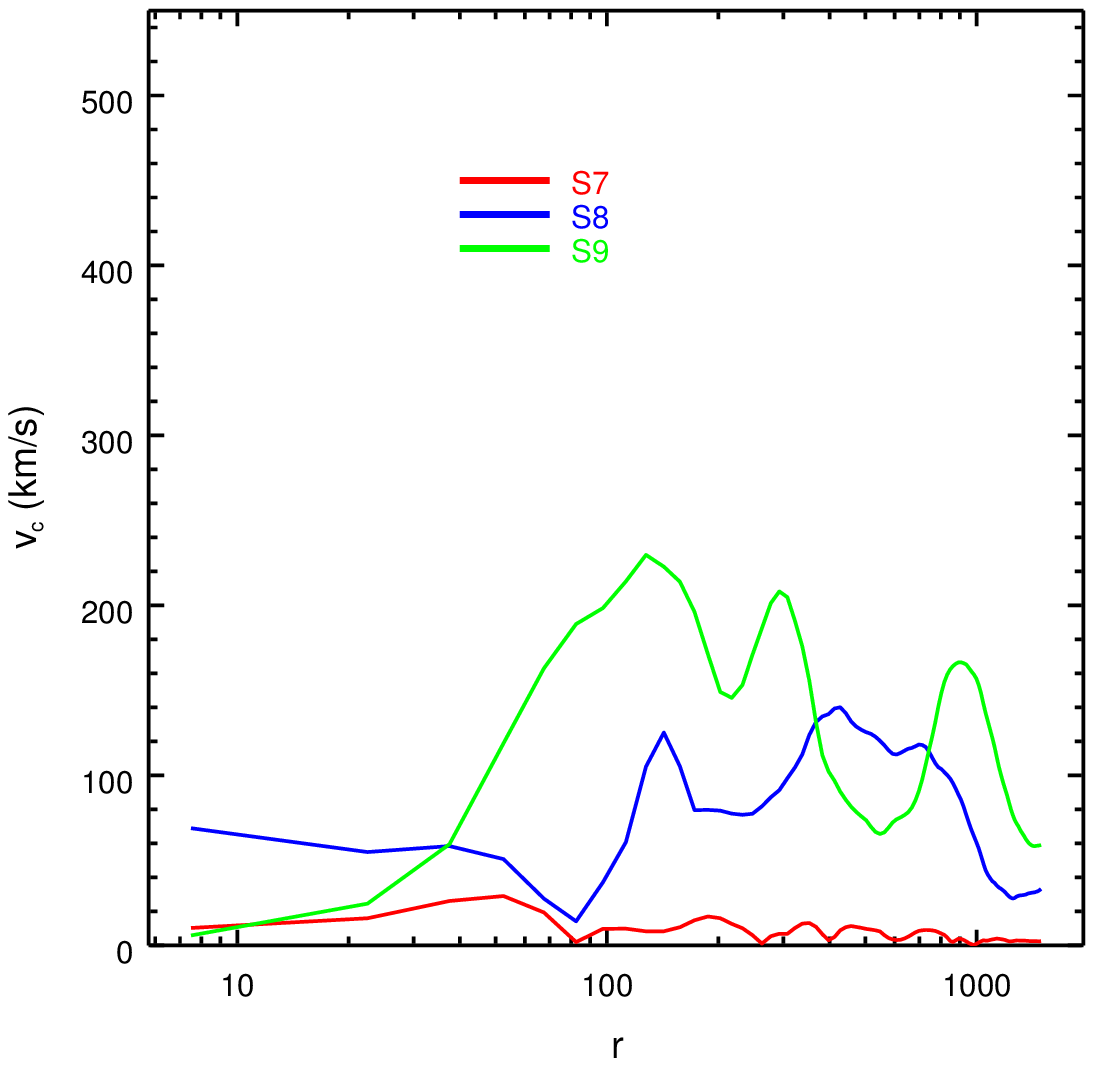}
\caption{Profiles of the gas circular velocity at $t$ = 10 Gyr. Left: Circular velocity profiles for the 1:1 mass-ratio simulations. Center: Circular velocity profiles for the 1:3 mass-ratio simulations. Right: Circular velocity profiles for the 1:10 mass-ratio simulations.\label{fig:circ_profile}}
\end{center}
\end{figure*}

\begin{figure*}
\begin{center}
\includegraphics[width=0.3\textwidth]{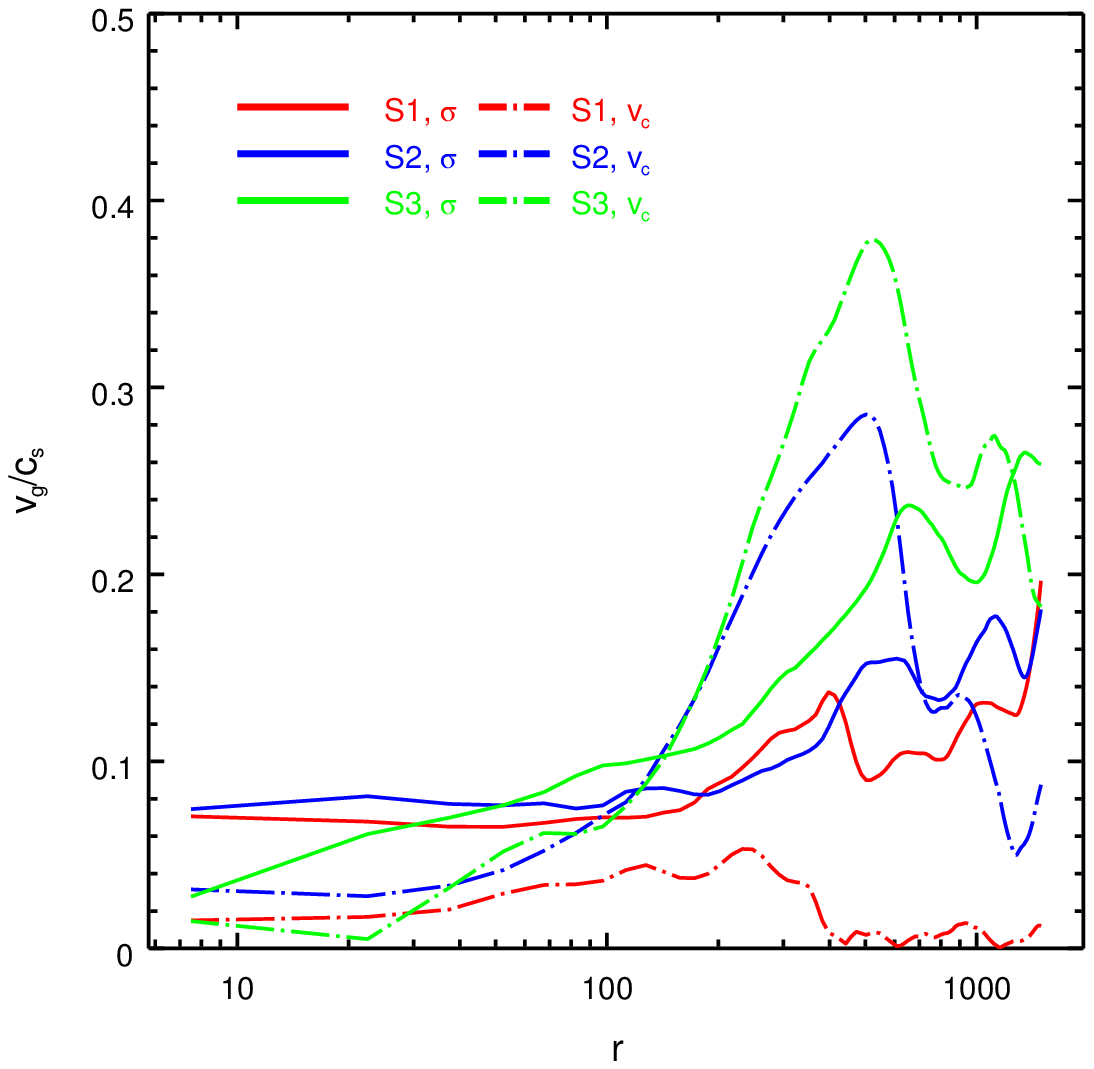}
\enspace
\includegraphics[width=0.3\textwidth]{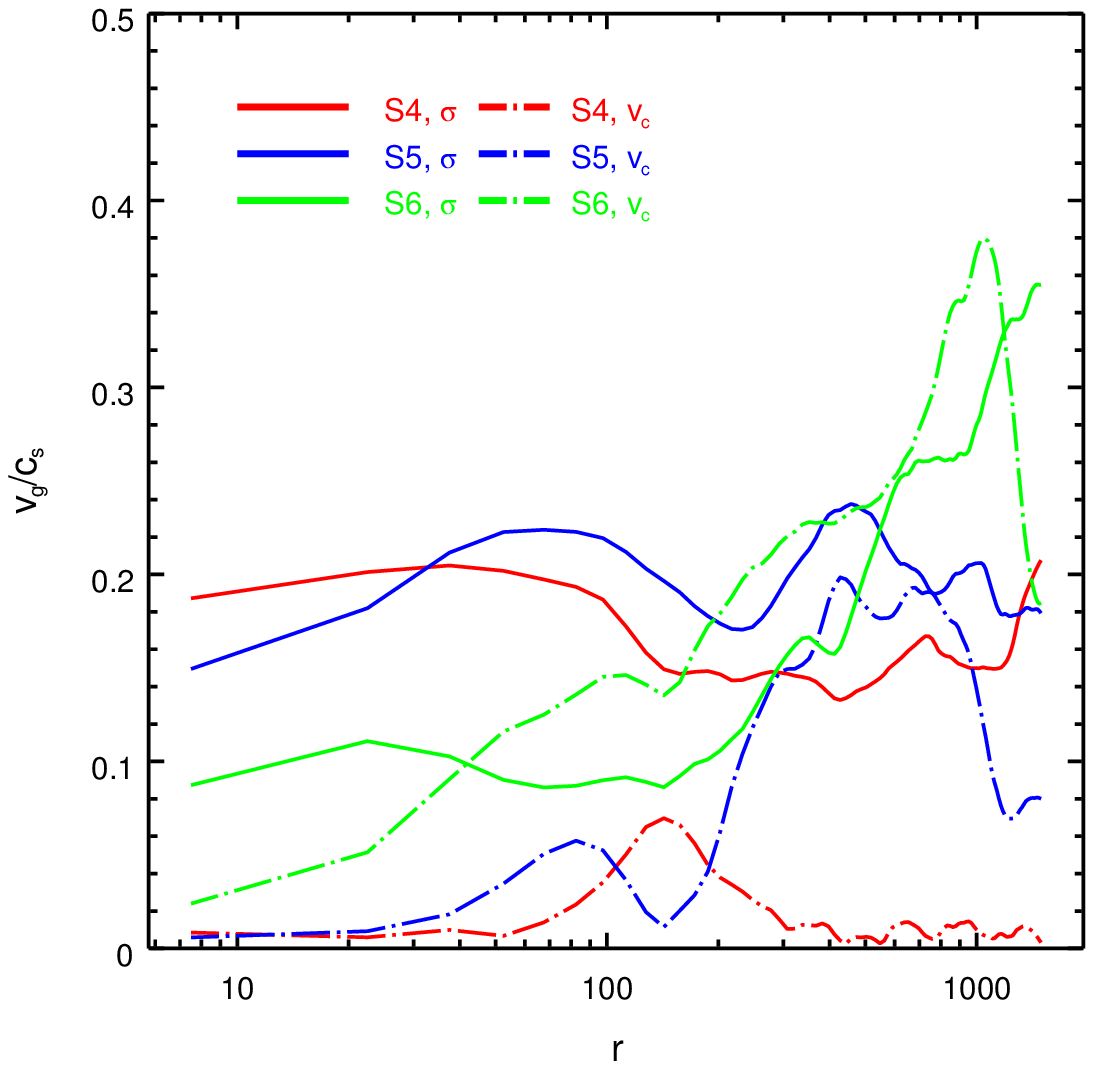}
\enspace
\includegraphics[width=0.3\textwidth]{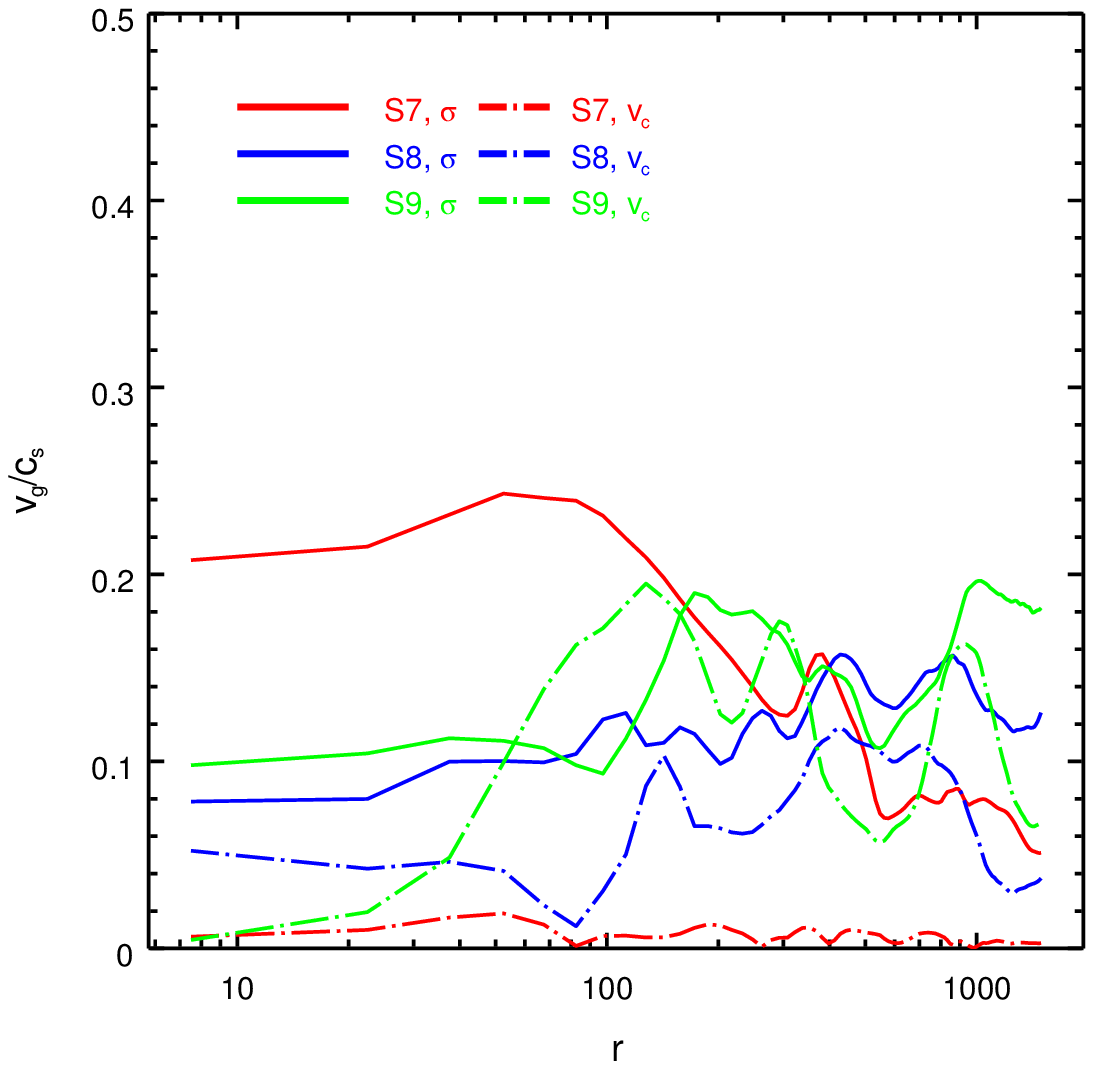}
\caption{Profiles of the ratio of the gas velocity to the local sound speed at $t$ = 10 Gyr, for both the velocity dispersion (solid lines) and the circular velocity (dot-dashed lines). Left: Profiles for the 1:1 mass-ratio simulations. Center: Profiles for the 1:3 mass-ratio simulations. Right: Profiles for the 1:10 mass-ratio simulations.\label{fig:mach_profile}}
\end{center}
\end{figure*}

\begin{figure*}
\begin{center}
\includegraphics[width=0.3\textwidth]{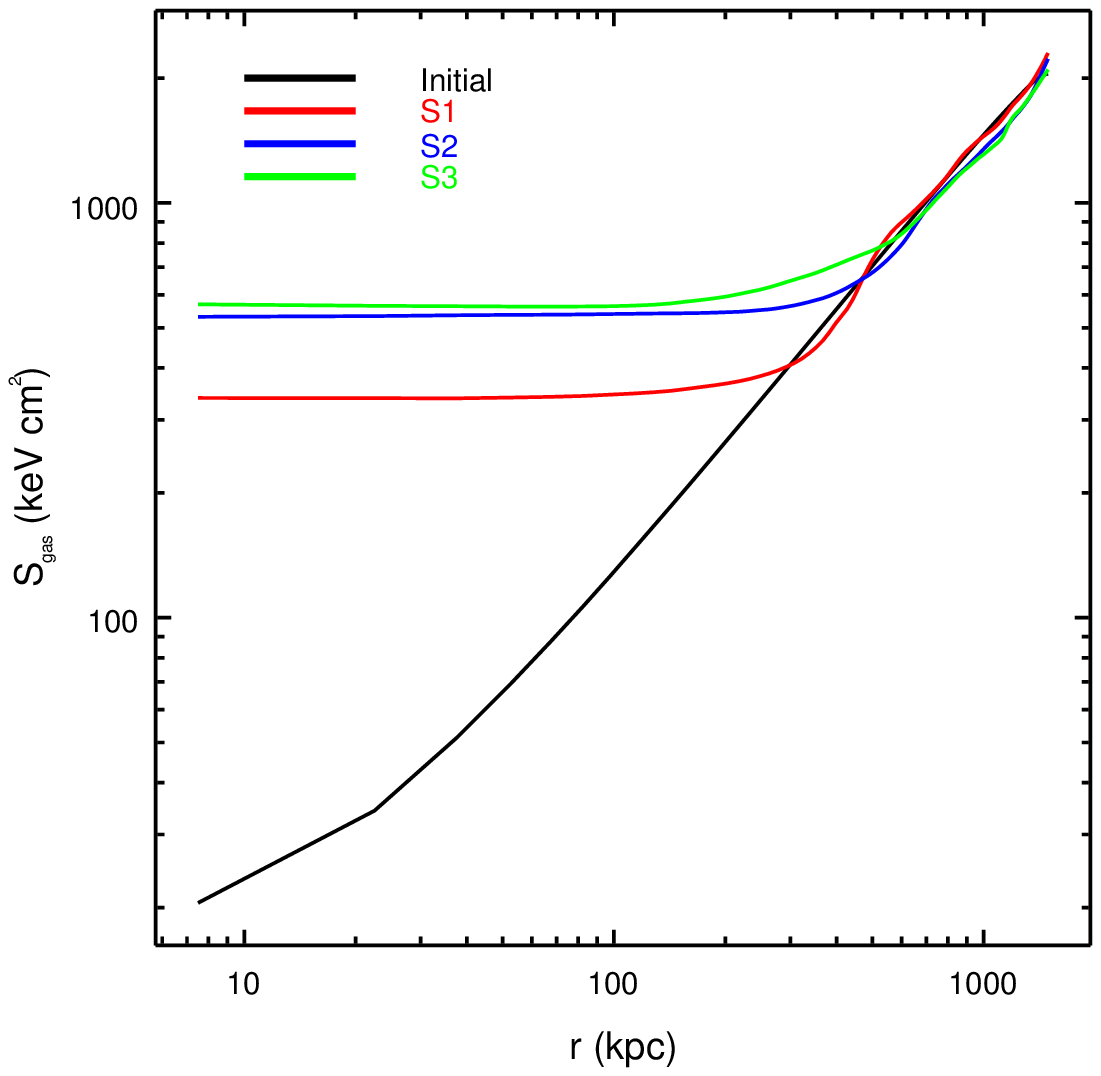}
\enspace
\includegraphics[width=0.3\textwidth]{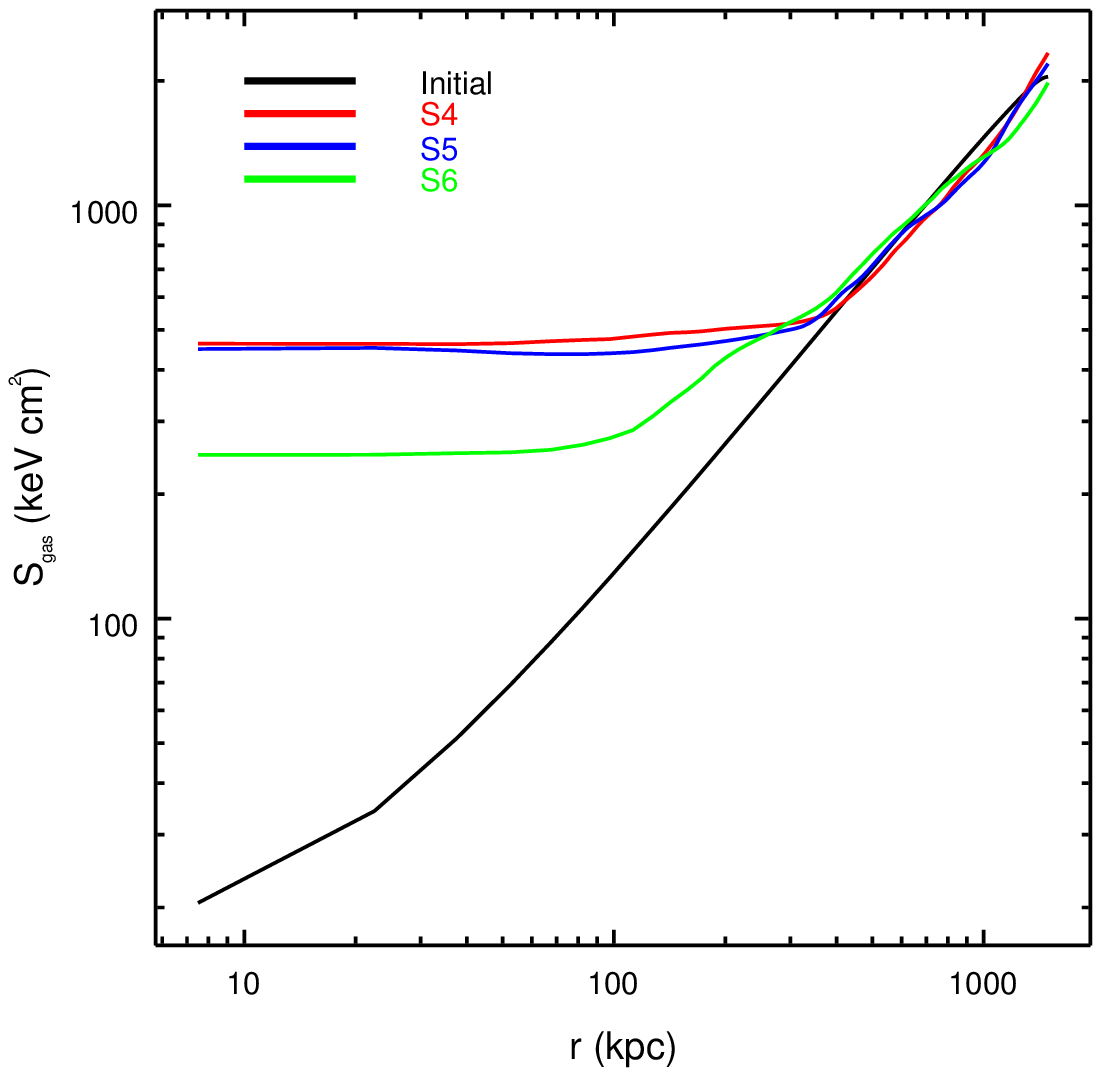}
\enspace
\includegraphics[width=0.3\textwidth]{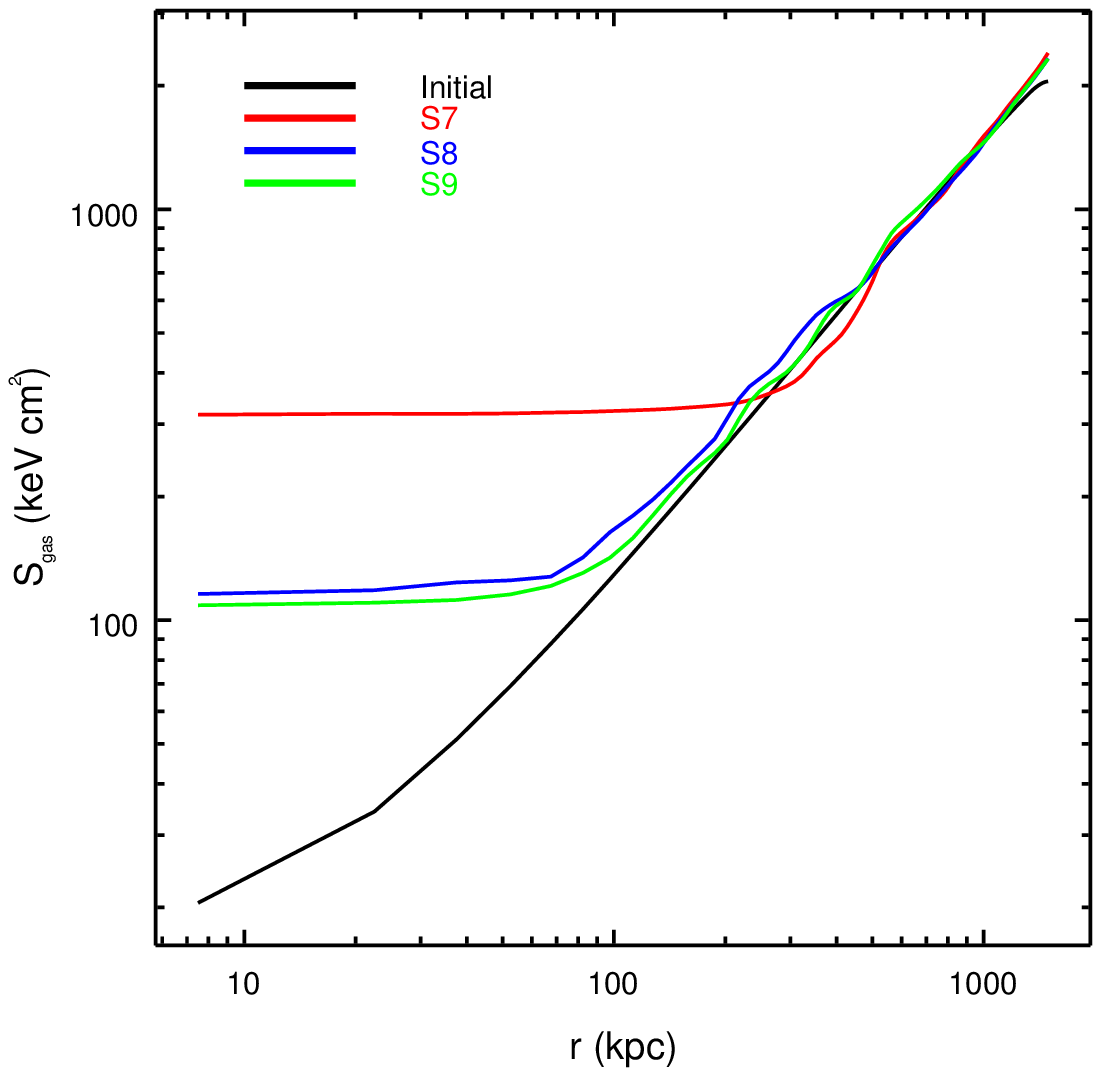}
\caption{Gas entropy profiles at $t$ = 10 Gyr, compared to the initial profile. Left: Entropy profiles for the 1:1 mass-ratio simulations. Center: Entropy profiles for the 1:3 mass-ratio simulations. Right: Entropy profiles for the 1:10 mass-ratio simulations.\label{fig:entr_profile}}
\end{center}
\end{figure*}

\begin{figure*}
\begin{center}
\includegraphics[width=0.3\textwidth]{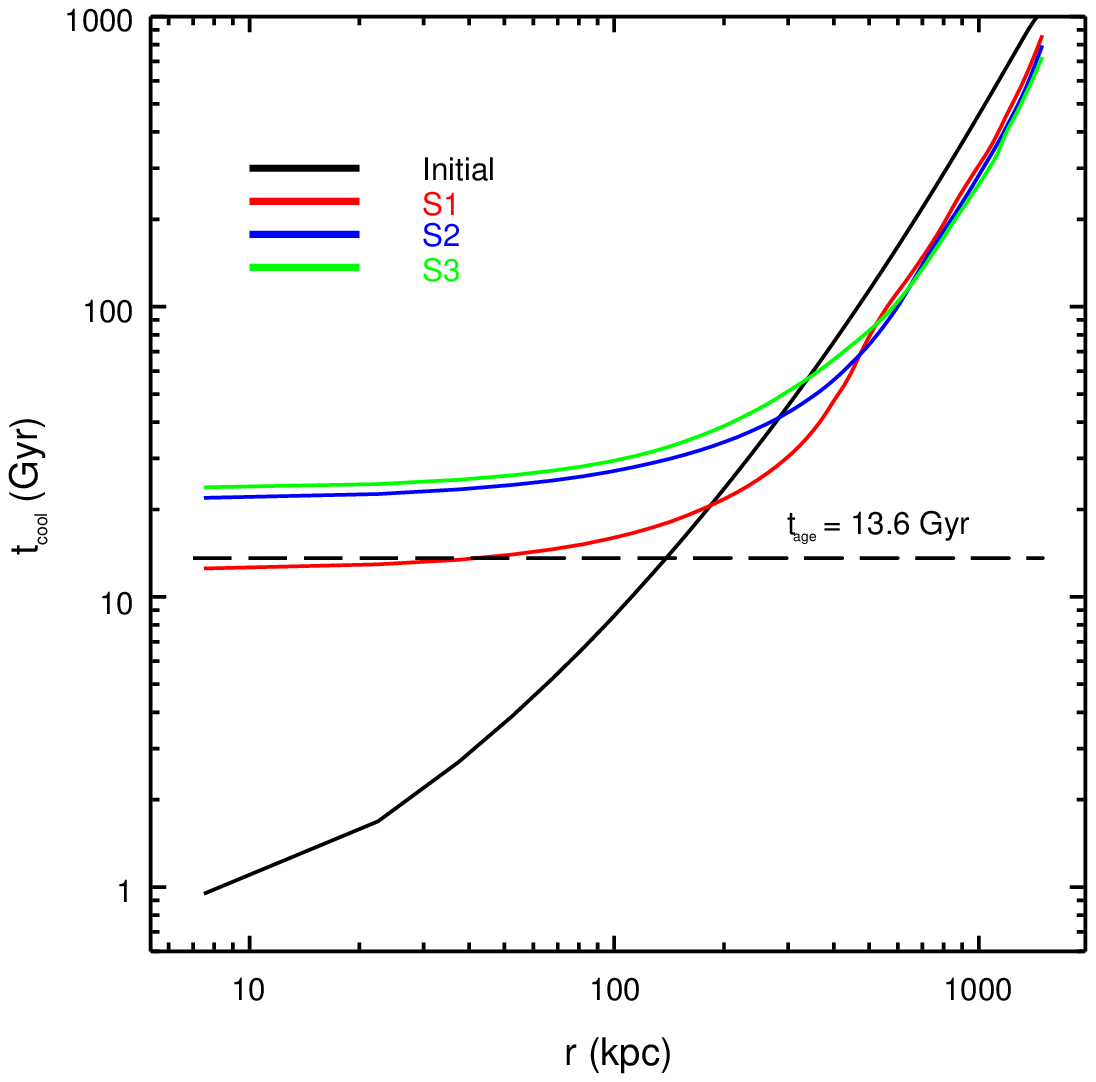}
\enspace
\includegraphics[width=0.3\textwidth]{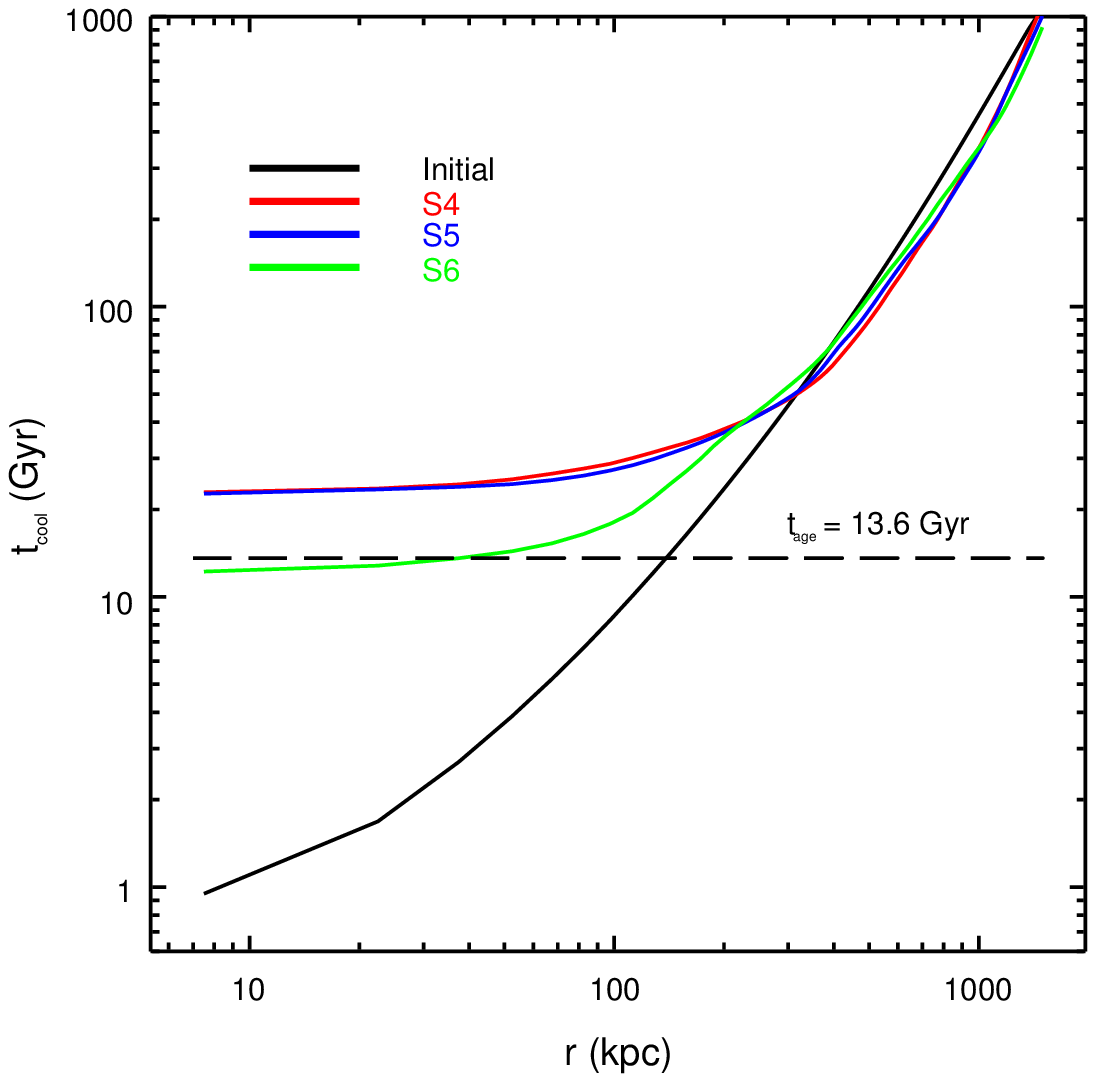}
\enspace
\includegraphics[width=0.3\textwidth]{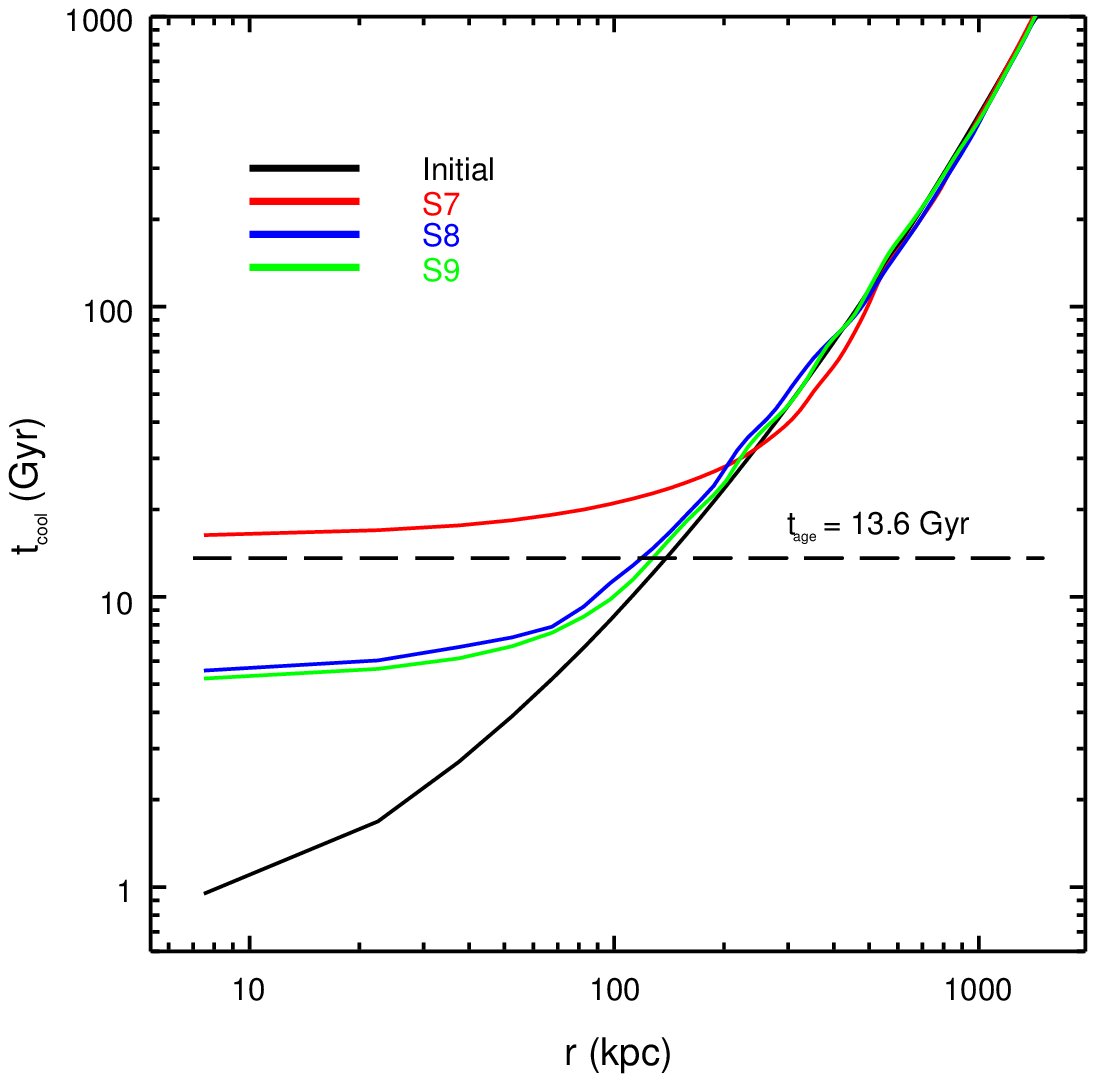}
\caption{Cooling time profiles at $t$ = 10 Gyr, compared to the initial profile. Left: Cooling time profiles for the 1:1 mass-ratio simulations. Center: Cooling time profiles for the 1:3 mass-ratio simulations. Right: Cooling time profiles for the 1:10 mass-ratio simulations.\label{fig:cool_profile}}
\end{center}
\end{figure*}

Our simulations demonstrate that significant mixing can occur as a result of major merging. Figures \ref{fig:entr_S1} through \ref{fig:entr_S9} demonstrate that fluid instabilities are ubiquitous throughout the duration of the merger, mixing the gas thoroughly. The presumed reason for the discrepancy is the choice of method to perform the simulations. The simulations of \citet{rit02} and \citet{poo08} were performed using SPH methods. A recent comparison of galaxy cluster merger simulations from a SPH code (GADGET-2) and a PPM code (FLASH) revealed that in the SPH simulations mixing of the two cluster components is inhibited, whereas in the PPM simulations the cluster components are thoroughly mixed \citep{mit08}. It has been demonstrated recently by \citet{age07} that spurious pressure forces can result at density gradients in SPH simulations, resulting from overestimations of the gas density of particles approaching a high-density region. As a result the growth of Rayleigh-Taylor and Kelvin-Helmholtz instabilities is inhibited. In addition, \citet{dol05} and \citet{wad08} showed that the artificial viscosity used in most SPH implementations acts to damp turbulent motions. Therefore, in a galaxy cluster merger simulation, both of these effects can inhibit the mixing of gas from the two cluster components. In contrast, PPM simulations such as ours produce the right results for this physical setup, noting the fact that in real clusters other physical mechanisms (such as viscosity and magnetic fields) may come into play. The results of our investigation over a parameter space of mergers were anticipated by the grid-based hydrodynamical simulations of \citet{tak05}. In their simplified setup of two rigid gravitational potentials, they showed that the gas from a subcluster falling into a large galaxy cluster will be ram-pressure stripped and mixed into the main cluster gas due to the action of Kelvin-Helmholtz and Rayleigh-Taylor instabilities. 

\begin{figure*}
\begin{center}
\plotone{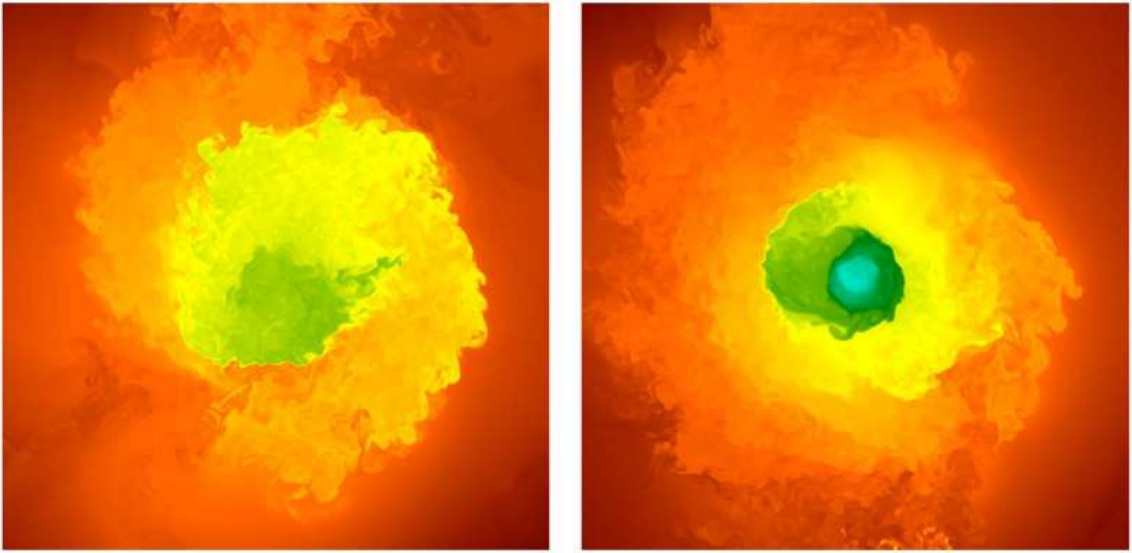}
\caption{Examples of core gas sloshing at late times in two of the offset mergers. The quantity plotted is gas entropy with the same color scale as Figures \ref{fig:entr_S1} through \ref{fig:entr_S9}. Each panel is 3~Mpc on a side. Left: Simulation S5. Right: Simulation S9.\label{fig:sloshing}}
\end{center}
\end{figure*}

Supporting this, we find a correspondence between the amount of mixing of the two cluster gas components and the velocity dispersion of the gas in the final merger remnants (Figures \ref{fig:mix_profile} and \ref{fig:turb_profile}). Where higher velocity dispersions have been generated in the cluster cores, there is a corresponding higher degree of mixing. Figures \ref{fig:entr_S1} through \ref{fig:entr_S9} showed the ubiquitous presence of instabilities as each merger proceeds, and it is these instabilities that seeded the random velocities that efficiently mix the gas. In corresponding SPH simulations the instabilities would be largely absent.

A caveat to these results is that despite the high resolution of these simulations, perturbations with wavelengths smaller than the smallest zone size ($\Delta{x} \approx 5h^{-1}$~kpc) are not resolved. Similarly, the turbulent cascade is not resolved below this length scale. It is possible that resolving these modes would enable more mixing of the gas than is present in these simulations. Simulations with higher resolution or subgrid models for turbulence \citep[see, e.g.][]{sca08, mai09}, would be necessary to determine whether or not this is the case.

\subsection{Destroying Cool Cores}\label{sec:bye_bye_cool_cores}

In all of the cases investigated here, the initially compact cool cores are completely destroyed as a result of the merger, and replaced with low-density, high-temperature, high-entropy cores. This is especially true in the cases where significant mixing of the cluster gases have occurred, such as each of the head-on mergers and all of the equal-mass cases. It is clear from Figures \ref{fig:entr_S1} through \ref{fig:entr_S9} that this entropy increase primarily occurs through the mixing of high and low-entropy gases caused by the onset of fluid instabilities. 

There is a correlation in our simulations between the height of the central entropy floor and the degree of mixing. In the equal-mass mergers, the inner $\sim$0.5~Mpc is very mixed, but there is a higher degree of mixing at larger radii in the off-center cases than in the head-on case. This is because in the off-center cases the cluster cores slip past each other and are stripped by instabilities, allowing for mixing over a large surface area, in contrast to the head-on case where the cores collide inelastically with each other and mixing only initially occurs at the boundary between the two clusters. Correspondingly, the entropy floor in simulations S2 and S3 is somewhat higher (nearly by a factor of 2) than in simulation S1, and the entropy floor has a larger extent. The differences in mixing are most evident in the 1:3 and the 1:10 mass-ratio mergers. In cases of zero impact parameter, the smaller, denser cluster (the secondary) penetrates the gas of the larger and allows for significant mixing of the central ICM of the two clusters, nearly 100$\%$ for the $R$ = 1:3, $b$ = 0~kpc case (simulation S4) and approximately 60$\%$ for the $R$ = 1:10, $b$ = 0~kpc case (simulation S7). It is in these cases where the greatest increase in core entropy is seen. However, in the off-center cases, most of the gas from the secondary is stripped as it proceeds along its trajectory, and so, though there is mixing in the cluster outskirts, the core of the primary is considerably less mixed with material from the other cluster. We still see a significant amount of mixing between the two components in simulation S5, so its entropy floor is comparable to that of S4, but the entropy floor of S6 is lower (by about a factor of two) where the core of the primary has been relatively unmixed with that of the secondary. In simulations S8 and S9 the inner $\sim$250~kpc has been relatively untouched, and they have roughly equal low entropy floors when compared to the merger product of simulation S7, which has been disrupted by the head-on collision.

It should be noted that our diagnostic for mixing, $M$, only indicates where material from the different clusters has mixed together. Even in the simulations where the central cores of the primary cluster remain relatively untouched by the gas from the secondary, mixing still plays an important role. The merger event sets of a process of ``sloshing'' of the center cluster gas in the dark matter potential \citep[as in][]{asc06}, bringing higher entropy gas (which is still from the main cluster) in from higher radii, which is then mixed with the lower entropy gas, raising the central regions to a higher adiabat (see Figure \ref{fig:sloshing} for examples of core gas sloshing from our simulations). An example from these simulations of this effect is seen in the 1:3 mass-ratio cases, where the entropy floors of simulations S4 and S5 have been raised to the same adiabat but the gas from the two clusters has mixed only $\sim$2/3 as much in the latter case. Although the mixing between the two cluster components was not complete, the central gas in this case still mixed with higher entropy gas from its own component, raising the entropy in the central regions. A detailed examination of this process in the context of an AMR code such as FLASH was carried out by \citet{zuh10}. 

In general, it has been shown that the results of non-radiative cosmological simulations depend on the numerical scheme adopted for hydrodynamics. Mesh-based Eulerian codes (such as FLASH) systematically produce higher entropy gas cores than do particle-based hydrodynamic approaches \citep[see, e.g.,][]{fre99,osh05,voit05c}. \citet{mit08} concluded that the extra gas mixing in the PPM simulations was responsible for this difference. In their simulation the time period of greatest difference in entropy generation between the PPM and SPH codes was the timescale of merger evolution during which large vortices and turbulent eddies were present in the PPM simulation, mixing the gas. Supporting this conclusion, works where the SPH algorithm has been modified to more properly resolve turbulent motions, instabilities, and mixing of the gas \citep[such as][]{dol05,kaw09} results in higher central entropy profiles in the merger remnants.

\citet{mit08} noted that the ability of mesh-based codes to produce high entropy floors may have implications for observations of cluster entropy profiles. They speculated that heating from recent merger events might be responsible for the observed clusters that do not possess cool, low entropy cores, but rather large high-entropy cores such as the final merger products seen in our simulations. Recent works \citep[e.g.][]{cav09,pra09} have clarified the existence of a bimodal distribution of cluster central entropies, with a large population of clusters possessing low-entropy cores ($S_0 \simlt 30$~keV~cm$^2$) and a similarly large population possessing high-entropy cores $S_0 \simgt 50$~keV~cm$^2$). The former set of clusters corresponds to cool-core systems much like our initial systems. The latter set of clusters may have had their core entropies raised by recent merging, and correspond to our resultant systems. \citet{pra09} pointed out a correlation between central gas mass fraction and central entropy, noting that the high-entropy floor clusters correspondingly had low central gas mass fractions. This is exactly what happens in each of our merger cases. This work indicates that mergers of clusters with even small subclusters could significantly raise the central entropy floor and central cooling time. Previous investigations using astrophysical codes where mixing was suppressed have indicated that merging was not able to sustain high central entropies and cooling times, but simulations with realistic mixing such as these suggest the opposite conclusion.

To demonstrate this conclusively, however, this work must be extended to more physically realistic setups. Mixing in the real ICM may not be as efficient or even more efficient than in our simulations due to additional physical processes that may be operational in the ICM. A real physical viscosity would act to dissipate turbulence and suppress fluid instabilities. Magnetic fields would only allow movement of the plasma along the field lines and would similarly partially suppress instabilities and mixing. Conduction, if efficient, would smooth out temperature, density, and velocity gradients, also suppressing instabilities \citep[see, e.g.][]{vie07}. On the other hand, a more accurate characterization of the turbulent cascade (by either performing simulations at higher resolution or by implementing a turbulent subgrid model) would likely increase the amount of mixing. Finally, to make a direct determination of the efficiency of heating generated from mergers cooling must be self-consistently included in the simulations. This would be necessary for a more direct comparison to the results of cosmological simulations. For example, \citet{bur08}, using a cosmological grid-based simulation with cooling and supernova feedback, have shown that cool cores are resilient to cluster mergers. In order to compare more directly with this result it would be necessary to include cooling in our simulations. Despite the limitations of the present work, it reinforces the conclusion of \citet{mit08} that merging remains a serious candidate for heating the ICM.

\subsection{Deriving Dark Matter Properties from Gas}\label{sec:dm_from_gas}

Depending on the degree to which dark matter can be understood to be ``collisionless'' (in the absence of other mechanisms such as self-interaction or annihilation), it is not expected to be subject to the fundamentally microscopic processes of heating and cooling that affects the plasma of the ICM. Consequently, though the form of the DM ``temperature'' radial profiles are remarkably stable and retain their essential form after a merger (Figure \ref{fig:tdm_profile}), the gas temperature profiles have been transformed from cool-core like profiles with positive temperature gradients into profiles with negative or flat temperature gradients after a merger (Figure \ref{fig:temp_profile}). It therefore can be expected that methods of determining dark matter properties by assuming equivalence between the dark matter ``temperature'' and the temperature of the gas will fail in the centers ($r \simlt 200$~kpc) of galaxy clusters. Figure \ref{fig:kappa_profile} confirms this expectation. In one sense, this reflects difficulties already known with using the thermodynamic state of the ICM to derive properties of clusters, in that the large differences in the central temperature and density structure between clusters (due to nongravitational heating of the gas such as AGN) must be accounted for in order to determine their effect on deriving quantities under the assumption of hydrostatic equilibrium. For example, global X-ray spectral temperatures of clusters are typically derived by excising the central emission out to some radius (typically $0.15r_{500}$, $0.1r_{200}$), in order to achieve a higher degree of similarity between temperature profiles \citep[e.g.,][]{vik06,vik09a}. However, we saw that as a result of merging the deviation of $\kappa$ from unity at the end of the simulation can be larger than 20\% even out to a radius of $r \approx 250$~kpc, depending on the strength of the merger. Attempts to derive the kinetic properties of dark matter from the properties of the X-ray-emitting gas must take this fact into account when applying this technique to clusters that show signs of recent merging. 

\section{Conclusions}\label{sec:conc}

We have performed a set of fiducial binary galaxy cluster merger simulations that explore a parameter space over initial mass ratio and impact parameter. In this first of a series of papers we focus on the mixing of the intracluster medium between the two clusters and the resulting effect on the thermodynamics of the gas of the merger remnant. In contrast to previous merger parameter space studies but in agreement with recent works focusing on mixing, we find that the gas of the merging clusters mixes very efficiently. The degree of mixing is dependent upon the mass ratio and impact parameter of the cluster merger, but all of the mergers investigated here demonstrate a significant amount of mixing in some regions of the final merger remnant. This is a result of ubiquitous presence of fluid instabilities that occur as the merger generates large shear flows and sharp density gradients, which are resolved in grid-based fluid codes such as the one used in this study but not in particle-based fluid codes used in many previous works. These instabilities result in the mixing of high and low-entropy gas and as a result the thermodynamic state of the primary cluster's core is dramatically affected. In each case the cluster has been changed from one with a high-density, low-entropy, low-temperature core with a short cooling time to a low-density, high-temperature, high-entropy core with a long cooling time. This is the case even for mergers with smaller subclusters. Further evidence that mixing is the culprit is indicated by the correlation between final merged systems with high central gas velocity dispersions and high central entropies, as the greater random motions in these systems allow for more efficient mixing. This result may also pose a difficulty for attempts to derive the kinetic properties of the dark matter from the thermodynamic properties of the X-ray emitting gas, which rely on the hypothesis of the near-equality of the gas temperature and dark matter velocity dispersion profiles. Additionally, such efficiency in heating the cluster core may be a primary factor in the generation of a population of galaxy clusters with high entropy floors and cooling times. Further simulation works including more realistic physics for the ICM (on the one hand, dissipative effects, magnetic fields, which would suppress mixing, and on the other hand, a more accurate characterization of the turbulence via higher resolution or a turbulent subgrid model which would enhance it) are needed to confirm this hypothesis. In spite of this, the main result of this work is that merging is still a serious candidate for the generation of heat to balance against cooling in the cores of galaxy clusters. 

\acknowledgements
JAZ is grateful to Don Lamb, Paul Ricker, Maxim Markevitch, Ryan Johnson, Eric Hallman, Ian Parrish, and Andrey Kravtsov for discussions and advice. Calculations were performed using the computational resources of Argonne National Laboratory and Lawrence Livermore National Laboratory.  JAZ is supported under {\it Chandra} grant GO8-9128X. The software used in this work was in part developed by the DOE-supported ASC / Alliance Center for Astrophysical Thermonuclear Flashes at the University of Chicago.

\end{document}